\documentclass[amsmath]{iucr}              
\usepackage{array}
\usepackage{amsmath}
\usepackage{longtable}
\usepackage{bm}
\usepackage{xr}
\usepackage[dvips]{graphics}
\usepackage{epic}
\usepackage{eepic}

\paperprodcode{a000000}      
\paperref{pz5003}            
\papertype{FA}               
\paperlang{english}          
\journalcode{A}              
\journalyr{2003}
\journaliss{0}
\journalvol{00}
\journalfirstpage{000}
\journallastpage{000}
\journalreceived{0 XXXXXXX 2003}
\journalaccepted{0 XXXXXXX 2003}
\journalonline{0 XXXXXXX 2003}

\newcommand\rf[1]{(\ref{#1})}       

\newcommand\kv{{\textbf k}}

\newcommand\zv{{\textbf z}}
\newcommand\cv{{\textbf c}}
\newcommand\hv{{\textbf h}}
\newcommand\az{{\textbf z+\textbf h}}

\renewcommand\b[1]{{\textbf b}^{(#1)}}
\newcommand\g{\gamma}
\renewcommand\d{\delta}
\newcommand\eps{\epsilon}
\newcommand\h{\eta}
\newcommand\G{\Gamma}
\newcommand\sv{{\textbf S}}
\newcommand\sx{{\textbf S}_x}
\newcommand\sy{{\textbf S}_y}
\newcommand\sz{{\textbf S}_z}
\newcommand\sxy{{\textbf S}_{xy}}
\newcommand\syz{{\textbf S}_{yz}}
\newcommand\szx{{\textbf S}_{zx}}
\newcommand\splus{{\textbf S}_{+}}
\newcommand\sminus{{\textbf S}_{-}}
\newcommand\GS{G_{\textbf S}}

\renewcommand\= {{\equiv}}

\newcommand\cross{\!\times\!}
\newcommand\plus{\!+\!}
\newcommand\xb{{\bar x}}
\newcommand\yb{{\bar y}}
\newcommand\zb{{\bar z}}

\newcommand\f[1]{{\Phi}_{r_#1}^{\delta}}
\newcommand\dg{\dagger}
\newcommand\fm{{\Phi}_{m}^{\mu}}
\newcommand\fd{{\Phi}_{d}^{\alpha}}
\newcommand\fh{{\Phi}_{h}^{\eta}}
\newcommand\fr{{\Phi}_{\bar{r}_8}^{\d}}

\newcommand\ppp{{P\plus p}}
\newcommand\ipp{{I\plus p}}
\newcommand\pcp{{P\cross p}}
\newcommand\icp{{I\cross p}}

\newcommand \bi{{\textbf b}^{(i)}}

\newcommand\feg{{\Phi}_{e}^{\gamma}}

\newcommand\rr{r_8}
\newcommand\rb{\bar{r}_8}
\newcommand\hf{\frac{1}{2}}
\newcommand\qt{\frac{1}{4}}

\newcommand\sr[1]{#1_\zb}
\newcommand\ssr[1]{{#1_\zb^*}}
\newcommand\ax[1]{#1^*2^\dagger2^{*\dagger}}

\newcommand{\PL}{P_{\cdots}^{\G_e}}
\newcommand{\SL}{S_{\cdots}^{\G_e}}
\newcommand{\fex}{\Phi_e^{2^*_\xb}(\bi\cv)}
\newcommand{\fey}{\Phi_e^{2^*_\yb}(\bi\cv)}
\newcommand{\fez}{\Phi_e^{2^*_\zb}(\bi\cv)}
\newcommand{\fezp}{\Phi_e^{2'_\zb}(\bi\cv)}
\newcommand{\fee}{\Phi_e^{\eps'}(\bi\cv)}
\newcommand{\fenz}{\Phi_e^{n^*_\zb}(\bi\cv)}
\newcommand{\mdmd}{\mu^{-1}\d\mu\d}
\newcommand{\adad}{\alpha^{-1}\d\alpha\d}
\newcommand{\dhdh}{\d^{-1}\h\d\h}
\newcommand{\hmhm}{\h^{-1}\mu\h\mu}
\newcommand{\dnd}{\d n_\zb\d^{-1}}
\newcommand{\dxd}{\d 2_\xb\d^{-1}}
\newcommand{\ana}{\alpha n_\zb\alpha^{-1}}
\newcommand{\axa}{\alpha 2_\xb\alpha^{-1}}
\newcommand{\al}{\alpha}
\newcommand{\ta}{\tilde{a}}
\newcommand{\tb}{\tilde{b}}

\begin{document}
\title{Symmetry of Magnetically Ordered Three Dimensional Octagonal
  Quasicrystals}
\shorttitle{Magnetic symmetry of octagonal quasicrystals}


\author{Shahar}{Even-Dar Mandel}
\cauthor{Ron}{Lifshitz}{ronlif@post.tau.ac.il}{}

\aff{School of Physics and Astronomy, Raymond and Beverly Sackler
  Faculty of Exact Sciences, Tel Aviv University, Tel Aviv 69978,
  \country{Israel}}


\shortauthor{Even-Dar Mandel \& Lifshitz}

\maketitle                        

\begin{synopsis}
  All 3-dimensional octagonal spin point groups and spin space-group
  types are enumerated and used to calculate selection rules for
  neutron scattering experiments.
\end{synopsis}

\begin{abstract}
  The theory of magnetic symmetry in quasicrystals, described in a
  companion paper [{\it Acta Crystallographica\/} A{\bf XX} (2003)
  xxx-xxx], is used to enumerate all 3-dimensional octagonal spin
  point groups and spin space-group types, and calculate the
  resulting selection rules for neutron diffraction experiments.
\end{abstract}

\section{Introduction}
\label{sec:3dintro}

We enumerate here all three-dimensional octagonal spin groups and
calculate their selection rules for neutron scattering, based on the
theory developed in a companion paper \cite{octa2d} where we have
provided the details for the extension to quasicrystals~\cite{prl98}
of Litvin and Opechowski's theory of spin
groups~\cite{litvin73,litvin74,litvin77}. We assume the reader is
familiar with the cmpanion paper, where we have also given
experimental as well as theoretical motivation for carrying out a
systematic enumeration of spin groups for quasicrystals.  This is the
first complete and rigorous enumeration of spin groups and calculation
of selection rules for any 3-dimensional quasicrystal.  Other than the
pedagogical example of 2-dimensional octagonal spin groups given in
the companion paper, past calculations are scarce.  Decagonal spin
point groups and spin space-group types in two dimensions have been
listed by \citeasnoun{icq5} without providing much detail regarding
the enumeration process. All possible lattice spin groups $\G_e$ for
icosahedral quasicrystals have been tabulated by \citeasnoun{prl98}
along with the selection rules that they impose, but a complete
enumeration of icosahedral spin groups was not given.  We intend to
continue the systematic enumeration of spin groups in future
publications by treating all the other common quasiperiodic crystal
systems (pentagonal, decagonal, dodecagonal, and icosahedral), though
we shall not provide complete details of the calculations as we do
here.

Although the enumeration in three dimensions is more elaborate, it
proceeds along the same lines as the two-dimensional example given in
the companion paper~\cite{octa2d}. Familiarity with the calculation of
ordinary (nonmagnetic) octagonal space groups~\cite{rmrw} may also
assist the reader in following the calculations performed here,
although knowledge of that calculation is not assumed. We begin in
Sec.~\ref{sec:background} with a description of the two rank-5
octagonal Bravais classes, a reminder of the octagonal point groups in
three dimensions (summarized in Table~\ref{tab:generators}), and a
summary of the effect of the different point group operations on the
generating vectors of the two lattice types
(Table~\ref{tab:symmetry}).  In Sec.~\ref{sec:point} we enumerate the
octagonal spin point groups by noting the restrictions (summarized in
Table~\ref{tab:normalsub}), imposed on their generators due to the
isomorphism between $G/G_\eps$ and $\G/\G_e$. We then proceed to
calculate the phase functions associated with the generators of the
spin point groups by making use of the group compatibility condition
\begin{equation}
  \label{eq:GCC}
  \forall (g,\g), (h,\eta)\in\GS:\ \Phi_{gh}^{\g\eta}(\kv) \=
  \Phi_g^\g(h\kv) + \Phi_h^\eta(\kv).
\end{equation}
We begin this in Sec.~\ref{sec:gammae} by calculating the
gauge-invariant phase functions $\feg(\kv)$ associated with the
lattice spin group $\G_e$. We then choose a gauge by a sequence of
gauge-transformations which we describe in Sec.~\ref{sec:gauge}. We
complete the calculation of the remaining phase functions, separately
for each octagonal point group, in Sec.~\ref{sec:fgs}. The resulting
spin space groups are listed, by point group and lattice type, in
Tables~\ref{tab:8V}--\ref{tab:8/mmmS}, using a generalized format in
which we do not explicitly identify the spin-space operations that are
paired with the generators of the point group. In
Sec.~\ref{sec:tables} we complete the enumeration, by making this
explicit identification, and introduce the notation used for octagonal
spin space groups. The actual identification of spin-space operations
is given in Appendix~\ref{sec:app-b} which is not included in this
publication but can be obtained via... [PRODUCTION EDITOR: please
indicate how one can obtain this appendix.]. We conclude in
Sec.~\ref{sec:selection_rules} by calculating the selection rules for
neutron diffraction experiments, imposed by the different octagonal
spin space-groups, which are summarized in
Tables~\ref{tab:sel-2}--\ref{tab:sel-n}.

\section{Background for the enumeration}
\label{sec:background}

\subsection{Three dimensional octagonal Bravais classes}

Recall that all rank-5 octagonal lattices in three dimensions fall
into two Bravais classes~\cite{mrrw}. Lattices of both types contain a
two-dimensional rank-4 {\it horizontal\/} sublattice in the plane
perpendicular to the unique {\it vertical\/} 8-fold axis. The
horizontal sublattice can be generated by four wave vectors
$\b1\ldots\b4$ of equal length, separated by angles of $\frac{\pi}{4}$
(as shown in Fig.~\ref{fig:star3d}). Throughout the paper we denote
horizontal generating vectors and their negatives by $\bi$, where the
index $i$ is taken modulo 8, and $\b{i}=-\b{i-4}$ if $i=5,6,7,$ or
$8$.

If the fifth generating vector $\cv$---which must be out of the
horizontal plane---is parallel to the 8-fold axis, the lattice is
called a {\it vertical lattice\/} or $V$-lattice (also called a
primitive or $P$-lattice). In this case we write $\cv=\zv$, to
emphasize that it is parallel to the 8-fold axis, and call $\cv$ a
{\it vertical stacking vector,} since the $V$-lattice can be viewed as
a vertical stacking of horizontal planes containing 2-dimensional
rank-4 octagonal lattices.

If the fifth generating vector $\cv$ contains both a vertical
component $\zv$ and a non-zero horizontal component $\hv$, then it is
called a {\it staggered stacking vector,} and the lattice is called a
{\it staggered lattice,} or $S$-lattice. One can show~\cite{mrrw}
that to within a rotation of the lattice, or the addition of a
horizontal lattice vector, the horizontal shift can be taken to have
the form
\begin{equation}
  \label{eq:shift}
  \hv=\frac{1}{2}\left(\b1+\b2+\b3-\b4\right).
\end{equation}
As shown in Fig.~\ref{fig:star3d}, $\hv$ lies halfway between the
generating vectors $\b1$ and $\b2$. Both lattice types are periodic
along the vertical axis with a period of one layer for $V$-lattices
and two layers for $S$-lattices.

\subsection{Three dimensional octagonal point groups and their generators}

There are seven octagonal point groups (geometric crystal classes) in
three dimensions, one of which ($\bar 8m2$) has two distinct
orientations with respect to both types of octagonal lattices, giving
rise to the eight octagonal arithmetic crystal classes, listed in the
first column of Table~\ref{tab:generators}. The point group generators,
listed in the second column of the Table, are an 8-fold rotation
$\rr$, an 8-fold roto-inversion $\rb = i\rr$ (where $i$ is the
three-dimensional inversion), a horizontal mirror $h$ whose invariant
plane is perpendicular to the the 8-fold axis, a vertical mirror $m$
whose invariant plane includes the 8-fold axis, and a 2-fold
(dihedral) axis $d$ perpendicular to the the 8-fold axis. The effect
of these point group operations on the generating vectors of the two
lattice types are summarized in Table~\ref{tab:symmetry}.

As shown in Fig.~\ref{fig:star3d}, the invariant planes of the
vertical mirrors and the axes of the 2-fold rotations can be oriented
either along (labeled $m$ and $d$) or between (labeled $m'$ and $d'$)
the 8-fold star, formed by the generating vectors of the octagonal
horizontal sublattice and their negatives. The point group $\bar 8m2$
has four mirrors and four 2-fold axes. If the mirrors are of type $m$
(along the star vectors) and the dihedral axes are of type $d'$
(between them) it is denoted by $\bar 8m2$. If the mirrors are of type
$m'$ (between the star vectors) and the dihedral axes are of type $d$
(along the star vectors) the point group is denoted by $\bar 82m$.

\section{Enumeration of spin point groups}
\label{sec:point}

As generators of the spin point group $\GS$ we take the generators of
the point group $G$, and combine each one with a representative
spin-space operation from the coset of $\G_e$ with which it is paired,
as listed in the second column of Table~\ref{tab:generators}.  The
spin-space operation paired in the spin point group with the 8-fold
generator $\rr$ or $\rb$ is denoted by $\d$, the operation paired with
the vertical mirror $m$ by $\mu$, the operation paired with
the horizontal mirror $h$ by $\eta$, and the spin-space operation
paired with the dihedral rotation $d$ is denoted by $\alpha$. To these
generators we add as many generators of the form $(e,\g_i)$ as
required, where $\g_i$ are the generators of $\G_e$ (three at the
most).

We begin by listing in Table~\ref{tab:normalsub} all the normal
subgroups $G_\eps$ of the seven octagonal point groups, along with the
resulting quotient groups $G/G_\eps$. The constraints on the
operations $\d$, $\mu$, $\eta$, and $\alpha$ due to the isomorphism
between $G/G_\eps$ and $\G/\G_e$, are summarized in the last column of
Table \ref{tab:normalsub}. The actual identification of these spin-space
operations is done at the last step of the enumeration process, as
described in Sec.~\ref{sec:tables}.

\section{Enumeration of lattice spin groups \textnormal{$\G_e$}}
\label{sec:gammae}

We calculate the phase functions $\Phi_e^\g(\kv)$, associated with
elements in the lattice spin group $\G_e$---keeping in mind that
$\G_e$ is abelian, and that no two phase functions $\Phi_e^\g(\kv)$
are identical---by finding the solutions to the constraints imposed on
these phase functions by all the other elements $\sigma\in\G$,
\begin{equation}\label{eq:gccGe}
  \forall \g\in\G_e,(g,\sigma)\in G_S:\quad
  \Phi_e^\g(\kv) \= \Phi_e^{\sigma\g\sigma^{-1}}(g\kv),
\end{equation}
by applying the group compatibility condition~\rf{eq:GCC} to the
relation $geg^{-1}=e$, where $g$ is one of the point group generators
$\rr$, $\rb$, $m$, $h$, and $d$.  The constraints imposed by the
different operations $\sigma$ are calculated in
Appendix~\ref{sec:app-a} of this paper which can be obtained via....
[PRODUCTION EDITOR: please indicate how one can obtain this
appendix.]. 


The results of these calculations are used below to enumerate the
lattice spin groups and calculate their phase functions. We also
introduce a notation---to be used in the final spin space-group
symbol---that encodes the the values of these phase functions. In this
notation the symbol of the lattice spin group $\G_e$ is added as a
superscript over the lattice symbol (unless $\G_e$ is $1$ or $1'$). In
addition, for each spin-space operation $\g$ in the symbol for $\G_e$
(with one exception when these operations are $2^*_\xb$ and $2^*_\yb$,
as noted below) we add a subscript to the lattice symbol describing
the sublattice of $L_\g$, defined by all wave vectors $\kv$ for which
$\feg(\kv)\=0$, with an additional index whenever this sublattice does
not uniquely describe the phase function. The results of this section
are summarized in the left-hand sides of
Tables~\ref{tab:sel-2}--\ref{tab:sel-2221'} and in the headings of
Table~\ref{tab:sel-n}.

There are a few cases in which the zeros of the phase functions,
associated with the operations $2^*_\xb$ and $2^*_\yb$, define a pair
of rank-5 {\it tetragonal\/} sublattices related by an 8-fold
rotation. Each wave vector in these tetragonal sublattices can be
decomposed into the sum of two vectors, one belonging to a
3-dimensional rank-3 tetragonal lattice of type $P$ or $I$, and the
second belonging to a 2-dimensional rank-2 square lattice. If the
squares of the rank-2 lattice are aligned along the same directions as
the squares of the rank-3 lattice then we denote the combined rank-5
lattice by $\ppp$ or $\ipp$; if they are aligned along the diagonals
of the rank-3 lattice we denote the combined rank-5 lattice by $\pcp$
(a lattice that does not occur here) or $\icp$. This is somewhat of an
{\it ad hoc\/} notation for rank-5 tetragonal lattices as compared,
for example, with the notation used by~\citeasnoun{hextrigarb} in
their enumeration of the analogous hexagonal and trigonal lattices of
arbitrary finite rank.  Nevertheless, it is more compact and
sufficient for our current purpose. Since the pair of sublattices,
associated with the operations $2^*_\xb$ and $2^*_\yb$, are identical
(to within an 8-fold rotation), and the different assignments of the
two sublattices to the two operations are scale equivelent, we denote
their symbol only once.

In all that follows we write $\Phi_g^\g(\bi)\=abcd$ instead of fully
writing $\Phi_g^\g(\b1)\=a$, $\Phi_g^\g(\b2)\=b$, $\Phi_g^\g(\b3)\=c$,
and $\Phi_g^\g(\b4)\=d$; we write $\Phi_g^\g(\bi\cv)\=abcde$ to
indicate in addition that $\Phi_g^\g(\cv)\=e$; and occasionally, if
the four phases on the horizontal generating vectors are equal to $a$
we write $\Phi_g^\g(\bi)\=a$ or $\Phi_g^\g(\bi\cv)\=a\ e$.

\subsection{\textnormal{$\G_e=1$}}

The trivial lattice spin group is always possible. In this case every
operation in $G$ is paired with a single operation in $\G$. The
lattice symbol remains the same as for nonmagnetic space groups: $P$
for vertical lattices, and $S$ for staggered lattices.

\subsection{\textnormal{$\G_e=2,2',1'$}}

Let $\g$ denote the single generator (2, $2'$, or $\eps'$) of $\G_e$.
Since $\g$ commutes with all elements of $\G$ we can infer the
possible values of the phase function $\feg$ from results R2, M1, D1,
and H1 of Appendix~\ref{sec:app-a}. We find that for all point groups
$G$
\begin{equation}
  \label{eq:2res}
  \feg(\bi\cv)\=
  \begin{cases}
    0\hf;\ \hf 0;\ \text{ or } \hf\hf & \text{$V$-lattice},\\
    0\hf & \text{$S$-lattice}.
  \end{cases}
\end{equation}
Because $\g$ is an operation of order 2, the zeros of its phase
function define a sublattice of index 2 in $L$. Let us express an
arbitrary wave vector in $L$ as $\kv=\sum_{i=1}^4 n_i\bi + l\cv$. The
zeros of the first solution $\feg(\bi\cv)\=0\hf$ define a vertical
octagonal sublattice, containing all wave vectors with even $l$, or
simply all the even layers of $L$, whether $L$ is a $V$-lattice or an
$S$-lattice. This solution is denoted on the two lattice types by
$P^\g_{2c}$ and $S^\g_P$, respectively. The second solution
$\feg(\bi\cv)\=\hf 0$ on the $V$-lattice defines a vertical octagonal
sublattice, containing all vectors with even $\sum_i n_i$, and is
denoted by $P^\g_P$. The third solution $\feg(\bi\cv)\=\hf\hf$ on the
$V$-lattice defines a staggered octagonal sublattice, containing all
vectors with even $\sum_i n_i + l$, and is denoted by $P^\g_S$. If
$\g=\eps'$ it is omitted from the symbols, yielding the same symbols
used for the so-called magnetic (or black-and-white) space
groups~\cite{rmp97}. The possible phase functions for $\G_e=2,2',1'$
are summarized in the left-hand side of Table~\ref{tab:sel-2}.

\subsection{\textnormal{$\G_e=21'$}}
\label{3dGe=21'}

Here $\G_e$ contains three operations of order 2 that commute with all
the elements of $\G$.  For $V$-lattices, there are $3!=6$
distinct ways of assigning the three different solutions \rf{eq:2res}
to the three phase functions of these operations. These are denoted by
$P^{21'}_{2c,P}$, $P^{21'}_{P,S}$, $P^{21'}_{S,2c}$, $P^{21'}_{P,2c}$,
$P^{21'}_{S,P}$, and $P^{21'}_{2c,S}$, where the first subscript
denotes the sublattice defined by the phase function $\Phi_e^2$, and
the second subscript denotes that of $\Phi_e^{\eps'}$. These solutions
are summarized in the left-hand side of Table~\ref{tab:sel-21'}.

For $S$-lattices $\G_e$ cannot be $21'$, because there is only one
solution in \rf{eq:2res}, and therefore there can be only one
operation in $\G_e$ that commutes with all elements of $\G$.

\subsection{\textnormal{$\G_e=222,2'2'2$}}

We generate $\G_e$ by $2^*_\xb$ ($2_\xb$ or $2'_\xb$) and $2^*_\yb$
($2_\yb$ or $2'_\yb$), noting that in the case of $\G_e=2'2'2$ the
unprimed 2-fold rotation defines the direction of the $\zb$ axis in
spin space.

For vertical lattices we see from results M3, D3, and H3 of
Appendix~\ref{sec:app-a} that the operations $\mu$, $\alpha$, and
$\eta$, paired with $m$, $d$, and $h$ if they are in the point group,
must all commute with $2^*_\xb$ and $2^*_\yb$. It only remains to
check whether $\d$, paired with the 8-fold generator, commutes with
the generators of $\G_e$.

If $\d$ commutes with $2^*_\xb$ and $2^*_\yb$ then again, as in the
previous section, we have three operations in $\G_e$ that commute with
with all the elements of $\G$, implying that $\G$ is orthorhombic and
not tetragonal. The difference between this case and that of the
previous section is that here different solutions can be related by
scale transformations that reorient the directions of the spin-space
axes. For $\G_e=222$, all three operations are equivalent so we can
define the directions of the spin-space axes such that
\begin{equation}
  \label{eq:222resV}
  \Phi_e^{2_\xb}(\bi\cv)\=0\frac12,\
  \Phi_e^{2_\yb}(\bi\cv)\=\frac12 0,\
  \Phi_e^{2_\zb}(\bi\cv)\=\frac12\frac12,
\end{equation}
and denote the lattice spin group and its associated phase functions
by $P^{222}_{2c,P,S}$. Any permutation of the three subscripts in the
symbol yields an alternative setting for the same scale-equivalence
class of solutions.

For $\G_e=2'2'2$, only the two primed rotations are equivalent, so we
have three distinct solutions, where we choose the directions of the
$\xb$ and $\yb$ axes in spin space such that the phase function
associated with $2'_\yb$ is $1/2$ on the horizontal generating
vectors,
\begin{eqnarray}
  \label{eq:22'2'resV}\nonumber
  1.\ \Phi_e^{2'_\xb}(\bi\cv)\=0\frac12,\
  \Phi_e^{2'_\yb}(\bi\cv)\=\frac12 0,\
  \Phi_e^{2_\zb}(\bi\cv)\=\frac12\frac12;\\
  2.\ \Phi_e^{2'_\xb}(\bi\cv)\=0\frac12,\
  \Phi_e^{2'_\yb}(\bi\cv)\=\frac12\frac12,\
  \Phi_e^{2_\zb}(\bi\cv)\=\frac12 0;\\\nonumber
  3.\ \Phi_e^{2'_\xb}(\bi\cv)\=\frac12 0,\
  \Phi_e^{2'_\yb}(\bi\cv)\=\frac12\frac12,\
  \Phi_e^{2_\zb}(\bi\cv)\=0\frac12.
\end{eqnarray}
These three solutions are denoted by $P^{2'2'2}_{2c,P,S}$,
$P^{222}_{2c,S,P}$, and $P^{222}_{P,S,2c}$, respectively. Their
scale-equivalent forms are obtained by exchanging the first two
subscripts: $P^{2'2'2}_{P,2c,S}$, $P^{222}_{S,2c,P}$, and
$P^{222}_{S,P,2c}$,

If $\d$ is a 4-fold rotation or a 2-fold diagonal rotation (requiring
$\G$ to be tetragonal) such that $\d{2^*_\xb}\d^{-1}=2^*_\yb$ then
according to result R4 there are two solutions for the phase functions
\begin{equation}
  \label{eq:222resVb}
  \Phi_e^{2^*_\xb}(\bi\cv)\=0\frac12 0\frac12 c,\
  \Phi_e^{2^*_\yb}(\bi\cv)\=\frac12 0\frac12 0 c,\quad
  c\=0 \text{ or } \hf.
\end{equation}
These are interesting solutions in which the horizontal planes of the
sublattices, defined by the zeros of the phase functions associated
with the operations $2^*_\xb$ and $2^*_\yb$, contain all vectors
$\kv=\sum_{i=1}^4 n_i\bi + l\cv$ from $L$ with even $n_2+n_4$, and
even $n_1+n_3$, respectively. Consequently the two sublattices are not
octagonal, but rather a pair of tetragonal lattices of rank 5, related
to each other by an 8-fold rotation. As described in the introduction
to this sections, we denote the lattice spin groups in this case by
$P^{2^*2^*2}_{\ppp,P}$ ($c\=0$) and $P^{2^*2^*2}_{\ipp,P}$
($c\=1/2$). Note that in both cases the sublattice defined by the
zeros of the phase function of $2_\zb$ is an octagonal $V$-lattice
with a thinned-out horizontal plane.

All these solutions on the $V$-lattice are valid for all octagonal
point groups because other generators of $\G$, if they exist, impose
no further restrictions on the phase functions above.

On staggered lattices (from results R2, M1, and D1 of
Appendix~\ref{sec:app-a}) only a single 2-fold operation can commute
with $\d$, $\mu$, or $\alpha$ that are paired with the 8-fold
generator, the mirror $m$, and the dihedral rotation $d$,
respectively. These operations therefore necessarily exchange the two
generators of $\G_e$, requiring $\G$ to be tetragonal. On the other
hand, (from result H3 of Appendix~\ref{sec:app-a}) $\eta$ which is
paired with $h$ necessarily commutes with the two generators of
$\G_e$. If these conditions are satisfied then for all point groups
the directions of the $\xb$ and $\yb$ axes in spin space can be chosen
so that there is a single solution for the phase functions,
\begin{equation}
  \label{eq:222resS}
  \Phi_e^{2^*_\xb}(\bi\cv)\=\frac12 0,\
  \Phi_e^{2^*_\yb}(\bi\cv)\=\frac12\frac12.
\end{equation}
The sublattices defined by the zeros of these two phase functions are
rank-5 tetragonal lattices of type $\icp$. One has the rank-3
$I$-lattice oriented in the directions of the $x$- and $y$-axes and
the other has the rank-3 $I$ lattice oriented along the diagonal
directions. The sublattice defined by the zeros of the phase function
of $2_\zb$ is an octagonal $V$-lattice. The lattice spin group is
therefore denoted by $S^{2^*2^*2}_{\icp,P}$.

The possible phase functions for $\G_e=222,2'2'2$
are summarized in the left-hand side of Table~\ref{tab:sel-222}.

\subsection{\textnormal{$\G_e=2'2'2'=2221'$}}

We choose to generate $\G_e$ using the three primed rotations and note
that since $\G_e$ has three 2-fold operations ($2_\zb$, $2'_\zb$, and
$\eps'$) that commute with all elements of $\G$, the lattice must be
vertical. Furthermore, since there can be no more than 3 operations
that commute with $\d$, all other 2-fold operations in $\G_e=2'2'2'$,
including the two generators $2'_\xb$ and $2'_\yb$, cannot commute
with $\d$ (requiring $\G$ to be tetragonal). The operations $\mu$,
$\eta$, and $\alpha$ must commute with all operations in $\G_e$.

If all these conditions are satisfied then the phase functions for
$2'_\xb$ and $2'_\yb$ have the same two solutions given in
Eq.~\rf{eq:222resVb} for $\G_e=2'2'2$. In both of these solutions
$\Phi_e^{2_\zb}(\bi\cv)\=\hf 0$. This leaves two possibilities
for the phase function of the third generator of $\G_e$,
\begin{equation}
  \label{eq:2'2'2'resS}
  \Phi_e^{2'_\zb}(\bi\cv)\=0\frac12 \text{ or } \frac12\frac12,
\end{equation}
giving a total of four distinct solutions for the phase functions of
$\G_e=2'2'2'$ on the $V$-lattice, denoted by $P^{2'2'2'}_{\ppp,2c}$,
$P^{2'2'2'}_{\ppp,S}$,$P^{2'2'2'}_{\ipp,2c}$, and
$P^{2'2'2'}_{\ipp,S}$ (and none on the $S$-lattice). These solutions
are summarized in the left-hand side of Table~\ref{tab:sel-2221'}.

\subsection{\textnormal{$\G_e=n,n',n1'\ (n>2)$}}

These lattice spin groups contain a single generator $\g$ of order
$N>2$, where $N=n$ unless $\G_e=n'$ and $n$ is odd, in which case
$N=2n$.  Note that if $n$ is odd then $n1'$ need not be considered
because it is the same as $n'$. For operations of order $N$ it follows
from the group compatibility condition~\rf{eq:GCC} that
\begin{equation}
  \label{eq:orderN}
  \Phi_e^{\g^N}(\kv) \= N\feg(\kv) \= 0 \Longrightarrow
  \feg(\kv) \=\frac{j}{N},\quad j=0, 1, \ldots N-1.
\end{equation}

For vertical lattices we find from result R3 of
Appendix~\ref{sec:app-a} that if the 8-fold generator is $\rr$ then
$\d$, paired with it, must commute with $\g$, and if it is $\rb$ then
$\d$ must be a perpendicular 2-fold rotation taking $\g$ to $\g^{-1}$.
Furthermore, we find from results M2, D2, and H2 of
Appendix~\ref{sec:app-a} that $\mu$ (paired with $m$) must commute
with $\g$, and $\alpha$ and $\eta$ (paired with $d$ and $h$) must both
be perpendicular 2-fold rotations. If these conditions are satisfied
whenever these operations are in the point group then
\begin{equation}
  \label{eq:resNVb}
  \Phi_e^\g(\bi)\=
  \begin{cases}
    aaaa & \text{$N$ even},\\
    0000 & \text{$N$ odd},
  \end{cases}
\end{equation}
where $a$ is either 0 or $1/2$. The only constraint on the phase
$\feg(\zv)$ comes from the requirement that $\g$ is an operation of
order $N$. If the in-plane phases are 0 then $\feg(\zv)\=j/N$, where
$j$ and $N$ must be co-prime, otherwise the true denominator is
smaller than $N$, and consequently the order of $\g$ is smaller than
$N$. If the in-plane phase $a\=1/2$ and $N$ is twice an even number,
then $j$ and $N$ must still be co-prime, but if $N$ is twice an odd
number then $j$ may be also even and $\g$ would still be an operation
of order $N$. This is so because even though the phase of $\g^{N/2}$
is zero on the stacking vector it is $1/2$ on the horizontal
generating vector, and therefore $\g^{N/2}\neq\eps$.

To summarize, for $\G_e=n$ (odd or even $n$) or $\G_e=n', n1'$ (even
$n$) the possible solutions for the phase function $\Phi_e^{n^*_\zb}$
are
\begin{equation}\label{eq:resnV}
    \Phi_e^{n^*_\zb}(\bi\cv)\=
    \begin{cases}
      0\frac{j}{n}
      & \text{$n$ odd},\\
      0\frac{j}{n}; \frac12\frac{j}{n}
      & \text{$n$ twice even},\\
      0\frac{j}{n}; \frac12\frac{j}{n};
      \frac12\frac{2j}{n}
      & \text{$n$ twice odd},
    \end{cases}
\end{equation}
where in all cases $j$ and $n$ are co-prime. The first subscript in
the symbols for the three distinct solutions for $\Phi_e^{n^*_\zb}$ are
$n_jc$ for $\Phi_e^{n^*_\zb}(\bi\cv)\=0\frac{j}{n}$, $n_jS$ for
$\Phi_e^{n^*_\zb}(\bi\cv)\=\hf\frac{j}{n}$, and $(\frac{n}{2})_jP$ for
$\Phi_e^{n^*_\zb}(\bi\cv)\=\hf\frac{2j}{n}$. The index $j$ is
necessary here because the sublattices defined by the zeros of these
phase functions do not depend on $j$, and therefore it must be
specified in addition to specifying the sublattice symbol.

For $\G_e=n'$ with odd $n$ the order of the generator is $2n$ and the
possible solutions are
\begin{equation}\label{eq:resnprimeV}
  \Phi_e^{n'_\zb}(\bi\cv)\= 0\frac{j}{2n}; \frac12\frac{j}{2n};
  \frac12\frac{j}{n}\qquad n \textnormal{ odd},
\end{equation}
where $j$ and $2n$ are co-prime.  The symbols for these three types of
lattice spin groups are $P^{n'}_{(2n)_jc}$, $P^{n'}_{(2n)_jS}$, and
$P^{n'}_{n_jP}$. Finally, for $\G_e=n1'$ ($n$
necessarily even), we need to find the possible solutions for the
phase function associated with the second generator $\eps'$. Since the
phase function $\Phi_e^{2_\zb}$ is determined by the phase function
$\Phi_e^{n_\zb}$ we are left with only two possible solutions for the
phase function $\Phi_e^{\eps'}$. If $n$ is twice even then we always
have
\begin{equation}
  \label{eq:resn1primeVa}
  \Phi_e^{\eps'}(\bi\cv)\= \hf 0 \text{ or } \hf\hf,
\end{equation}
thus obtaining a total of four solutions denoted $P^{n1'}_{n_jc,P}$,
$P^{n1'}_{n_jc,S}$, $P^{n1'}_{n_jS,P}$, and $P^{n1'}_{n_jS,S}$.
If $n$ is twice odd then the possible solutions depend on
$\Phi_e^{n_\zb}$ as follows:
\begin{equation}\label{eq:resn1primeVb}
  \Phi_e^{\eps'}(\bi\cv)\=
  \begin{cases}
    \hf 0 \text{ or } \hf\hf & \Phi_e^{n^*_\zb}(\bi\cv)\= 0\frac{j}{n},\\
    0\hf \text{ or } \hf 0
    & \Phi_e^{n^*_\zb}(\bi\cv)\= \frac12\frac{j}{n},\\
    \hf\hf \text{ or } 0\hf  & \Phi_e^{n^*_\zb}(\bi\cv)\= \frac12\frac{2j}{n},
  \end{cases}
\end{equation}
giving a total of six solutions denoted by $P^{n1'}_{n_jc,P}$,
$P^{n1'}_{n_jc,S}$, $P^{n1'}_{n_jS,2c}$, $P^{n1'}_{n_jS,P}$,
$P^{n1'}_{(\frac{n}{2})_jP,S}$, and $P^{n1'}_{(\frac{n}{2})_jP,2c}$.

On staggered lattices we first note that $\G_e$ cannot be $n1'$ which
contains three operations of order 2 that commute with all the
elements of $\G$. Furthermore, we find from the results of the
previous sections that $\eta$ (paired with $h$ if it is in $G$) must
be a 2-fold rotation perpendicular to the axis of $\g$ and that two
possibilities exist for the remaining generators of $\G$:

\noindent (1) If the 8-fold generator is $\rr$ then $\d$, paired with
it, must commute with $\g$, and if it is $\rb$ then $\d$ must be a
perpendicular 2-fold rotation taking $\g$ to $\g^{-1}$. The operation
$\mu$ (paired with $m$) must commute with $\g$, and $\alpha$ (paired
with $d$) must be a perpendicular 2-fold rotation. If these conditions
are satisfied whenever these operations are in $G$ then the in-plane
phases are all 0. For $\G_e=n$ (either parity) and $\G_e=n'$ (even
$n$) the possible solutions for the phase function $\Phi_e^{n^*_\zb}$
are therefore
\begin{equation}\label{eq:resnS1}
    \Phi_e^{n^*_\zb}(\bi\cv)\= 0\frac{j}{n},
\end{equation}
where $j$ and $n$ are co-prime. These are denoted by $S^{n^*}_{n_jc}$.
For $\G_e=n'$ with odd $n$ the order of the generator is $2n$ and the
possible solutions are
\begin{equation}\label{eq:resnprimeS}
  \Phi_e^{n'_\zb}(\bi\cv)\= 0\frac{j}{2n},
\end{equation}
where $j$ and $2n$ are co-prime. These are denoted by $S^{n'}_{(2n)_jc}$.

\noindent (2) If the 8-fold generator is $\rb$ then $\d$, paired with
it, must commute with $\g$, and if it is $\rr$ then $\d$ must be a
perpendicular 2-fold rotation taking $\g$ to $\g^{-1}$. The operation
$\alpha$ (paired with $d$) must commute with $\g$, and $\mu$ (paired
with $m$) must be a perpendicular 2-fold rotation. If these conditions
are satisfied whenever these operations are in $G$ then $n=4$ and the
possible solutions for the phase function $\Phi_e^{4^*_\zb}$ are
\begin{equation}\label{eq:resnS2}
    \Phi_e^{4^*_\zb}(\bi\cv)\= \hf\frac14 \text{ or } \hf\frac34,
\end{equation}
denoted by $S^{4^*}_{4_1S}$ and $S^{4^*}_{4_3S}$

The possible phase functions for $\G_e=n,n',n1'\ (n>2)$
are summarized in the headings of Table~\ref{tab:sel-n}.

\section{Initial choice of gauge}
\label{sec:gauge}

Before starting the actual calculation of phase functions for the
generators of the different octagonal point groups, we make an initial
choice of gauge that (i) sets the phases $\Phi_{g_8}^\d(\bi)$ to zero,
where $g_8$ is the 8-fold generator---$\rr$ for point groups $8$,
$8mm$, $822$, $8/m$, and $8/mmm$, or $\rb$ for $\bar8$, $\bar8m2$, and
$\bar82m$; and (ii) sets the phase $\Phi_{g}^\sigma(\cv)$ to zero for
one generator satisfying $g\zv=-\zv$.

\subsection{Setting \textnormal{$\Phi_{g_8}^\d(\bi)$} to zero}
\label{sec:chi1}

We can make $\Phi_{g_8}^\d(\bi)\=0$ with a gauge transformation given
by
\begin{equation}\label{eq:chi1}
\chi_1(\bi)\= \frac{1}{2}\Phi_{g_8}^\d\left(\bi \pm \b{i+1} + \b{i+2}
  \pm \b{i+3}\right),
\end{equation}
where the upper signs are for $g_8=\rr$ and the lower signs for
$g_8=\rb$, and the value of $\chi_1(\cv)$ is unimportant. Using this
gauge function we obtain
\begin{eqnarray}
  \Delta\Phi_{g_8}^\d(\bi) \= \chi_1\left(g_8\bi - \bi\right) \=
  \chi_1\left(\pm\b{i+1} - \bi\right)\nonumber\\
    \= \frac{1}{2}\Phi_{g_8}^\d\left(\b{i+4} - \bi\right)\equiv
    -\Phi_{g_8}^\d(\bi)
\end{eqnarray}
thereby setting $\Phi_{g_8}^\d(\bi)$ to zero.

\subsection{Setting \textnormal{$\Phi_g^\sigma(\cv)$} to zero when
  \textnormal{$g\zv=-\zv$}}
\label{sec:chi2}

We apply a second gauge transformation to set $\Phi_g^\sigma(\cv)$ to
zero for a single $g\in G$ for which $g\zv=-\zv$. We use this
transformation for the operation $\rb$ when $G$ is $\bar8$, $\bar
82m$, or $\bar 8m2$, and for the operations $d$ when $G$ is 822, and
$h$ when $G$ is $8/m$ or $8/mmm$. For both lattice types we take
\begin{equation}\label{eq:chi2}
  \chi_2(\bi)\=0,\quad \chi_2(\cv)=\frac{1}{2}\Phi_g^\sigma(\cv),
\end{equation}
from which we get, for vertical lattices
\begin{equation}
  \Delta\Phi_g^\sigma(\zv)\= \chi_2(g\zv-\zv)\= \chi_2(-2\zv) \=
  -\Phi_g^\sigma(\zv)
\end{equation}
and for staggered lattices
\begin{equation}
  \begin{split}
    \Delta\Phi_g^\sigma(\az)\equiv \chi_2(g(\az)-(\az))\qquad\qquad\\
    \= \chi_2\left((-2(\az) + (g\hv + \hv)\right) \=
    -\Phi_g^\sigma(\az),
  \end{split}
\end{equation}
where the last equality is due to the fact that the vector $g\hv+\hv$
is in the horizontal plane where $\chi_2$ is zero. Also note that
since $\chi_2$ is zero on the horizontal sublattice it does not affect
the phases that were set to zero in the previous section.

\subsection{Remaining gauge freedom}
\label{sec:chi3}

We still remain with some freedom to perform a third gauge
transformation without undermining the effect of the previous two
gauge transformations. Such a transformation can be performed with a
gauge function $\chi_3$ satisfying
\begin{equation}\label{eq:remain1}
  \Delta\Phi_{g_8}^\d(\bi)\=\chi_3(g_8\bi-\bi)\=0,
\end{equation}
and
\begin{equation}\label{eq:remain2}
  \Delta\Phi_g^\sigma(\cv)\=\chi_3(g\cv - \cv)\=0,
\end{equation}
where $g$ is the operation chosen in the previous section. The first
condition, for all point groups and both lattice types, requires
$\chi_3$ to have the same value of 0 or $1/2$ on all the horizontal
generating vectors.  For point groups 8 and $8mm$, where the second
gauge transformation $\chi_2$ is not used, there is no constraint on
the value of $\chi_3(\cv)$. For the remaining point groups we must
consider the two lattice types separately. For $V$-lattices $g\cv$ in
Eq.~\rf{eq:remain2} is $-\cv$ and therefore $\chi_3(\cv)\=0$ or $1/2$
independently of its value on the horizontal generating vectors. Thus,
on $V$-lattices one must check for gauge equivalence of solutions to
the group compatibility conditions using the three non-trivial gauge
functions
\begin{equation}\label{eq:chi3V}
  \chi_3(\bi\cv)\=0\hf;\ \hf 0; \text{ or } \hf\hf.
\end{equation}
For $S$-lattices, using Table~\ref{tab:symmetry}, and the fact that
$\chi_3(2\hv)\=0$, we find that we may still perform
gauge-transformations using the gauge functions
\begin{equation}\label{eq:chi3S}
  \chi_3(\bi\cv)\=
  \begin{cases}
    0\hf;\ \hf 0; \text{ or } \hf\hf;
    & G=8/m, 8/mmm,\\
    0\hf;\ \hf \frac14; \text{ or } \hf\frac34;
    & G=\bar8, \bar8m2, \bar82m, 822.
  \end{cases}
\end{equation}

\section{Determination of phase functions for the different
  octagonal point groups}
\label{sec:fgs}

Further calculations of phase functions are carried out for each point
group $G$ separately, based on its specific generating relations. Two
typical relations appear in most of the point groups.  All 2-fold
generators $(g,\sigma)\in G_S$ satisfy a generating relation of the
form $g^2=e$, and therefore impose, through the group compatibility
condition~\rf{eq:GCC}, equations of the form
\begin{equation}\label{eq:gccg^2}
  \Phi_e^{\sigma^2}(\kv)\=\Phi_g^{\sigma}(g\kv+\kv).
\end{equation}
In general $\sigma^2$ may be different from $\eps$ and therefore the
phase $\Phi_e^{\sigma^2}(\kv)$ is not necessarily zero. But note that
for any $\sigma\in\G$ the operation $\sigma^2$ is either $\eps$, or a pure
rotation about the $\zb$-axis in spin space, and therefore
$\Phi_e^{\sigma^2}(\bi)\=0000$ or $\hf\hf\hf\hf$.

Generating relations involving pairs of generators are of the form
$ghg=h$, where $(g,\sigma)$ and $(h,\tau)$ are in $G_S$. Application
of the group compatibility condition~\rf{eq:GCC} to these generating
relations yields
\begin{equation}\label{eq:gccghg}
  \Phi_h^{\sigma\tau\sigma}(\kv)\=
  \Phi_{ghg}^{\sigma\tau\sigma}(\kv)\=
  \Phi_g^\sigma(\kv+hg\kv)+\Phi_h^\tau(g\kv).
\end{equation}
We expand the left hand side by
\begin{equation}
  \Phi_h^{\sigma\tau\sigma}(\kv)\=
  \Phi_{h}^{\tau\tau^{-1}\sigma\tau\sigma}(\kv)\=
  \Phi_h^\tau(\kv)+\Phi_e^{\tau^{1}\sigma\tau\sigma}(\kv)
\end{equation}
to obtain
\begin{equation}\label{eq:gccghge}
  \Phi_e^{\tau^{-1}\sigma\tau\sigma}(\kv)\=
  \Phi_g^\sigma(\kv+hg\kv)+\Phi_h^\tau(g\kv-\kv).
\end{equation}
This last form, as well as Eq.~\rf{eq:gccg^2} for the 2-fold
generators, emphasizes the fact that the new phase functions that are
yet to be determined may depend on the phase functions $\feg(\kv)$
that were calculated in section~\ref{sec:gammae}. Note that
$\tau^{-1}\sigma\tau\sigma$ is the product of two conjugate operations
in $\G$ and is therefore either the identity $\eps$ or a pure rotation
about the $\zb$-axis in spin space, and therefore
$\Phi_e^{\tau^{-1}\sigma\tau\sigma}(\bi)\=0000$ or $\hf\hf\hf\hf$.

\subsection{Point group \textnormal{$G=8$} (generator \textnormal{$\rr$})}
\label{G=8}

The only phase  to be determined is $\f8(\cv)$ because the in-plane
phases of $\f8$ where set to zero by the gauge transformation of
section \ref{sec:chi1}. Successive applications of the group
compatibility condition to the generating relation $\rr^8=e$ yield
\begin{equation}\label{d^8}
  \Phi_e^{\d^8}(\cv)\=\Phi_{\rr}^\d(\cv+\rr\cv+\ldots+\rr^7\cv),
\end{equation}
where in general, $\d^8$ may be different from $\eps$ and therefore the
phase $\Phi_e^{\d^8}(\cv)$ is not necessarily zero. For vertical
lattices $\cv=\zv$ and  Eq.~\rf{d^8} becomes
\begin{equation}\label{d^8V}
  \Phi_e^{\d^8}(\zv)\=8\f8(\zv),
\end{equation}
with gauge-invariant solutions of the form
\begin{equation}\label{8Vsol}
  \f8(\zv)\=\frac{1}{8}\Phi_e^{\d^8}(\zv)+\frac{c}{8},\qquad
  c=0\ldots 7.
\end{equation}
For staggered lattices $\cv=\az$ and Eq.~\rf{d^8} becomes
\begin{equation}\label{d^8S}
  \Phi_e^{\d^8}(\az)\=\f8(8\zv)\=8\f8(\az),
\end{equation}
where the second equality follows from the fact that $8\hv$ is a
lattice vector in the horizontal plane for which $\f8(8\hv)\=0$.
The solutions of Eq.~\rf{d^8S} are
\begin{equation}
  \f8(\az)\=\frac{1}{8}\Phi_e^{\d^8}(\az)+\frac{c}{8},\qquad
  c=0\ldots 7,
\end{equation}
just as for the vertical stacking vector, but unlike the vertical
stacking vector, the phase $\f8(\az)$ is not gauge invariant because
\mbox{$\rr(\az)\neq(\az)$}. We need to check whether any of the
solutions are gauge-equivalent through the remaining gauge freedom
given by one of the gauge functions \rf{eq:chi3S}. Taking
$\chi_3(\bi)\=1/2$ changes the phase $\f8(\az)$ by
\begin{equation}
  \Delta\f8(\az)\= \chi_3(\rr(\az)-(\az))\= \chi_3(\b4)\= \frac12.
\end{equation}
Therefore, on the $S$-lattice two solutions differing by $1/2$ are
gauge-equivalent, so the distinct solutions are
\begin{equation}\label{8Ssol}
  \f8(\az)\= \frac18\Phi_e^{\d^8}(\az)+\frac{c'}{8}\qquad
  c'=0\ldots 3.
\end{equation}

The phase functions for point group $G=8$ are summarized in
Table~\ref{tab:8V} for $V$-lattices, and Table~\ref{tab:8S} for
$S$-lattices.

\subsection{Point group \textnormal{$G=\bar 8$} (generator
  \textnormal{$\rb$})}
\label{G=8bar}

The only phase function to be determined, $\fr$, is zero everywhere on
both lattice types due to the choice of gauge in sections
\ref{sec:chi1} and \ref{sec:chi2}. Note that in this case
\begin{equation}
  \Phi_e^{\d^8}(\kv)\=\Phi_{\rb^8}^{\d^8}(\kv)\=
  \fr(\kv+\rb\kv+\ldots\rb^7\kv)\=0,
\end{equation}
and therefore $\d^8=\eps$ whenever $\rb$ is a generator of $G$.

The phase functions for point group $G=\bar 8$ are summarized in
Table~\ref{tab:bar8V} for $V$-lattices, and Table~\ref{tab:bar8S} for
$S$-lattices.

\subsection{Point group \textnormal{$G=8mm$} (generators \textnormal{$\rr$}
and \textnormal{$m$})}
\label{G=8mm}

We need to determine the phase $\f8(\cv)$ and the phase function
$\fm(\kv)$. We use the generating relations $\rr^8=m^2=e$ and
$r_8mr_8=m$, which impose Eq.~\rf{d^8} for the 8-fold generator,
Eq.~\rf{eq:gccg^2} for the 2-fold generator, and Eq.~\rf{eq:gccghge}
for the additional generating relation $r_8mr_8=m$. We begin by noting
that if $m$ is the mirror that leaves $\b1$ invariant, then
application of Eq.~\rf{eq:gccg^2} to $\b3$ which is perpendicular to
$m$ (\mbox{$m\b3=-\b3$}) yields
\begin{equation}\label{8mm-mb3}
\Phi_e^{\mu^2}(\b3)\=\Phi_m^\mu(m\b3+\b3)\=0,
\end{equation}
implying that $\Phi_e^{\mu^2}(\bi)\=0000$. Application of
Eq.~\rf{eq:gccg^2} to $\b1$ then yields
\begin{equation}\label{8mm-mb1}
0\=2\Phi_m^\mu(\b1) \Longrightarrow
  \Phi_m^\mu(\b1) \= 0 {\rm \ or\ } \frac12,
\end{equation}
and application of Eq.~\rf{eq:gccg^2} to $\b2$ and $\b4$ shows that
$\Phi_m^\mu(\b2)\=\Phi_m^\mu(\b4)$, but provides no further
information regarding the actual values of these phases. Next,
we apply Eq.~\rf{eq:gccghge} to the horizontal generating
vectors to obtain
\begin{equation}\label{8mm-rmrb}
\Phi_m^\mu(\b{i+1})\= \Phi_m^\mu(\bi) + \Phi_e^{\mu^{-1}\d\mu\d}(\bi).
\end{equation}
Thus, the value of $\Phi_m^\mu$ on $\b1$ determines the values of
$\Phi_m^\mu$ on the remaining horizontal generating vectors through
the phase function $\Phi_e^{\mu^{-1}\d\mu\d}$:
\begin{equation}\label{8mm-mbi}
\Phi_m^\mu(\b{i}) \=
\begin{cases}
  0000 \text{\ or\ } \hf\hf\hf\hf
  & \text{if $\Phi_e^{\mu^{-1}\d\mu\d}(\bi)\=0000$},\\
  0\hf0\hf \text{\ or\ } \hf0\hf0
  & \text{if $\Phi_e^{\mu^{-1}\d\mu\d}(\bi)\=\hf\hf\hf\hf$}.
\end{cases}
\end{equation}

For the vertical stacking vector in $V$-lattices, for which
\mbox{$m\zv=\rr\zv=\zv$}, Eq.~\rf{d^8V} remains unchanged, and
Eqs.~\rf{eq:gccghge} and \rf{eq:gccg^2} become
\begin{subequations}\label{8mmV}
\begin{equation}\label{8mmV1}
\Phi_e^{\mu^{-1}\d\mu\d}(\zv)\equiv 2\f8(\zv)
\end{equation}
\begin{equation}\label{8mmV2}
\Phi_e^{\mu^2}(\zv)\equiv 2\Phi_m^\mu(\zv)
\end{equation}
\end{subequations}
The solutions to these equations are
\begin{subequations}\label{8mmVsol}
\begin{equation}\label{8mmVsol1}
\f8(\zv) \= \hf\Phi_e^{\mu^{-1}\d\mu\d}(\zv) + a,
\end{equation}
\begin{equation}\label{8mmVsol2}
\Phi_m^\mu(\zv) \= \hf \Phi_e^{\mu^2}(\zv) + b,
\end{equation}
\end{subequations}
where $a$ and $b$ are independently 0 or $1/2$.

For the staggered stacking vector in $S$-lattices, for which
\mbox{$\rr(\az)=\az+\b4$} and \mbox{$m(\az)=\az-\b3$} we obtain
Eq.~\rf{d^8S} together with
\begin{subequations}\label{8mmS}
\begin{equation}\label{8mmS1}
\Phi_e^{\mu^{-1}\d\mu\d}(\az)\equiv 2\f8(\az)+\fm(\b4)\\
\end{equation}
\begin{equation}\label{8mmS2}
\Phi_e^{\mu^2}(\az)\equiv 2\Phi_m^\mu(\az)-\fm(\b3)
\end{equation}
\end{subequations}
Noting that the signs of the phases $\fm(\bi)$ are unimportant, the
solutions to these equations are
\begin{subequations}\label{8mmSsol}
\begin{equation}\label{8mmSsol1}
\f8(\az) \= \hf \fm(\b4) + \hf\Phi_e^{\mu^{-1}\d\mu\d}(\az) + a,
\end{equation}
\begin{equation}\label{8mmSsol2}
\Phi_m^\mu(\az) \= \hf \fm(\b3) + \hf \Phi_e^{\mu^2}(\az) + b,
\end{equation}
\end{subequations}
where $a$ and $b$ are independently 0 or $1/2$, but we still need to check for gauge-eqivalence using the remaining gauge freedom given by the gauge
functions~\rf{eq:chi3S}. A gauge transformation with $\chi_3(\bi)\=1/2$ changes both of the phases in Eqs.~\rf{8mmSsol} by $1/2$, implying that the
two solutions with $ab\=00$ and $\hf\hf$ are gauge-equivalent and the two solutions $ab\=0\hf$ and $\hf 0$ are also gauge-equivalent. As
representatives of the gauge-equivalence classes we take the solutions with $a\=0$.

We finally note that Eqs.~\rf{d^8V} and \rf{d^8S} further imply that
for both lattice types
\begin{equation}\label{8mmVSsol3}
\Phi_e^{\d^8}(\cv) \= 4\Phi_e^{\mu^{-1}\d\mu\d}(\cv),
\end{equation}
which is true for the horizontal generating vectors as well and
therefore $\d^8=(\mu^{-1}\d\mu\d)^4$.

The phase functions for point group $G=8mm$ are summarized in
Table~\ref{tab:8mmV} for $V$-lattices, and Table~\ref{tab:8mmS} for
$S$-lattices.

\subsection{Point group \textnormal{$G=\bar 8m2$} (generators
  \textnormal{$\rb$} and \textnormal{$m$})}

Here we only need to determine the phase function $\fm(\kv)$ because
by the initial choice of gauge $\fr(\kv)\=0$.  We use the generating
relations $m^2=e$ and $\rb m\rb=m$, which through group compatibility
conditions of the form~\rf{eq:gccg^2} and \rf{eq:gccghge} yield
equations that resemble those for $G=8mm$ with $\rb$ replacing $\rr$.

For the horizontal generating vectors we again find that $\Phi_m^\mu$
has two possible solutions given by Eq.~\rf{8mm-mbi}.  For the
vertical stacking vector, for which $\rb\zv=-\zv$ and $m\zv=\zv$,
Eqs.~\rf{eq:gccg^2} and \rf{eq:gccghge} become
\begin{subequations}\label{8m2v}
\begin{equation}\label{8m2v1}
  \Phi_e^{\mu^{-1}\d\mu\d}(\zv)\=-2\fm(\zv),
\end{equation}
\begin{equation}\label{8m2v2}
  \Phi_e^{\mu^2}(\zv)\equiv 2\Phi_m^\mu(\zv).
\end{equation}
\end{subequations}
The solutions to these equations are
\begin{subequations}\label{8m2Vsol}
\begin{equation}\label{8m2Vsol1}
  \Phi_m^\mu(\zv) \= \hf \Phi_e^{\mu^2}(\zv) + a,\quad a\=0 \text{ or } \hf,
\end{equation}
with the additional condition that
\begin{equation}\label{8m2Vsol2}
  \Phi_e^{\mu^2}(\zv) + \Phi_e^{\mu^{-1}\d\mu\d}(\zv) \= 0.
\end{equation}
\end{subequations}

For the staggered stacking vector, for which $\rb(\az)=-(\az) - \b4$
and $m(\az)=(\az)-\b3$, Eqs.~\rf{eq:gccg^2} and \rf{eq:gccghge} become
\begin{subequations}\label{8m2S}
\begin{equation}\label{8m2S1}
  \Phi_e^{\mu^{-1}\d\mu\d}(\az)\equiv -2\fm(\az) -\fm(\b4)\\
\end{equation}
\begin{equation}\label{8m2S2}
  \Phi_e^{\mu^2}(\az)\equiv 2\Phi_m^\mu(\az)-\fm(\b3)
\end{equation}
\end{subequations}
The solutions to these equations are
\begin{subequations}\label{8m2Ssol}
\begin{equation}\label{8m2Ssol1}
  \Phi_m^\mu(\az) \= \hf\fm(\b3) + \hf \Phi_e^{\mu^2}(\az) + a,\quad
  a\=0 \text{ or } \hf,
\end{equation}
with the additional condition that
\begin{equation}\label{8m2Ssol2}
  \Phi_e^{\mu^2}(\az) + \Phi_e^{\mu^{-1}\d\mu\d}(\az) \= \fm(\b3) + \fm(\b4).
\end{equation}
\end{subequations}
A gauge transformation~\rf{eq:chi3S} with $\chi_3(\bi)\=1/2$
changes the phase $\Phi_m^\mu(\az)$ by $1/2$ and therefore the two
solutions in Eq.~\rf{8m2Ssol1} are gauge-equivalent. We take the one
with $a\=0$.

The phase functions for point group $G=\bar8m2$ are summarized in
Table~\ref{tab:bar8m2V} for $V$-lattices, and Table~\ref{tab:bar8m2S}
for $S$-lattices.

\subsection{Point group \textnormal{$G=822$} (generators \textnormal{$\rr$}
  and \textnormal{$d$})}

We need to determine the phase $\f8(\cv)$ and the phases $\fd(\bi)$---since
$\f8(\bi)\=\fd(\cv)\=0$ by our choice of gauge---using the generating
relations $\rr^8=d^2=e$ and $\rr d\rr=d$. The relation $\rr^8=e$
yields the same equation for $\f8(\cv)$ as in the case of point group
$G=8$, giving rise to the same solutions as those given by Eqs.~\rf{8Vsol}
and \rf{8Ssol}.

The determination of $\fd(\bi)$ is similar to that of $\fm(\bi)$ in
the case of point group $G=8mm$. If $d$ is the dihedral rotation that
leaves $\b1$ invariant, then application of Eq.~\rf{eq:gccg^2} to
$\b3$ which is perpendicular to $\b1$ yields
\begin{equation}\label{822-db3}
\Phi_e^{\alpha^2}(\b3)\=\fd(d\b3+\b3)\=0,
\end{equation}
implying that $\Phi_e^{\alpha^2}(\bi)\=0000$. Application of
Eq.~\rf{eq:gccg^2} to $\b1$ then yields
\begin{equation}\label{822-db1}
0\=2\fd(\b1) \Longrightarrow
  \fd(\b1) \= 0 {\rm \ or\ } \frac12,
\end{equation}
and application of Eq.~\rf{eq:gccg^2} to $\b2$ and $\b4$ shows that
$\fd(\b2)\=\fd(\b4)$, but provides no further
information regarding the actual values of these phases. Next,
we apply Eq.~\rf{eq:gccghge} to the horizontal generating
vectors to obtain
\begin{equation}\label{822-rdrb}
\fd(\b{i+1})\= \fd(\bi) + \Phi_e^{\alpha^{-1}\d\alpha\d}(\bi).
\end{equation}
Thus, $\fd(\b1)$ determines the values of $\fd$ on the remaining
horizontal generating vectors through the phase function
$\Phi_e^{\alpha^{-1}\d\alpha\d}$:
\begin{equation}\label{822-dbi}
\fd(\b{i}) \=
\begin{cases}
  0000 \text{\ or\ } \hf\hf\hf\hf
  & \text{if $\Phi_e^{\alpha^{-1}\d\alpha\d}(\bi)\=0000$},\\
  0\hf0\hf \text{\ or\ } \hf0\hf0
  & \text{if $\Phi_e^{\alpha^{-1}\d\alpha\d}(\bi)\=\hf\hf\hf\hf$}.
\end{cases}
\end{equation}

For the vertical stacking vector, for which $ d\zv=-\zv$, we obtain
\begin{subequations}
\begin{equation}\label{822V1}
\Phi_e^{\alpha^{-1}\d\alpha\d}(\zv)\=0,
\end{equation}
\begin{equation}\label{822V2}
\Phi_e^{\alpha^2}(\zv)\=0.
\end{equation}
\end{subequations}
The latter of the two, together with Eq.~\rf{822-db3}, implies that on
$V$-lattices $\alpha^2=\eps$.

For the staggered stacking vector, for which $d(\az)= -(\az) +2\hv
-\b3$, we obtain the additional constraints
\begin{subequations}\label{822S}
\begin{equation}\label{822S1}
\fd(\b4)\=\Phi_e^{\alpha^{-1}\d\alpha\d}(\az),
\end{equation}
\begin{equation}\label{822S2}
\fd(\b3)\=\Phi_e^{\alpha^2}(\az).
\end{equation}
\end{subequations}
>From Eq.~\rf{822S1} we find that $\Phi_e^{\alpha^{-1}\d\alpha\d}(\az)$
must be either 0 or $1/2$ and therefore that $\alpha^{-1}\d\alpha\d$
is either $\eps$ or $2_\zb$. In either case
$\Phi_e^{\alpha^{-1}\d\alpha\d}(\bi)\=0$ so only the first of
the two cases given in Eq.~\rf{822-dbi} is possible. This then implies
that the left-hand sides of Eqs.~\rf{822S} are equal and therefore
that $\alpha^{-1}\d\alpha\d=\alpha^2$. The values of $\fd$ on the
horizontal generating vectors for $S$-lattices are thus given by
\begin{equation}\label{822-dbiS}
\fd(\b{i}) \=
\begin{cases}
  0000
  & \text{if $\alpha^{-1}\d\alpha\d=\alpha^2=\eps$},\\
  \hf\hf\hf\hf
  & \text{if $\alpha^{-1}\d\alpha\d=\alpha^2=2_\zb$}.
\end{cases}
\end{equation}

The phase functions for point group $G=822$ are summarized in
Table~\ref{tab:822V} for $V$-lattices, and Table~\ref{tab:822S} for
$S$-lattices.

\subsection{Point group \textnormal{$G=\bar 82m$} (generators
  \textnormal{$\rb$} and \textnormal{$d$})}

We need to determine the phase function $\fd(\kv)$ because
we have chosen a gauge in which $\fr(\kv)\=0$.  We use the generating
relations $d^2=e$ and $\rb d\rb=d$, which through group compatibility
conditions of the form~\rf{eq:gccg^2} and \rf{eq:gccghge} yield
equations that resemble those for $G=822$ with $\rb$ replacing $\rr$.

For the horizontal generating vectors we again find that $\fd$ has two
possible solutions given by Eq.~\rf{822-dbi}, and that
$\Phi_e^{\alpha^2}(\bi)\=0000$.  For the vertical stacking vector, for
which $\rb\zv=d\zv=-\zv$, Eqs.~\rf{eq:gccg^2} and \rf{eq:gccghge}
become
\begin{subequations}\label{82mV}
\begin{equation}\label{82mV1}
  \Phi_e^{\alpha^{-1}\d\alpha\d}(\zv)\=-2\fd(\zv),
\end{equation}
\begin{equation}\label{82mV2}
  \Phi_e^{\alpha^2}(\zv)\equiv 0.
\end{equation}
\end{subequations}
The solutions to Eq.~\rf{82mV1} are
\begin{equation}\label{82mVsol1}
  \fd(\zv) \= -\hf \Phi_e^{\alpha^{-1}\d\alpha\d}(\zv) + a,\quad
  a\=0 \text{ or } \hf,
\end{equation}
and Eq.~\rf{82mV2} implies that on $V$-lattices $\alpha^2=\eps$.

For the staggered stacking vector, for which $\rb(\az)=-(\az) - \b4$
and $d(\az)=-(\az) +2\hv -\b3$, Eqs.~\rf{eq:gccg^2} and
\rf{eq:gccghge} become
\begin{subequations}\label{82mS}
\begin{equation}\label{82mS1}
  \Phi_e^{\alpha^{-1}\d\alpha\d}(\az)\equiv -2\fd(\az) -\fd(\b4),
\end{equation}
\begin{equation}\label{82mS2}
  \Phi_e^{\alpha^2}(\az)\equiv -\fd(\b3).
\end{equation}
\end{subequations}
The solutions to Eq.~\rf{82mS1} are
\begin{equation}\label{82mSsol1}
  \fd(\az) \= \hf\fd(\b4) - \hf \Phi_e^{\alpha^{-1}\d\alpha\d}(\az) + a,\quad
  a\=0 \text{ or } \hf,
\end{equation}
with the additional condition, given by Eq.~\rf{82mS2}, that
$\fd(\b3)$ is determind by $\alpha^2$, reducing the possible solutions
in~\rf{822-dbi} to one. Note that none of the gauge transformations
$\chi_3$ in Eq.~\rf{eq:chi3S} can change the phase $\fd(\az)$.

The phase functions for point group $G=\bar82m$ are summarized in
Table~\ref{tab:bar82mV} for $V$-lattices, and Table~\ref{tab:bar82mS}
for $S$-lattices.

\subsection{Point group \textnormal{$G=8/m$} (generators \textnormal{$\rr$}
  and \textnormal{$h$})}

We are using a gauge in which $\f8(\bi)\=\fh(\cv)\=0$, and therefore
need to determine the phases $\fh(\bi)$ and $\f8(\cv)$ using the
generating relations $\rr^8=h^2=e$ and $h\rr h=\rr$.
Equations~\rf{eq:gccghge} and \rf{eq:gccg^2} for the horizontal
generating vectors are
\begin{subequations}\label{8/mip}
\begin{equation}\label{8/mip1}
  \fh(\bi+\b{i+1})\=\Phi_e^{\d^{-1}\eta\d\eta}(\bi),
\end{equation}
\begin{equation}\label{8/mip2}
  2\fh(\bi)\=\Phi_e^{\eta^2}(\bi).
\end{equation}
\end{subequations}
Due to the fact that $\Phi_e^{\d^{-1}\eta\d\eta}(\bi)\=0000$ or
$\hf\hf\hf\hf$, application of Eq.~\rf{8/mip1} to successive
horizontal generating vectors establishes that $\fh(-\bi)\=\fh(\bi)$,
and therefore that the in-plane phases are given by
\begin{equation}\label{8/m-hbi}
  \fh(\b{i}) \=
  \begin{cases}
    0000 \text{\ or\ } \hf\hf\hf\hf
    & \text{if $\Phi_e^{\d^{-1}\eta\d\eta}(\bi)\=0000$},\\
    0\hf0\hf \text{\ or\ } \hf0\hf0
    & \text{if $\Phi_e^{\d^{-1}\eta\d\eta}(\bi)\=\hf\hf\hf\hf$}.
  \end{cases}
\end{equation}
Eq.~\rf{8/mip2} then implies that $\Phi_e^{\eta^2}(\bi)\=0000$.
For the vertical stacking vector, for which $h\zv=-\zv$,
the equations are
\begin{subequations}\label{8/mV}
\begin{equation}\label{8/mV1}
  -2\f8(\zv)\=\Phi_e^{\d^{-1}\eta\d\eta}(\zv)
\end{equation}
\begin{equation}\label{8/mV2}
  \Phi_e^{\eta^2}(\zv)\=0
\end{equation}
\end{subequations}
The solutions to Eq.~\rf{8/mV1} are
\begin{equation}\label{8/mVsol1}
  \f8(\zv) \= -\hf \Phi_e^{\d^{-1}\eta\d\eta}(\zv) + c,\quad
  c\=0 \text{ or } \hf,
\end{equation}
and Eq.~\rf{8/mV2} implies that on $V$-lattices $\eta^2=\eps$.

For the staggered stacking vector, for which $h(\az)=-(\az) + 2\hv$,
and using the fact that $\f8(2\hv)\=\fh(2\rr\hv)\=0$, the equations are
\begin{subequations}\label{8/mS}
\begin{equation}\label{8/mS1}
  -\fh(\b4)-2\f8(\az)\=\Phi_e^{\d^{-1}\eta\d\eta}(\az)
\end{equation}
\begin{equation}\label{8/mS2}
  \Phi_e^{\eta^2}(\az)\=0.
\end{equation}
\end{subequations}
The solutions to Eq.~\rf{8/mS1} are
\begin{equation}\label{8/mSsol1}
  \f8(\az) \= \hf\fh(\b4) - \hf \Phi_e^{\d^{-1}\eta\d\eta}(\az) + c,\quad
  c\=0 \text{ or } \hf,
\end{equation}
but a gauge transformation~\rf{eq:chi3S} with $\chi_3(\bi)\=1/2$ shows
that the two solutions are gauge-equivalent, allowing us to take
$c\=0$. Eq.~\rf{8/mS2} implies that $\eta^2=\eps$ on $S$-lattices, as
well.  Finally, the generating relation $\rr^8=e$ imposes Eq.~\rf{d^8}
as in the case of point group $G=8$, for both lattice types. Together
with Eqs.~\rf{8/mV1} and \rf{8/mS1} this implies that
\begin{equation}\label{8/mVSsol}
\Phi_e^{\d^8}(\cv) \= -4\Phi_e^{\d^{-1}\eta\d\eta}(\cv),
\end{equation}
which is true for the horizontal generating vectors as well, and
therefore $\d^8=(\d^{-1}\eta\d\eta)^{-4}$.

The phase functions for point group $G=8/m$ are summarized in
Table~\ref{tab:8/mV} for $V$-lattices, and Table~\ref{tab:8/mS} for
$S$-lattices.

\subsection{Point group \textnormal{$G=8/mmm$} (generators
  \textnormal{$\rr$}, \textnormal{$m$}, and \textnormal{$h$})}

We need to determine the phase $\f8(\cv)$, the phases $\fh(\bi)$, and
the complete phase function $\fm(\kv)$. The generating relations and
the equations they impose are the same as those for $G=8mm$ and those
for $G=8/m$, with the additional condition imposed by the generating
relation $mhm=h$ which yields
\begin{subequations}\label{8/mmm}
\begin{equation}\label{8/mmm1}
    \Phi_e^{\eta^{-1}\mu\eta\mu}(\bi)\=2\fm(\bi)\=0,
\end{equation}
\begin{equation}\label{8/mmm2}
  \Phi_e^{\eta^{-1}\mu\eta\mu}(\zv)\=0,
\end{equation}
\begin{equation}\label{8/mmm3}
  \Phi_e^{\eta^{-1}\mu\eta\mu}(\az)\=\fm(\b1) - \fh(\b3),
\end{equation}
\end{subequations}
for the horizontal generating vectors, vertical stacking vector, and
staggered stacking vector, respectively.

For $V$-lattices Eqs.~\rf{8/mmm1} and \rf{8/mmm2} imply that
$\eta^{-1}\mu\eta\mu=\eps$ but do not impose any additional
constraints on the phase functions already determined for point groups
$G=8mm$ and $G=8/m$. Thus the solutions for point group $G=8/mmm$ are
simply the combination of the two, so $\fh(\bi)$ is given by
Eq.~\rf{8/m-hbi}, $\fm(\bi)$ by Eq.~\rf{8mm-mbi}, and $\f8(\zv)$ and
$\fm(\zv)$ by Eqs.~\rf{8mmVsol}, with the additional conditions that
\begin{equation}\label{8/mmmVcond}
\Phi_e^{\mu^{-1}\d\mu\d}(\zv) + \Phi_e^{\d^{-1}\eta\d\eta}(\zv) \=0,
\end{equation}
$\Phi_e^{\mu^2}(\bi)\=0000$, $\eta^2=\eta^{-1}\mu\eta\mu=\eps$, and $\d^8 =
(\mu^{-1}\d\mu\d)^4 = (\d^{-1}\eta\d\eta)^{-4}$.

For $S$-lattices Eq.~\rf{8/mmm3} implies that
$\Phi_e^{\eta^{-1}\mu\eta\mu}(\az)$ can be either 0 or $1/2$ and
therefore that $\eta^{-1}\mu\eta\mu$ is either $\eps$ or $2_\zb$. It
also imposes a constraint between the two phase functions $\fm$ and
$\fh$. Therefore, on $S$-lattices $\fh(\bi)$ is again given by
Eq.~\rf{8/m-hbi}, but
\begin{equation}\label{8/mmmSmbi}
  \fm(\bi) \= \fh(\bi) + \Phi_e^{\eta^{-1}\mu\eta\mu}(\az),
\end{equation}
and $\f8(\az)$ and $\fm(\az)$ are given by Eqs.~\rf{8mmSsol}, with the
additional conditions that
\begin{subequations}
\begin{equation}\label{8/mmmScond1}
  \Phi_e^{\mu^{-1}\d\mu\d}(\bi) \= \Phi_e^{\d^{-1}\eta\d\eta}(\bi),
\end{equation}
\begin{equation}\label{8/mmmScond2}
  \Phi_e^{\mu^{-1}\d\mu\d}(\az) + \Phi_e^{\d^{-1}\eta\d\eta}(\az)
  \= \Phi_e^{\eta^{-1}\mu\eta\mu}(\az),
\end{equation}
$\Phi_e^{\mu^2}(\bi)\=\Phi_e^{\eta^{-1}\mu\eta\mu}(\bi)\=0000$,
$\Phi_e^{\eta^{-1}\mu\eta\mu}(\az)\=0$ or $1/2$ depending on whether
$\eta^{-1}\mu\eta\mu$ is $\eps$ or $2_\zb$, $\eta^2=\eps$, and $\d^8 =
(\mu^{-1}\d\mu\d)^4 = (\d^{-1}\eta\d\eta)^{-4}$.
\end{subequations}

The phase functions for point group $G=8/mmm$ are summarized in
Table~\ref{tab:8/mmmV} for $V$-lattices, and Table~\ref{tab:8/mmmS}
for $S$-lattices.

\section{Spin group tables}
\label{sec:tables}

Tables~\ref{tab:8V}-\ref{tab:8/mmmS} contain in compact form all the
information needed to generate the complete list of octagonal spin
space-group types. All that is still required is to explicitly
identify the spin-space operations $\d$, and whenever necessary $\mu$,
$\alpha$, and $\eta$, recalling that these operations must also
satisfy the constraints, summarized in Table~\ref{tab:normalsub}, that
are due to the isomorphism between $G/G_\eps$ and $\G/\G_e$. Once
these operations are identified their different combinations that
determine the phase functions---$\d^8$, $\d\g\d^{-1}$,
$\mu^{-1}\d\mu\d$, etc.---can be calculated to give the different spin
space-group types.  The explicit identification of spin-space
operations is listed in Appendix~\ref{sec:app-b} which can be obtained
via... [PRODUCTION EDITOR: please indicate how one can obtain this
appendix.]. For each spin space-group Table
(Tables~\ref{tab:8V}-\ref{tab:8/mmmS}) there is a corresponding Table
in Appendix~\ref{sec:app-b} which lists all the possible
identifications of spin-space operations, and for each one, indicates
the line in the spin space-group Table to which it corresponds.

To each spin space-group type we give a unique symbol based on the
familiar International (Hermann-Mauguin) symbols for the regular
(nonmagnetic) space groups.  To incorporate all the spin space-group
information we augment the regular symbol, which for the case of
octagonal quasicrystals is explained in detail by~\citeasnoun{rmrw},
in the following ways: (1) The symbol for the lattice spin group
$\G_e$ is added as a superscript over the lattice symbol, unless
$\G_e=1$ or $1'$.  (2) The values of the phase functions, associated
with the elements of $\G_e$, are encoded by subscripts under the
lattice symbol, describing the sublattices defined by the zeros of the
phase functions for the operations in the symbol for $\G_e$ (as
explained in Sec.~\ref{sec:gammae}).  (3) To each generator of
the point group $G$ we add a superscript listing an operation from the
coset of $\G_e$ with which it is paired (if that operation can be
$\eps$ we omit it, if it can be $\eps'$ we simply add a prime, and we
omit the axis about which rotations are performed if it is the
$\zb$-axis).  (4) The values of phase functions for the spin point
group generators $\f8$, $\fr$, $\fm$, $\fd$, and $\fh$ are encoded by
making changes and adding subscripts to the point group symbol
(similar to the way it is done for the regular space groups), as
described in the captions of Tables~\ref{tab:8V}-\ref{tab:8/mmmS},
where we use the same notation, simply without explicitly listing
$\G_e$ and its associated phase functions, and without explicitly
identifying the spin-space operations paired with the point group
generators. Specific examples of spin space-group symbols can be found
in the captions of Tables \ref{tab:8V}, \ref{tab:8S}, \ref{tab:8mmV}
and \ref{tab:8/mmmS}.

\section{Magnetic selection rules}
\label{sec:selection_rules}

We calculate the symmetry-imposed constraints on $\sv(\kv)$, for any
wave vector $\kv=\sum_{i=1}^4 n_i\bi+l\cv$ in the magnetic lattice
$L$, as described in the companion paper~\cite{octa2d}, by considering
all spin point group operations $(g,\g)$ for which $g\kv=\kv$. The
point group condition for each of these operations provides an
eigenvalue equation
\begin{equation}
  \label{eq:eigenvalue}
  \g\sv(\kv)=e^{-2\pi i\Phi_g^\g(\kv)} \sv(\kv)
\end{equation}
from which we obtain the constraints on $\sv$. These constraints may
require $\sv(\kv)$ to vanish or to take a particular form which
transforms under the operations $\g$ in~\rf{eq:eigenvalue} according
to the 1-dimensional representation dictated by the phases
$\Phi_g^\g(\kv)$. When there are no constraints then
$\sv=(S_\xb,S_\yb,S_\zb)$ is an arbitrary 3-component axial vector.
When there are constraints, $\sv(\kv)$ takes one of the following
forms:
\begin{equation}
  \label{eq:sforms}
  \begin{array}{lll}
    \sx=(S_\xb,0,0), & \sy=(0,S_\yb,0), & \sz=(0,0,S_\zb),\\
    \syz=(0,S_\yb,S_\zb), &  \szx=(S_\xb,0,S_\zb), &  \sxy=(S_\xb,S_\yb,0),\\
    \splus=(S_\xb,+iS_\xb,0), & \sminus=(S_\xb,-iS_\xb,0), &\\
  \end{array}
\end{equation}
as explained below.

In section~\ref{sel-gammae} we determine the selection rules due
elements in $\G_e$. These affect all wave vectors $\kv\in L$, and are
summarized in Tables~\ref{tab:sel-2}--\ref{tab:sel-n}. In
section~\ref{sel-invariant} we determine the remaining selection rules
for wave vectors that lie in the invariant subspaces of the different
point group elements. These selection rules are expressed in a general
manner in terms of the spin-space operations paired with these point
group elements, without identifying them explicitly.

\subsection{Calculation of selection rules due to \textnormal{$\G_e$}}
\label{sel-gammae}

If $\g$ is an operation of order 2 then its phases $\feg(\kv)$ are
either 0 or $1/2$. In this case Eq.~\rf{eq:eigenvalue} reduces to
\begin{equation}
  \label{eq:sel-2}
  \g\sv(\kv)=
  \begin{cases}
    \sv(\kv) & \text{if $\feg(\kv)\=0$},\\
    -\sv(\kv) & \text{if $\feg(\kv)\=\hf$},
  \end{cases}
\end{equation}
so if $\feg(\kv)\=0$, $\sv(\kv)$ must be invariant under $\g$, and
if $\feg(\kv)\=1/2$ it must change its sign under $\g$. This implies
different constraints on the form of $\sv(\kv)$ depending on the
particular type of operation of order 2:
\begin{enumerate}
\item If $\g$ is the time inversion $\eps'$, then $\sv(\kv)=0$ if
  $\feg(\kv)\=0$ and there are no constraints on $\sv(\kv)$ if
  $\feg(\kv)\=1/2$.
\item If $\g$ is a 2-fold rotation $2_\zb$, then $\sv(\kv)=\sz$ if
  $\feg(\kv)\=0$ and $\sv(\kv)=\sxy$ if $\feg(\kv)\=1/2$. Similar
  constraints are obtained for rotations about the other two axes.
\item If $\g$ is a 2-fold rotation followed by time inversion
  $2'_\zb$, then $\sv(\kv)=\sxy$ if $\feg(\kv)\=0$ and $\sv(\kv)=\sz$
  if $\feg(\kv)\=1/2$. Similar constraints are obtained for primed
  rotations about the other two axes.
\end{enumerate}
Thus, for lattice spin groups $\G_e$ containing only operations of
order 2 the form of $\sv(\kv)$ depends on whether the phases at
$\kv=\sum_{i=1}^4 n_i\bi+l\cv$, associated with all the operations,
are 0 or $1/2$. This can easily be calculated for each $\G_e$ from the
values of its phase functions on the lattice generating vectors. The
outcome depends, at most, on the parities of $n_1+n_3$, $n_2+n_4$,
and $l$. Tables~\ref{tab:sel-2} - \ref{tab:sel-2221'} summarize the
results for all such $\G_e$.

For operations of order greater than 2 we find that
\begin{enumerate}
\item[4.] If $\g$ is an $n$-fold rotation $n_\zb$ ($n>2$) then
  Eq.~\rf{eq:eigenvalue} requires $\sv(\kv)$ to acquire the phase
  $2\pi\Phi_e^{n_\zb}(\kv)$ upon application of the $n$-fold rotation. One
  can directly verify that the only possible forms that $\sv$ can have
  that satisfy this requirement are
  \begin{equation}\label{eq:irrepn}
  \sv(\kv)= \left\{
    \begin{array}{l l}
      \sz & \textnormal{if } \Phi_e^{n_\zb}(\kv)\=0,\\
      \sv_{\pm} & \textnormal{if } \Phi_e^{n_\zb}(\kv)\=\pm\frac{1}{n},\\
      0 & \textnormal{otherwise}.
    \end{array}\right.
  \end{equation}
\item[5.] If $\g$ is an $n$-fold rotation followed by time inversion
  $n'_\zb$ then one can obtain the constraints on the form of $\sv$
  from the constraints~\rf{eq:irrepn} by adding $1/2$ to the phases,
  thus
  \begin{equation}\label{eq:irrepn'}
  \sv(\kv)= \left\{
    \begin{array}{l l}
      \sz & \textnormal{if } \Phi_e^{n'_\zb}(\kv)\=\hf,\\
      \sv_{\pm} & \textnormal{if } \Phi_e^{n'_\zb}(\kv)\=\hf \pm\frac{1}{n},\\
      0 & \textnormal{otherwise}.
    \end{array}\right.
  \end{equation}
 \end{enumerate}
Table~\ref{tab:sel-n} summarizes the results for lattice spin groups
$\G_e=n,n'$ and $n1'$, containing operations of order greater than 2.

\subsection{Additional selection rules on invariant subspaces of
  nontrivial point group operations}
\label{sel-invariant}

In addition to the selection rules arising from $\G_e$ there are also
selection rules that occur when $\kv$ lies along one of the rotation
axes or within one of the mirror planes and is therefore invariant
under additional operations $(g,\g)$ with non-trivial $g$. In this
case the eigenvalue equation \rf{eq:eigenvalue} imposes further
restrictions on the Fourier coefficients of the spin density field.

\subsubsection{Selection rules along the $z$-axis:}
When the eightfold rotation $\rr$ is in the point group $G$ it leaves
all the wave vectors along the $z$-axis invariant. These wave vectors
are given by $\kv=l\zv$, where $l$ is any integer if the lattice is
vertical, and $l$ is even if the lattice is staggered. For both
lattice types Eq.~\rf{eq:eigenvalue}, determining the selection rules
for these wave vectors, becomes
\begin{equation}
  \d\sv(l\zv)\=e^{-2\pi il\f8(\cv)}\sv(l\zv),
\end{equation}
where $\cv$ is the appropriate stacking vector.  When $G$ is generated
by $\rb$, and does not contain $\rr$, the wave vectors along the
$z$-axis are left invariant by $\rb^2$ and its powers. The phase
$\Phi_{\rb^2}^{\bar\d^2}(l\zv)\=\Phi_{\rb}^{\bar\d}(\rb l\zv) +
\Phi_{\rb}^{\bar\d}(l\zv)$ is necessarily zero because $\rb
l\zv=-l\zv$. Eq.~\rf{eq:eigenvalue} can therefore be written as
\begin{equation}
  \bar\d^2\sv(l\zv)=\sv(l\zv),
\end{equation}
requiring $\sv(l\zv)$ to be invariant under $\bar\d^2$, where $\bar\d$
is the spin-space operation paired with $\rb$.

\subsubsection{Selection rules within the horizontal mirror $h$:}

If the horizontal mirror $h$ is present in the point group then all
the Fourier coefficients of the spin density filed associated with
wave vectors $\kv_h=\sum_i n_i\bi$ in the horizontal sublattice are
subject to selection rules, dictated by the values of the phase
function $\fh$. Eq.~\rf{eq:eigenvalue}, for both lattice types, is
simply
\begin{equation}
  \eta\sv(\kv_h)\=e^{-2\pi i \sum_{i=1}^4 n_i\fh(\bi)}\sv(\kv_h).
\end{equation}
Thus, if a horizontal mirror is present in the point group then
$\sv(\kv_h)$, for $\kv_h=\sum_i n_i\bi$ in the horizontal sublattice,
must remain invariant under $\eta$, unless (a) $\fh(\bi)\=\hf 0\hf 0$
and $n_1+n_3$ is odd; or (b) $\fh(\bi)\=0\hf 0\hf$ and $n_2+n_4$ is
odd; or (c) $\fh(\bi)\=\hf\hf\hf\hf$ and $\sum_i n_i$ is odd; in
which case $\sv(\kv_h)$ must change its sign under $\eta$.

\subsubsection{Selection rules within vertical mirrors and along dihedral
  axes:}

There are two sets of conjugate vertical mirrors which we have labeled
$m$ and $m'$ and two sets of conjugate dihedral axes $d$ and $d'$. It
is sufficient to determine the selection rules for a single member of
each of these four sets because the selection rules for the remaining
conjugate operations can be inferred from the 8-fold rotational
symmetry of the spin density field. To see this, take for example the
operation $(m,\mu)$ and examine all the wave
vectors $\kv$, satisfying $m\kv=\kv$, to obtain the general selection
rules for the invariant subspace of $m$,
\begin{equation}
  \mu\sv(\kv)\=e^{-2\pi i \fm(\kv)}\sv(\kv).
\end{equation}
If $g$ is the 8-fold generator of the point group ($\rr$ or $\rb$),
paired with the spin space opertaion $\d$, then the selection rules on
the invariant subspaces of the remaining three conjugate operations
are simply given by
\begin{equation}
  \d^n\mu\d^{-n}\sv(g^n\kv)\=e^{-2\pi i \fm(\kv)}\sv(g^n\kv),
  \quad n=1,2,3,
\end{equation}
where we have used that fact that, since $m\kv=\kv$, it follows from
successive applications of the group compatibility
condition~\rf{eq:GCC} that
\begin{equation}
  \Phi_{gmg^{-1}}^{\d\mu\d^{-1}}(g\kv) \= \fm(\kv).
\end{equation}
Similar expressions can be derived for the other operations and so we
proceed below to obtain the general selection rules only for the
vertical mirrors and dihedral axes that are oriented either along the
generating vector $\b1$ or between the two generating vectors $\b1$
and $\b2$ (as depicted in Fig.~\ref{fig:star3d}).

Wave vectors along the dihedral axis $d$, containing the generating
vector $\b1$, on either lattice type, can be expressed as
$\kv_d=n_1\b1 + n_2(\b2-\b4)$ for any two integers $n_1$ and $n_2$.
Since it is always the case that $\fd(\b2-\b4)\=0$, the selection
rules for such wave vectors are determined by
\begin{equation}
  \alpha\sv(\kv_d)\=e^{-2\pi i n_1\fd(\b1)}\sv(\kv_d).
\end{equation}
Thus, if a dihedral operation $(d,\alpha)$ is present in the spin
point group then $\sv(\kv_d)$, for $\kv_d$ along the axis of $d$, must
remain invariant under $\alpha$, unless $\fd(\b1)\=\hf$ and $n_1$ is
odd, in which case $\sv(\kv_d)$ must change its sign under $\alpha$.

Wave vectors along the dihedral axis $d'$, between the generating
vectors $\b1$ and $\b2$, on either lattice type, can be expressed as
$\kv_{d'}=n_1(\b1+\b2) + n_3(\b3-\b4)$ for any two integers $n_1$ and
$n_3$. Since it is always the case that
$\Phi_{d'}^{\alpha'}(\b1-\b3)\=\Phi_{d'}^{\alpha'}(\b2+\b4)\=0$, the
selection rules are determined by
\begin{equation}
  \alpha'\sv(\kv_{d'}) \=
  e^{-2\pi i (n_1+n_3)\Phi_{d'}^{\alpha'}(\b1+\b2)}\sv(\kv_{d'}).
\end{equation}
Thus, if a dihedral operation $(d',\alpha')$ is present in the spin
point group then $\sv(\kv_{d'})$, for $\kv_{d'}$ along the axis of
$d'$, must remain invariant under $\alpha'$, unless
$\Phi_{d'}^{\alpha'}(\bi) \= 0\hf 0\hf$ or $\hf 0\hf 0$, and $n_1+n_3$
is odd, in which case $\sv(\kv_{d'})$ must change its sign under
$\alpha'$. The operation $\alpha'$, as well as the values of the phase
function $\Phi_{d'}^{\alpha'}$, are determined for each separate spin
point group according to the generating relations for that group,
namely, $(d',\alpha')=(\rr,\d)(d,\alpha)$ for point group $822$,
$(d',\alpha')=(\rb,\d)(m,\mu)$ for point group $\bar8m2$, and
$(d',\alpha')=(\rr,\d)(m,\mu)(h,\eta)$ for point group $8/mmm$.

Wave vectors within the vertical mirror $m$, containing the generating
vector $\b1$, can be expressed as $\kv_m=n_1\b1 + n_2(\b2-\b4) +l\zv$,
where $l$ is any integer if the lattice is vertical and $l$ is even if
the lattice is staggered. This is due to the fact that there are no
wave vectors in odd layers of $S$-lattices that are invariant under
mirrors of type $m$~\cite[footnote 46]{rmrw}. Since it is always the
case that $\fm(\b2-\b4)\=0$, the selection rules for such wave vectors
are determined by
\begin{equation}
  \mu\sv(\kv_m)\=e^{-2\pi i \left[n_1\fm(\b1) + l\fm(\cv)\right]}\sv(\kv_m).
\end{equation}

Finally, wave vectors within the vertical mirror $m'$, between the
generating vectors $\b1$ and $\b2$, on both lattice types, can be
expressed as $\kv_{m'}=n_1(\b1+\b2) + n_3(\b3-\b4) + l\cv$ for any
three integers $n_1$, $n_3$, and $l$, where $\cv$ is the appropriate
stacking vector. Since it is always the case that
$\Phi_{m'}^{\mu'}(\b1-\b3)\=\Phi_{m'}^{\mu'}(\b2+\b4)\=0$, the
selection rules are determined by
\begin{equation}
  \mu'\sv(\kv_{m'}) \=
  e^{-2\pi i \left[(n_1+n_3)\Phi_{m'}^{\mu'}(\b1+\b2) +
  l\Phi_{m'}^{\mu'}(\cv)\right]} \sv(\kv_{d'}).
\end{equation}

This completes the general calculation of selection rules on invariant
subspaces of point group operations. One can apply these rules in a
straightforward manner to each spin space group once the spin space
operations $\d$, $\eta$, $\mu$, and $\alpha$ are explicitly identified.

\ack{This research was funded by the Israel Science Foundation through
  Grant No.~278/00.}

\referencelist[spin]

\onecolumn


\begin{table}
\caption{Three-dimensional octagonal point groups. There are seven
  octagonal point groups (geometric crystal classes) in three
  dimensions, one of which ($\bar 8m2$) has two distinct orientations
  with respect to both types of octagonal lattices, giving the eight octagonal
  arithmetic crystal classes, listed in the first column. The set of
  generators for each point group, used  throughout the paper, are listed
  in the second column along with the symbols, used to denote the
  spin-space operations with which they are paired in the spin point groups. }
\label{tab:generators}
\begin{center}
\begin{tabular}{l|l}
\cline{1-2}
Point group &Generators\\
\cline{1-2}
8&$(\rr,\d)$\\
$\bar 8$&$(\rb,\d)$\\
$8mm$&$(\rr,\d),(m,\mu)$\\
$\bar 8m2$&$(\rb,\d),(m,\mu)$\\
$822$&$(\rr,\d),(d,\alpha)$\\
$\bar 82m$&$(\rb,\d),(d,\alpha)$\\
$8/m$ & $(\rr,\d),(h,\h)$\\
$8/mmm$&$(\rr,\d),(m,\mu),(h,\h)$\\
\end{tabular}
\end{center}
\end{table}


\begin{table}
\caption{Effect of the point group generators on the lattice
  generating vectors. The horizontal shift $\hv$ associated with the
  staggered stacking vector is defined in Eq.~\rf{eq:shift}.}
\label{tab:symmetry}
\setlength{\extrarowheight}{2pt}
\begin{center}
\begin{tabular}{>{$}c<{$}|>{$}c<{$}|>{$}c<{$}|>{$}c<{$}|>{$}c<{$}|>{$}c<{$}|>{$}c<{$}}
& \b1 & \b2 & \b3 & \b4 & \zv & \az\\
\cline{1-7}
r_8 & \b2 & \b3 & \b4 & -\b1 & \zv & (\az) + \b4\\
\bar r_8 & -\b2 & -\b3 & -\b4 & \b1 & -\zv & -(\az) - \b4\\
m & \b1 & -\b4 & -\b3 & -\b2 & \zv & (\az) - \b3\\
d & \b1 & -\b4 & -\b3 & -\b2 & -\zv & -(\az) +2\hv - \b3\\
h & \b1 & \b2 & \b3 & \b4 & -\zv & -(\az) +2\hv \\
\end{tabular}
\end{center}
\end{table}


\begin{table}
\caption{Normal subgroups $G_\eps$ of the seven octagonal point
  groups. The resulting quotient group $G/G_\eps$ is represented in
  the seventh column by a point group, isomorphic to it. Isomorphic
  groups $G$ are listed in the same section of the Table. Constraints on
  the spin-space orperations $\d$, $\mu$, $\alpha$, and $\eta$,
  paired with the
  generators $\rr$, $m$, $d$, and $h$ of $G$ are listed in the last
  column. In each line the first power of $\d$ that is in $\G_e$ is
  given. $\mu^2$, $\alpha^2$, and $\eta^2$ are always in $\G_e$,
  therefore we only note whether $\mu$, $\alpha$, and $\eta$ are
  themselves in $\G_e$. If a spin space operation is in $\G_e$ it is
  taken to be $\eps$.}
\label{tab:normalsub}
\setlength{\extrarowheight}{2pt}
\begin{center}
\begin{tabular}{>{$}c<{$}>{$}c<{$}|>{$}c<{$}>{$}c<{$}|>{$}c<{$}>{$}c<{$}||>{$}c<{$}|>{$}c<{$}}
\cline{1-8}
G &G_\eps &G &G_\eps &G &G_\eps &G/G_\eps &\textnormal{Constraints}\\
\cline{1-8}
&&8  &8 & \bar 8 &\bar 8 &1 &\d=\eps\\
&&   &4 &&4 &2 &\d^2\in\G_e\\
&&&2 &&2 &4 &\d^4\in\G_e\\
&&&1 &&1 &8 &\d^8\in\G_e\\
\cline{1-8}
8mm &8mm &822 &822 &\bar8m2 &\bar8m2 &1&\d=\eps,\mu=\eps\\
&8 &&8 &&8 &2 &\d=\eps,\mu\notin\G_e\\
&4mm &&422 && 4mm &2&\d^2\in\G_e,\mu=\eps\\
&4m'm' &&42'2' && 42'2' &2&\d=\mu\notin\G_e\\
&4 &&4 && 4 &222&\d^2\in\G_e,\mu\notin\G_e,\d\G_e\neq\mu\G_e\\
&2 && 2 && 2 &422 &\d^4\in\G_e,\mu\notin\G_e,\d^2\G_e\neq\mu\G_e\\
&1 && 1 && 1 &822 &\d^8\in\G_e,\mu\notin\G_e,\d^4\G_e\neq\mu\G_e\\
\cline{1-8}
&&&&8/m & 8/m&1&\d=\eps,\h=\eps\\
&&&&& 8&2&\d=\eps,\h\notin\G_e\\
&&&&& \bar 8&2&\d=\h\notin\G_e\\
&&&&& 4/m&2&\d^2\in\G_e,\h=\eps\\
&&&&& 4&2/m&\d^2\in\G_e,\h\notin\G_e,\d\notin\h\G_e\\
&&&&& \bar 4&4&\d^4\in\G_e,\h\in\d^2\G_e\\
&&&&& 2/m&4&\d^4\in\G_e,\h=\eps\\
&&&&& 2&4/m&\d^4\in\G_e,\h\notin\G_e,\h\notin\d^2\G_e\\
&&&&& m &8&\d^8\in\G_e,\h=\eps\\
&&&&& \bar 1&8&\d^8\in\G_e,\h\in\d^4\G_e\\
&&&&& 1&8/m&\d^8\in\G_e,\h\notin\G_e,\h\notin\d^4\G_e\\
\cline{1-8}
&&&& 8/mmm & 8/mmm & 1& \d=\mu=\h=\eps\\
&&&&& 8mm&2&\d=\mu=\eps,\h\notin\G_e\\
&&&&& 822&2&\d=\eps,\h=\mu\notin\G_e\\
&&&&& 8/m&2&\d=\h=\eps,\mu\notin\G_e\\
&&&&&\bar 8m2&2&\d=\h\notin\G_e,\mu=\eps\\
&&&&&\bar 82m&2&\d=\mu=\h\notin\G_e\\
&&&&& 4/mmm&2& \d^2\in\G_e,\h=\mu=\eps\\
&&&&& 4/mm'm'&2&\d=\mu\notin\G_e,\h=\eps\\
&&&&& 8&222&\d=\eps,\h\notin\G_e,\mu\notin\G_e,\h\notin\mu\G_e\\
&&&&&\bar 8&222&\d=\h\notin\G_e,\mu\notin\G_e,\mu\notin\h\G_e\\
&&&&& 4mm&222&\d^2\in\G_e,\mu=\eps,\h\notin\G_e,\h\notin\d\G_e\\
&&&&& 4m'm'&222&\d=\mu\notin\G_e,\h\notin\G_e,\h\notin\mu\G_e\\
&&&&& 422&222&\d^2\in\G_e,\mu=\h\notin\G_e,\h\notin\d\G_e\\
&&&&& 42'2'&222&\d^2\in\G_e,\mu\notin\G_e,\h\notin\G_e,\h\in\d\mu\G_e\\
&&&&& 4/m&222&\d^2\in\G_e,m\notin\G_e,m\notin\d\G_e,\h=\eps\\
&&&&& 4&mmm&\d^2\in\G_e,\mu\notin\G_e,\h\notin\G_e,\d\G_e\neq\h\G_e\neq\mu\G_e\\
&&&&& \bar 4&422&\d^4\in\G_e,\mu\notin\G_e,\mu\notin\d\G_e,\h\in\d^2\G_e\\
&&&&& 2/m&422&\d^4\in\G_e,\mu\notin\G_e,\mu\notin\d\G_e,\h=\eps\\
&&&&& 2&4/mmm&\d^4\in\G_e,\mu\notin\G_e,\h\notin\G_e,\d^2\G_e\neq\h\G_e\neq\mu\G_e\\
&&&&& m &822&\d^8\in\G_e,\mu\notin\G_e,\mu\notin\d^4\G_e,\h=\eps\\
&&&&&\bar 1&822&\d^8\in\G_e,\mu\notin\G_e,\mu\notin\d^4\G_e,\h\in\d^4\G_e\\
&&&&& 1&8/mmm&\d^8\in\G_e,\mu\notin\G_e,\h\notin\G_e,\d^4\G_e\neq\mu\G_e\neq\h\G_e\\
\cline{1-8}
\end{tabular}
\end{center}
\end{table}


\begin{table}
\caption{Spin space-group types on $V$-lattices with $G=8$. In this,
  and in the following tables $a$, $b$, $a'$, and $b'$ are independently 0 or $\hf$,
  as long as there are no two operations in $\G_e$ with identical phase functions, and $c$ is
  any integer between 0 and 7.
  For $\G_e=n$, $n'$, or $n1'$ the integer $j$ is co-prime with $N$,
  where $N=n$ unless $\G_e=n'$ and $n$ is odd, in which case
  $N=2n$. If $N$ is odd 
  $a$ is necessarily 0.  The integer $d=1$ unless $N$ is twice an odd
  number and $\Phi_e^{n^*_\zb}(\bi)\equiv\hf$, in which case  $d=1$ or 2.
  $A_0$ denotes the values $0\hf0\hf$ of a phase function on the
  horizontal generating vectors, and $A_1$ denotes the values
  $\hf0\hf0$ on the same generators.
  Lines $3a$ and $3b$ refer to distinct spin space-group types if $\G_e=2'2'2$,
  but give scale-equivalent solutions if $\G_e=222$, or $2'2'2'$,
  where $3a$ is taken as the representative solution.
  The spin space group symbols for all groups in this table are of the
  form $\PL8^\d_c$. For example, if $\G_e=222$, $G_\eps=4$ and $\G=2'2'2'$, then $\d$ can be chosen to
  be $\eps'$, the corresponding line in the Table is $3a$ or $3b$, but since for
  $\G_e=222$ they are scale-equivalent only $3a$ is taken. If, in
  addition, $\f8(\zv)\equiv\frac{5}{8}$ the spin space group symbol
  is $P^{222}_{2c,P,S}8'_5$.} 
\label{tab:8V}
\setlength{\extrarowheight}{5pt}
\begin{center}
\begin{tabular}{>{$}l<{$}|>{$}l<{$}>{$}l<{$}>{$}l<{$}>{$}l<{$}>{$}l<{$}}
\cline{1-6}
\multicolumn{6}{l}{$\G_e=1,1',2,2',21'$}\\
\cline{1-6}
&  \d^8   & \fez  & \fee   & \f8(\zv)\\
\cline{1-6}
1& \eps   & ab    & a'b'   &\frac{c}{8}\\
2& 2_\zb  &0\hf   &--      &\frac{c}{8}+\frac{1}{16}\\
\cline{1-6}
\multicolumn{6}{l}{$\G_e=222,2'2'2,2'2'2'\hfill
(\d\in4221'\Rightarrow\d^8=\eps)$}\\
\cline{1-6}
&  \dxd   & \fex  & \fey   & \fezp & \f8(\zv)\\
\cline{1-6}
3a&2_\xb  & 0\hf  &  \hf0  &  --   &  \frac{c}{8}\\
3b&       &  ab   & \hf\hf &  --   &  \frac{c}{8}\\
4& 2_\yb  &  A_0b   &  A_1b  &  a\hf &  \frac{c}{8}\\
\cline{1-6}
\multicolumn{6}{l}{$\G_e=n,n',n1'\hfill (\d n_\zb\d^{-1}=n_\zb)$}\\
\cline{1-6}
&  \d^8    & \fenz            & \fee &\f8(\zv)\\
\cline{1-6}
5& \eps    & a\frac{dj}{N}    & a'b' &\frac{c}{8}      \\
6& n_\zb   & a\frac{dj}{n}    & a'b' &\frac{c}{8}+\frac{dj}{8n}\\
7& n^2_\zb & a\frac{dj}{N}    & a'b' &\frac{c}{8}+\frac{dj}{4N}\\
8& n_\zb^3 & a\frac{dj}{n}    & a'b' &\frac{c}{8}+\frac{3dj}{8n}\\
9& n^4_\zb & a\frac{dj}{N}    & a'b' &\frac{c}{8}+\frac{dj}{2N}\\
\cline{1-6}
\end{tabular}
\end{center}
\end{table}


\begin{table}
\caption{Spin space-group types on $S$-lattices with $G=8$.
 In this and in the following tables $c'$ is any
  integer between 0 and 3. For $\G_e=n$, or $n'$, the integer $j$ is
  co-prime with $N$, where $N=n$ unless $\G_e=n'$ and $n$ is odd, in
  which case $N=2n$.
  The spin space-group symbols for all groups in this table are of the
  form $\SL8^\d_{c'}$. For example, if $\G_e=n$, $G_\eps=4$ and
  $\G=n22$ then $\d$ can be chosen to be $2_\xb$ and the corresponding
  line in the Table
  is 9 (and $n$ is necessarily 4). If, in addition, $j=3$ and $\f8(\az)\equiv\qt$
  the spin space-group symbol is $S^4_{4_3S}8^{2_\xb}_2$.}
\label{tab:8S}
\setlength{\extrarowheight}{5pt}
\begin{center}
\begin{tabular}{>{$}l<{$}|>{$}l<{$}>{$}l<{$}>{$}l<{$}>{$}l<{$}}
\cline{1-5}
\multicolumn{5}{l}{$\G_e=1,1',2,2'$}\\
\cline{1-5}
  &\d^8   &\feg(\kv) &  \f8(\az)\\
\cline{1-5}
1 &\eps   &   0\hf   &  \frac{c'}{8}       \\
2 &2_\zb  &   0\hf   &  \frac{c'}{8}+\frac{1}{16}\\
\cline{1-5}
\multicolumn{5}{l}{$\G_e=222,2'2'2\hfill
(\d\in4221'\Rightarrow\d^8=\eps,\d 2_\xb\d^{-1}=2_\yb)$}\\
\cline{1-5}
  &\fex    &\fey    &  \f8(\az)\\
\cline{1-5}
3 &\hf0    &\hf\hf  &\frac{c'}{8}\\
\cline{1-5}
\multicolumn{5}{l}{$\G_e=n,n',n1'$}\\
\cline{1-5}
  &\dnd        &\d^8    & \fenz              &\f8(\az)\\
\cline{1-5}
4 &n_\zb       &\eps    & 0\frac{j}{N}       &\frac{c'}{8}       \\
5 &n_\zb       &n_\zb   & 0\frac{j}{n}       &\frac{c'}{8}+\frac{j}{8n}\\
6 &n_\zb       &n^2_\zb & 0\frac{j}{N}       &\frac{c'}{8}+\frac{j}{4N}\\
7 &n_\zb       &n^3_\zb & 0\frac{j}{n}       &\frac{c'}{8}+\frac{3j}{8n}\\
8 &n_\zb       &n^4_\zb & 0\frac{j}{N}       &\frac{c'}{8}+\frac{j}{2N}\\
9 &n_\zb^{-1}  &\eps    & \hf\frac{j}{n}(n=4)&\frac{c'}{8}\\
\cline{1-5}
\end{tabular}
\end{center}
\end{table}


\begin{table}
\caption{Spin space-group types on $V$-lattices with
  $G=\bar{8}$.
  The possible values of $a$, $b$, $a'$, $b'$, $N$, $j$, and $d$, as well as the
  notations $A_0$ and $A_1$ are as explained in the caption of Table \ref{tab:8V}.
  Recall that since $\rb$ is a generator of $G$, $\d^8$ is necessarily $\eps$.
  Lines $2a$ and $2b$ refer to distinct spin space-group types if $\G_e=2'2'2$
  but are scale-equivalent if $\G_e=222$, or $2'2'2'$, for which line
  $2a$ suffices. Spin space-group symbols for all groups in this
  table are of the form $\PL\bar{8}^\d$.}
\label{tab:bar8V}
\setlength{\extrarowheight}{5pt}
\begin{center}
\begin{tabular}{>{$}l<{$}|>{$}l<{$}>{$}l<{$}>{$}l<{$}>{$}l<{$}}
\cline{1-5}
\multicolumn{5}{l}{$\G_e=1,1',2,2',21'$}\\
\cline{1-5}
  &\fez   &\fee\\
\cline{1-5}
1 &ab     &a'b'\\
\cline{1-5}
\multicolumn{5}{l}{$\G_e=222,2'2'2,2'2'2'$}\\
\cline{1-5}
  &\dxd   &\fex   & \fey    & \fezp\\
\cline{1-5}
2a&2_\xb  &0\hf   &\hf0     &--    \\
2b&2_\xb  &ab     &\hf\hf   &--    \\
3 &2_\yb  &A_0b     &A_1b     &a\hf \\
\cline{1-5}
\multicolumn{5}{l}{$\G_e=n,n',n1'\hfill (\d n_\zb\d^{-1}=n^{-1}_\zb)$}\\
\cline{1-5}
  &\d^8    & \fenz          & \fee\\
\cline{1-5}
4 &\eps    & a\frac{dj}{N}  & a'b'\\
\cline{1-5}
\end{tabular}
\end{center}
\end{table}


\begin{table}
\caption{Spin space group-types on $S$-lattices with
  $G=\bar{8}$.
  The possible values of $N$ and $j$ are as explained in
  the caption of Table~\ref{tab:8S}.
  Recall that since $\rb$ is a generator of $G$, $\d^8$ is
  necessarily $\eps$.
  Spin space-group symbols for all groups in this
  table are of the form $\SL\bar{8}^\d$.}
\label{tab:bar8S}
\setlength{\extrarowheight}{5pt}
\begin{center}
\begin{tabular}{>{$}l<{$}|>{$}l<{$}>{$}l<{$}>{$}l<{$}}
\cline{1-4}
\multicolumn{4}{l}{$\G_e=1,1',2,2',21'$}\\
\cline{1-4}
  &\feg(\bi\cv)\\
\cline{1-4}
1 &0\hf        \\
\cline{1-4}
\multicolumn{4}{l}{$\G_e=222,2'2'2$}\\
\cline{1-4}
  &\dxd  &  \fex & \fey\\
\cline{1-4}
2 &2_\yb &  \hf0 &\hf\hf\\
\cline{1-4}
\multicolumn{4}{l}{$\G_e=n,n'$}\\
\cline{1-4}
  &\dnd         &\fenz   \\
\cline{1-4}
3 &n_\zb^{-1}   &0\frac{j}{N} \\
4 &n_\zb        &\hf\frac{j}{n}(n=4)\\
\cline{1-4}
\end{tabular}
\end{center}
\end{table}


\begin{table}
\caption{Spin space-group types on $V$-lattices with
  $G=8mm$.
  The possible values of $a$, $b$, $a'$, $b'$, $N$, $j$, and $d$, as well as the
  notations $A_0$ and $A_1$ are as explained in the caption of Table \ref{tab:8V}.
  Lines $5a$ and $5b$ refer to distinct spin space-group types if $\G_e=2'2'2$
  but are scale-equivalent if $\G_e=222$, or $2'2'2'$, for which line $3a$ suffices.
  $\hat{a},\ \hat{b}$ and $\ta$ denote either 0 or  $\hf$.
  Spin space-group symbols are of the form $\PL8^\d m^\mu
  m^{\d\mu}$ where the primary $8^\d$ is replaced by $8_4^\d$ if
  $\ta=\hf$ and the secondary $m^\mu$ is replaced according to the values
  of $\hat{a}$ and $\hat{b}$: $\hat{a}=\hat{b}=0\Rightarrow
  m^\mu\rightarrow m^\mu$, $\hat{a}=0,\hat{b}=\hf\Rightarrow
  m^\mu\rightarrow c^\mu$, $\hat{a}=\hf,\hat{b}=0\Rightarrow
  m^\mu\rightarrow b^\mu$, $\hat{a}=\hat{b}=\hf\Rightarrow
  m^\mu\rightarrow n^\mu$. The tertiary $m^{\d\mu}$ is replaced by $c^{\d\mu}$ if
  either $\hat{b}$ or $\ta$ (but not both) is $\hf$. Furtheremore, a
  subscript $a$  is added to
  the secondary $m^\mu$ when $\fm(\bi)\equiv A_0+\hat{a}$. For example, if $\G_e=2'2'2$, $G_\eps=4mm$,
  $\G=2'2'2'$, then $\d$ can be chosen to be $\eps'$ and the spin
  space group is described by line $5a$ or $5b$, if $ab=\hf0$, 
  $\hat{a}=\hf$, $\hat{b}=0$ and $\ta=0$ the spin space group symbol will be $P^{2'2'2}_{P,S,2c}8'nc'$.}
\label{tab:8mmV}
\setlength{\extrarowheight}{5pt}
\begin{center}
\begin{tabular}{>{$}l<{$}|>{$}l<{$}>{$}l<{$}>{$}l<{$}>{$}l<{$}>{$}l<{$}>{$}l<{$}>{$}l<{$}>{$}l<{$}}
\cline{1-9}
\multicolumn{9}{l}{$\G_e=1,1',2,2',21'\hfill (\d^8=\eps)$}\\
\cline{1-9}
  &\mu^{-1}\d\mu\d & \mu^2 &\fez    & \fee       & \fm(\bi) & \fm(\zv) & \f8(\zv)\\
\cline{1-9}
1 &\eps  &\eps   &ab      &a'b'    &\hat{a}      &\hat{b} &\tilde{a}\\
2 &\eps  &2_\zb  &0\hf    &\hf a   &\hat{a}      &\hat{b} &\tilde{a}\\
3 &2_\zb &\eps   &0\hf    &\hf a   &\hat{a}      &\hat{b} &\tilde{a}\\
  &      &       &\hf a   &a'b'    &\hat{a}+A_0  &\hat{b} &\tilde{a}+\frac{a}{2}\\
4 &2_\zb &2_\zb  &0\hf    &\hf b   &\hat{a}      &\hat{b}+\qt &\tilde{a}+\qt\\
\cline{1-9}
\multicolumn{9}{l}{$\G_e=222,2'2'2,2'2'2'\hfill
(\mu 2_\xb\mu^{-1}=2_\xb\Rightarrow \mu^2=\eps)$}\\
\cline{1-9}
  &\d2_\xb\d^{-1}&\mu^{-1}\d\mu\d & \fex & \fey     & \fezp & \fm(\bi)
  & \fm(\zv) & \f8(\zv)\\
\cline{1-9}
5a&2_\xb &\eps  &0\hf  &\hf0   &--     &\hat{a}     &\hat{b}   &\tilde{a}\\
5b&2_\xb &\eps  &ab    &\hf\hf &--     &\hat{a}     &\hat{b}   &\tilde{a}\\
6 &2_\yb &2_\zb &A_0b  &A_1b   &a\hf   &A_0+\hat{a} &\hat{b}   &\tilde{a}\\
\cline{1-9}
\multicolumn{9}{l}{$\G_e=n,n',n1'\hfill
(\d n_\zb\d^{-1}=\mu n_\zb\mu^{-1}=n_\zb,\d^8=(\mu^{-1}\d\mu\d)^4)$}\\
\cline{1-9}
  &\mu^{-1}\d\mu\d & \mu^2 & \fenz & \fee &\fm(\bi) &\fm(\zv) &\f8(\zv)\\
\cline{1-9}
7 &\eps &\eps & a\frac{dj}{N} &a'b' &\hat{a} &\hat{b}              &\tilde{a}\\
8 &\eps &n_\zb  &0\frac{j}{n} & ab  &\hat{a} &\hat{b}+\frac{j}{2n} &\tilde{a}\\
9 &n_\zb &\eps  &0\frac{j}{n} &ab   &\hat{a} &\hat{b} &\tilde{a}+\frac{j}{2n}\\
  &      &      &\hf\frac{dj}{n}&ab &\hat{a}+A_0 &\hat{b}
    &\tilde{a}+\frac{dj}{2n}\\
10&n_\zb &n_\zb &0\frac{j}{n} &ab   &\hat{a} &\hat{b}+\frac{j}{2n}
    &\tilde{a}+\frac{j}{2n}\\
\cline{1-9}
\end{tabular}
\end{center}
\end{table}


\begin{table}
\caption{Spin space-group types on $S$-lattices with
  $G=8mm$.
  The possible values of $N$ and $j$ are as explained in the caption
  of Table \ref{tab:8S}, and the notation $A_0$ is explained in the caption of Table \ref{tab:8V}.
  $\hat{a}$ and $\ta$ denote either 0 or $\hf$.
  For any choice of spin-space operations $(\mu^{-1}\d\mu\d)^4=\d^8$.
  Spin space-group symbols are of the form $\SL 8^\d m^\mu m^{\d\mu}$
  where the 
  primary $8^\d$ is replaced by $8_2^\d$ if $\hat{a}=\hf$ and the
  secondary $m^\mu$ is replaced by $d^\mu$ if $\hat{a}=\hf$. The tertiary $m^{\d\mu}$
  is replaced by $c^{\d\mu}$ if $\ta=\hf$. A subscript $a$  is added to the secondary $m$
  if $\fm(\bi)\equiv\hat{a}+A_0$.}
\label{tab:8mmS}
\setlength{\extrarowheight}{5pt}
\begin{center}
\begin{tabular}{>{$}l<{$}|>{$}l<{$}>{$}l<{$}>{$}l<{$}>{$}l<{$}>{$}l<{$}>{$}l<{$}>{$}l<{$}>{$}l<{$}}
\cline{1-9}
\multicolumn{9}{l}{$\G_e=1,1',2,2'\hfill (\d^8=\eps)$}\\
\cline{1-9}
  &\mu^{-1}\d\mu\d & \mu^2 &\feg(\kv) & \fm(\bi) & \fm(\az) & \f8(\az)\\
\cline{1-9}
1 &\eps  &\eps   &0\hf & \hat{a}  &\ta+\hf\hat{a}     & \hf\hat{a}\\
2 &\eps  &2_\zb  &0\hf & \hat{a}  &\ta+\hf\hat{a}+\qt & \hf\hat{a}\\
3 &2_\zb &\eps   &0\hf & \hat{a}  &\ta+\hf\hat{a}     & \hf\hat{a}+\qt\\
4 &2_\zb &2_\zb  &0\hf & \hat{a}  &\ta+\hf\hat{a}+\qt & \hf\hat{a}+\qt\\
\cline{1-9}
\multicolumn{9}{l}{$\G_e=222,2'2'2\hfill
(\mu 2_\xb\mu^{-1}=2_\yb\Rightarrow \mu^2=2_\zb)$}\\
\cline{1-9}
  &\d 2_\xb\d^{-1} & \mu^{-1}\d\mu\d & \fex & \fey & \fm(\bi)
  &\fm(\az)  & \f8(\az)\\
\cline{1-9} 5 &2_\yb & 2_\zb & \hf0 & \hf\hf & \hat{a} &\ta+\hf\hat{a}+\qt
   &\hf\hat{a}+\qt\\
\cline{1-9}
\multicolumn{9}{l}{$\G_e=n,n'\hfill (\d n_\zb\d^{-1}=\mu n_\zb\mu^{-1})$}\\
\cline{1-9}
  &\d n_\zb\d^{-1}& \mu^{-1}\d\mu\d& \mu^2 & \d^8  & \fenz & \fm(\bi)
  & \fm(\az) & \f8(\az) \\
\cline{1-9} 6 &n_\zb & \eps & \eps  & \eps & 0\frac{j}{N} & \hat{a} &\ta+\hf\hat{a}
     & \hf\hat{a}\\
7 &n_\zb & \eps & n_\zb & \eps & 0\frac{j}{n} & \hat{a}
     &\ta+\hf\hat{a}+\frac{j}{2n}  &\hf\hat{a}\\
8 &n_\zb & n_\zb & \eps &n_\zb^4& 0\frac{j}{n} & \hat{a} &\ta+\hf\hat{a}
     &\hf\hat{a}+\frac{j}{2n}\\
9 &n_\zb & n_\zb & n_\zb &n_\zb^4& 0\frac{j}{n} & \hat{a}
     &\ta+\hf\hat{a}+\frac{j}{2n}  &\hf\hat{a}+\frac{j}{2n}\\
10&n_\zb^{-1}    & \eps  & \eps  & \eps  &\hf\frac{j}{n}(n=4)&
\hat{a}
     &\ta+\hf\hat{a}  &\hf\hat{a}\\
11&n_\zb^{-1} & n_\zb^{-1} & \eps & \eps
&\hf\frac{j}{n}(n=4)&\hat{a}+A_0
     &\ta+ \hf\hat{a}  &\hf(\hf-\hat{a})-\frac{j}{8}\\
\cline{1-9}
\end{tabular}
\end{center}
\end{table}


\begin{table}
\caption{Spin space-group types on $V$-lattices with
  $G=\bar{8}m2$.
  The possible values of $a$, $b$, $a'$, $b'$, $N$, $j$, and $d$, as well as the
  notations $A_0$ and $A_1$ are as explained in the caption of Table \ref{tab:8V}.
  Lines $4a$ and $4b$ refer to distinct spin space-group types if $\G_e=2'2'2$
  but are scale-equivalent if $\G_e=222$, or $2'2'2'$, for which line
  $4a$ suffices. $\hat{a}$ and $\hat{b}$ denote either 0 or $\hf$.
  Note that since $\bar{8}$ is a generator of $G$  $\d^8=\eps$.
  Spin space-group symbols are of the form $\PL\bar{8}^\d m^\mu 2^{\d\mu}$
  where the secondary $m^\mu$ is replaced as in Table
  \ref{tab:8mmV} above.}
\label{tab:bar8m2V}
\setlength{\extrarowheight}{5pt}
\begin{center}
\begin{tabular}{>{$}l<{$}|>{$}l<{$}>{$}l<{$}>{$}l<{$}>{$}l<{$}>{$}l<{$}>{$}l<{$}>{$}l<{$}}
\cline{1-8}
\multicolumn{8}{l}{$\G_e=1,1',2,2',21'$}\\
\cline{1-8}
  &\mu^{-1}\d\mu\d & \mu^2 & \fez & \fee & \fm(\bi)   & \fm(\zv) \\
\cline{1-8}
1 &\eps     &\eps   &ab    &a'b'  &\hat{a}     &\hat{b}\\
2 &2_\zb    &\eps   &\hf0  &a\hf  &\hat{a}+A_0 &\hat{b}\\
3 &2_\zb    &2_\zb  &0\hf  &\hf b &\hat{a}     &\hat{b}+\qt\\
\cline{1-8}
\multicolumn{8}{l}{$\G_e=222,2'2'2,2'2'2'\hfill
(\mu^2=\eps,\mu2_\xb\mu^{-1}=2_\xb)$}\\
\cline{1-8}
  &\d2_\xb\d^{-1}&\mu^{-1}\d\mu\d&\fex &\fey   &\fezp &\fm(\bi) &\fm(\zv)\\
\cline{1-8}
4a&2_\xb  &\eps     &0\hf &\hf0   &--    &\hat{a}     &\hat{b}\\
4b&2_\xb  &\eps     &ab   &\hf\hf &--    &\hat{a}     &\hat{b}\\
5 &2_\yb  &2_\zb    &A_0b   &A_1b   &a\hf  &\hat{a}+A_0 &\hat{b}\\
\cline{1-8}
\multicolumn{8}{l}{$\G_e=n,n',n1'\hfill
 (\d n_\zb\d^{-1}=n_\zb^{-1},\mu n_\zb\mu^{-1}=n_\zb,
  \mu^2=(\mu^{-1}\d\mu\d)^{-1})$}\\
\cline{1-8}
  &\mu^{-1}\d\mu\d & \fenz   & \fee &\fm(\bi) &\fm(\zv)\\
\cline{1-8}
6 &\eps   & a\frac{dj}{N} & a'b' &\hat{a}  &\hat{b}\\
7 &\eps   & 0\frac{j}{n}  & ab   &\hat{a}  &\hat{b}+\frac{j}{2n}\\
\cline{1-8}
\end{tabular}
\end{center}
\end{table}


\begin{table}
\caption{Spin space-group types on $S$-lattices with $G=\bar{8}m2$.
   The possible values of $N$ and $j$ are as explained in
  the caption of Table \ref{tab:8S}.
   $\hat{a}$ denotes either 0 or $\hf$.
   Note that since $\bar{8}$ is a generator of $G$
  $\d^8=\eps$. In addition $\mu^2=(\mu^{-1}\d\mu\d)^{-1}$.
  Spin space-group symbols are of the form
  $\SL\bar{8}^\d m^\mu 2^{\d\mu}$ where the secondary $m^\mu$ is
  replaced as in Table \ref{tab:8mmS} above.}
\label{tab:bar8m2S}
\setlength{\extrarowheight}{5pt}
\begin{center}
\begin{tabular}{>{$}l<{$}|>{$}l<{$}>{$}l<{$}>{$}l<{$}>{$}l<{$}>{$}l<{$}}
\cline{1-6}
\multicolumn{6}{l}{$\G_e=1,1',2,2'$}\\
\cline{1-6}
  &\mu^{-1}\d\mu\d & \feg(\kv) & \fm(\bi) & \fm(\az) \\
\cline{1-6}
1 &\eps            & 0\hf      & \hat{a}  &\hf\hat{a}\\
2 &2_\zb           & 0\hf      & \hat{a}  &\hf\hat{a}+\qt\\
\cline{1-6}
\multicolumn{6}{l}{$\G_e=222,2'2'2\hfill
 (\d2_\xb\d^{-1}=\mu2_\xb\mu^{-1}=2_\yb)$}\\
\cline{1-6}
  &\mu^{-1}\d\mu\d & \fex & \fey   &\fm(\bi) & \fm(\az) \\
\cline{1-6}
3 &2_\zb           & \hf0 & \hf\hf & \hat{a} &\hf\hat{a}+\qt       \\
\cline{1-6}
\multicolumn{6}{l}{$\G_e=n,n'\hfill
 (\d n_\zb d^{-1}=\mu n_\zb^{-1}\mu^{-1})$}\\
\cline{1-6}
  &\dnd      & \mdmd      & \fenz          &\fm(\bi) & \fm(\az) \\
\cline{1-6}
4 &n_\zb^{-1}& \eps       & 0\frac{j}{N}   &\hat{a}  &\hf\hat{a}\\
5 &n_\zb^{-1}& n_\zb^{-1} & 0\frac{j}{n}   &\hat{a}  &\hf\hat{a}+\frac{j}{2n}\\
6 &n_\zb     & \eps       & \hf\frac{j}{n}(n=4) &\hat{a}  &\hf\hat{a}\\
\cline{1-6}
\end{tabular}
\end{center}
\end{table}


\begin{table}
\caption{Spin space-group types for $V$-lattices with $G=822$.
  The possible values of $a$, $b$, $a'$, $b'$, $c$, $N$, $j$, and $d$, as well as the
  notations $A_0$ and $A_1$ are as explained in the caption of Table~\ref{tab:8V}.
  Lines $4a$ and $4b$ refer to distinct spin space-group types if $\G_e=2'2'2$
  but are scale-equivalent if $\G_e=222$, or $2'2'2'$, for which line $4a$ suffices.
  $\grave{a}$ denotes either 0 or  $\hf$.   Note that $\alpha^2=\eps$.
  Spin space-group symbols are of the form $\PL8^\d_c 2^\al 2^{\d\al}$,
  where the secondary $2^\al$ is replaced by $2_1^\al$ if
  $\grave{a}=\hf$. An additional subscript $a$ is added to the
  secondary $2^\al$ if $\fd(\bi)=\grave{a}+A_0$.}
\label{tab:822V}
\setlength{\extrarowheight}{5pt}
\begin{center}
\begin{tabular}{>{$}l<{$}|>{$}l<{$}>{$}l<{$}>{$}l<{$}>{$}l<{$}>{$}l<{$}>{$}l<{$}>{$}l<{$}}
\cline{1-8}
\multicolumn{8}{l}{$\G_e=1,1',2,2',21'$}\\
\cline{1-8}
  &\adad  &\d^8  &\fez  &\fee   &\fd(\bi)      &\f8(\zv)\\
\cline{1-8}
1 &\eps   &\eps  &ab    &a'b'   &\grave{a}     &\frac{c}{8}\\
2 &\eps   &2_\zb &0\hf  &\hf a  &\grave{a}     &\frac{c}{8}+\frac{1}{16}\\
3 &2_\zb  &\eps  &\hf0  &a\hf   &\grave{a}+A_0 &\frac{c}{8}\\
\cline{1-8}
\multicolumn{8}{l}{$\G_e=222,2'2'2,2'2'2'\hfill
 (\alpha 2_\xb\alpha^{-1}=2_\xb)$}\\
\cline{1-8}
  &\dxd &\adad  & \fex    & \fey    &\fezp   &\fd(\bi)      &\f8(\zv)\\
\cline{1-8}
4a&2_\xb &\eps   & 0\hf   &\hf0     &--      &\grave{a}     &\frac{c}{8}\\
4b&      &       & ab     &\hf\hf   &--      &\grave{a}     &\frac{c}{8}\\
5 &2_\yb &2_\zb  & A_0b   &A_1b     &a\hf    &\grave{a}+A_0 &\frac{c}{8}\\
\cline{1-8}
\multicolumn{8}{l}{$\G_e=n,n',n1'\hfill
 (\d n_\zb\d^{-1}=n_\zb,\alpha n_\zb\alpha^{-1}=n^{-1}_\zb,
  \alpha^{-1}\d\alpha\d=\eps)$}\\
\cline{1-8}
  &\d^8    & \fenz          & \fee &\fd(\bi)    &\f8(\zv)\\
\cline{1-8}
6 &\eps    & a\frac{dj}{N}  & a'b' &\grave{a}   &\frac{c}{8}\\
7 &n_\zb   & 0\frac{j}{n}   & ab   &\grave{a}   &\frac{c}{8}+\frac{j}{8n}\\
8 &n^3_\zb & 0\frac{j}{n}   & ab   &\grave{a}   &\frac{c}{8}+\frac{3j}{8n}\\
9 &n^2_\zb & a\frac{dj}{N}  & a'b' &\grave{a}   &\frac{c}{8}+\frac{dj}{4N}\\
10&n^4_\zb & a\frac{dj}{N}  & a'b' &\grave{a}   &\frac{c}{8}+\frac{dj}{2N}\\
\cline{1-8}
\end{tabular}
\end{center}
\end{table}


\begin{table}
\caption{Spin space-group types for $S$-lattices with $G=822$.
  The possible values of $c'$, $N$ and $j$ are explained in
  the caption Table \ref{tab:8S}.
  Note that $\alpha^{-1}\d\alpha\d=\alpha^2$. Spin space-group symbols are
  of the form $\SL8^\d_{c'}2^\al2^{\d\al}$, where the secondary
  $2^\al$ is replaced by $2_1^\al$ if $\fd(\bi)\equiv\hf$.}
\label{tab:822S}
\setlength{\extrarowheight}{8pt}
\begin{center}
\begin{tabular}{>{$}l<{$}|>{$}l<{$}>{$}l<{$}>{$}l<{$}>{$}l<{$}>{$}l<{$}}
\cline{1-6}
\multicolumn{6}{l}{$\G_e=1,1',2,2'$}\\
\cline{1-6}
  &\adad  & \d^8  &\feg(\bi\cv)  &\fd(\bi)&\f8(\az)\\
\cline{1-6}
1 &\eps   &\eps   &0\hf   &0       &\frac{c'}{8}     \\
2 &\eps   &2_\zb  &0\hf   &0       &\frac{c'}{8}+\frac{1}{16} \\
3 &2_\zb  &\eps   &0\hf   &\hf     &\frac{c'}{8}     \\
\cline{1-6}
\multicolumn{6}{l}{$\G_e=222,2'2'2\hfill
 (\d\in4221'\Rightarrow\d^8=\eps,
  \d2_\xb\d^{-1}=\alpha2_\xb\alpha^{-1}=2_\yb$)}\\
\cline{1-6}
  &\adad     & \fez & \fey    & \fd(\bi) & \f8(\az)\\
\cline{1-6}
4 &2_\zb     & \hf0 & \hf\hf  & \hf      &\frac{c'}{8}\\
\cline{1-6}
\multicolumn{6}{l}{$\G_e=n,n'\hfill
 (\d n_\zb\d^{-1}=\alpha n_\zb^{-1}\alpha^{-1},\adad=\al^2=\eps)$}\\
\cline{1-6}
  &\dnd     & \d^8    & \fenz          & \fd(\bi)   &\f8(\az)\\
\cline{1-6}
5 &n_\zb    & \eps    & 0\frac{j}{N}   &0   &\frac{c'}{8}       \\
6 &n_\zb    & n_\zb   & 0\frac{j}{n}   &0   &\frac{c'}{8}+\frac{j}{8n} \\
7 &n_\zb    & n^2_\zb & 0\frac{j}{N}   &0   &\frac{c'}{8}+\frac{j}{4N} \\
8 &n_\zb    & n^3_\zb & 0\frac{j}{n}   &0   &\frac{c'}{8}+\frac{3j}{8n}\\
9 &n_\zb    & n^4_\zb & 0\frac{j}{N}   &0   &\frac{c'}{8}+\frac{j}{2N} \\
10&n_\zb^{-1} & \eps  &\hf\frac{j}{n}(n=4)  &0   &\frac{c'}{8}\\
\cline{1-6}
\end{tabular}
\end{center}
\end{table}


\begin{table}
\caption{Spin space-group types on $V$-lattices with
  $G=\bar{8}2m$.
  The possible values of $a$, $b$, $a'$, $b'$, $N$, $j$, and $d$, as well as the
  notations $A_0$ and $A_1$ are as explained in the caption of Table \ref{tab:8V}.
  Lines $3a$ and $3b$ refer to distinct spin space-group types if $\G_e=2'2'2$
  but are scale-equivalent if $\G_e=222$, or $2'2'2'$, for which line
  $3a$ suffices. $\grave{a}$ and $\grave{b}$ denote either 0 or $\hf$.
  Note that for all choices of spin-space operations $\d^8=\eps$ and $\al^2=\eps$.
  Spin space-group symbols are of the form $\PL \bar{8}^\d 2^\al m^{\d\al}$ where
  $2^\al$ is replaced by $2_1^\al$ if $\grave{a}=\hf$, and $m^{\d\al}$
  is replaced by $c^{\d\al}$ if $\grave{b}=\hf$. A subscript $a$ is
  added to the secondary 2 if $\fd(\bi)\equiv\grave{a}+A_0$.}
\label{tab:bar82mV}
\setlength{\extrarowheight}{7pt}
\begin{center}
\begin{tabular}{>{$}l<{$}|>{$}l<{$}>{$}l<{$}>{$}l<{$}>{$}l<{$}>{$}l<{$}>{$}l<{$}>{$}l<{$}}
\cline{1-8}
\multicolumn{8}{l}{$\G_e=1,1',2,2',21'$}\\
\cline{1-8}
  &\adad    & \fez  & \fee  & \fd(\bi)    & \fd(\zv)\\
\cline{1-8}
1 &\eps     & ab    & a'b'  & \grave{a}   & \grave{b}\\
2 &2_\zb    &0\hf   &a'b'   &\grave{a}    & \grave{b}+\qt\\
  &         &\hf b  &a'b'   &\grave{a}+A_0& \grave{b}+\hf a\\
\cline{1-8}
\multicolumn{8}{l}{$\G_e=222,2'2'2,2'2'2'\hfill
 (\al 2_\xb\al^{-1}=2_\xb)$}\\
\cline{1-8}
  &\dxd     &\adad  &\fex   &\fey    &\fezp  &\fd(\bi)     &\fd(\zv)\\
\cline{1-8}
3a&2_\xb    &\eps   &0\hf   &\hf0    &--     &\grave{a}    &\grave{b}\\
3b&         &       &ab     &\hf\hf  &--     &\grave{a}    &\grave{b}\\
4 &2_\yb    &2_\zb  &A_0b   &A_1b    &a\hf   &\grave{a}+A_0&\grave{b}\\
\cline{1-8}
\multicolumn{8}{l}{$\G_e=n,n',n1'\hfill
 (\dnd=\ana=n_\zb^{-1})$}\\
\cline{1-8}
  &\adad  & \fenz          & \fee   & \fd(\bi)      &\fd(\zv)\\
\cline{1-8}
5 &\eps   & a\frac{dj}{N}  & a'b'   & \grave{a}     &\grave{b}\\
6 &n_\zb  & 0\frac{j}{n}   & a'b'   & \grave{a}     &\grave{b}-\frac{j}{2n}\\
  &       & \hf\frac{dj}{n}& a'b'   & \grave{a}+A_0 &\grave{b}-\frac{dj}{2n}\\
\cline{1-8}
\end{tabular}
\end{center}
\end{table}


\begin{table}
\caption{Spin space-group types for $S$-lattices with $G=\bar{8}2m$.
  The possible values of $N$ and $j$ are explained in
  the caption of Table \ref{tab:8S}, and the notation $A_0$ is as
  explained in the caption of Table~\ref{tab:8V}.
  $\grave{b}$ denotes either 0 or $\hf$.
   Note that since $\bar{8}$ is a generator of $G$
  $\d^8=\eps$. Spin space-group symbols are $\SL\bar{8}^\d 2^\al
  m^{\d\al}$ if $\grave{b}=0$ and $\SL\bar{8}^\d 2^\al c^{\d\al}$ if
  $\grave{b}=\hf$. The secondary 2 is replaced by $2_1$ if $\fd(\bi)\equiv\hf$
  and by $2_a$ if $\fd(\bi)\equiv0\hf0\hf$.}
\label{tab:bar82mS}
\setlength{\extrarowheight}{7pt}
\begin{center}
\begin{tabular}{>{$}l<{$}|>{$}l<{$}>{$}l<{$}>{$}l<{$}>{$}l<{$}>{$}l<{$}}
\cline{1-6}
\multicolumn{6}{l}{$\G_e=1,1',2,2'$}\\
\cline{1-6}
  &\adad     &\al^2  &\feg(\bi\cv)  &  \fd(\bi)  &  \fd(\az)\\
\cline{1-6}
1 &\eps      &\eps   &0\hf          &0           &\grave{b}\\
2 &\eps      &2_\zb  &0\hf          &\hf         &\grave{b}+\qt\\
3 &2_\zb     &\eps   &0\hf          &0           &\grave{b}-\qt\\
4 &2_\zb     &2_\zb  &0\hf          &\hf         &\grave{b}\\
\cline{1-6}
\multicolumn{6}{l}{$\G_e=222,2'2'2\hfill
 (\dxd=\axa=2_\yb\Rightarrow\adad=\al^2=\eps)$}\\
\cline{1-6}
  &\fex   &   \fey   &  \fd(\bi)   &  \fd(\az)   \\
\cline{1-6}
5 &\hf0   & \hf\hf   & \hf         & \grave{b}\\
\cline{1-6}
\multicolumn{6}{l}{$\G_e=n,n'\hfill (\dnd=\ana,\al^2=\eps)$}\\
\cline{1-6}
  &\dnd        & \adad    &\fenz         &\fd(\bi)   &\fd(\az)\\
\cline{1-6}
6 &n_\zb^{-1}  & \eps     &0\frac{j}{N}  &0          &\grave{b}-\frac{j}{2N}\\
7 &n_\zb^{-1}  & n_\zb    &0\frac{j}{n}  &0          &\grave{b}-\frac{j}{2n}\\
8 &n_\zb       &\eps      &\hf\frac{j}{n}(n=4)&A_0        &\grave{b}\\
9 &n_\zb       &n_\zb     &\hf\frac{j}{n}(n=4)&A_0        &\grave{b}-\frac{j}{8}\\
\cline{1-6}
\end{tabular}
\end{center}
\end{table}


\begin{table}
\caption{Spin space-group types on $V$-lattices with $G=8/m$.
  The possible values of $a$, $b$, $a'$, $b'$, $N$, $j$, and $d$, as well as the
  notations $A_0$ and $A_1$ are as explained in the caption of Table \ref{tab:8V}.
  Lines $3a$ and $3b$ refer to distinct spin space-group types if $\G_e=2'2'2$
  but are scale-equivalent if $\G_e=222$, or $2'2'2'$, for which line $3a$ suffices.
  $\check{a}$ and $\ta$ denote either 0 or $\hf$.
  Note that $\h^2=\eps$. Spin space-group symbols are of the form $\PL8^\d/m^\h$ where
  $8^\d$ is replaced by $8^\d_4$ if $\ta=\hf$ and $m^\h$ is replaced
  by $n^\h$ if $\check{a}=\hf$, a subscript $a$ is added if
  $\fh(\bi)\equiv \check{a}+A_0$.}
\label{tab:8/mV}
\setlength{\extrarowheight}{7pt}
\begin{center}
\begin{tabular}{>{$}l<{$}|>{$}l<{$}>{$}l<{$}>{$}l<{$}>{$}l<{$}>{$}l<{$}>{$}l<{$}}
\cline{1-7}
\multicolumn{7}{l}{$\G_e=1,1',2,2',21'\hfill (\d^8=\eps)$}\\
\cline{1-7}
  &\dhdh   & \fez  & \fee & \fh(\bi)         & \f8(\zv)\\
\cline{1-7}
1 &\eps    &ab     &a'b'  & \check{a}        & \tilde{a}\\
2 &2_\zb   &0\hf   &\hf b & \check{a}        &\tilde{a}+\qt\\
  &        &\hf b  &a'b'  & \check{a}+A_0    &\tilde{a}+\frac{b}{2}\\
\cline{1-7}
\multicolumn{7}{l}{$\G_e=222,2'2'2,2'2'2'\hfill
 (\h 2_\xb\h^{-1}=2_\xb,\d^{-1}\h\d\h=\d^8=\eps)$}\\
\cline{1-7}
  &\dxd      & \fex & \fey & \fezp & \fh(\bi) & \f8(\zv)\\
\cline{1-7}
3a&2_\xb     &0\hf  &\hf0  & --    & \check{a}&\tilde{a} \\
3b&          &ab    &\hf\hf& --    & \check{a}&\tilde{a} \\

4 &2_\yb     &A_0b    &A_1b  &a\hf   & \check{a}& \tilde{a}\\
\cline{1-7}
\multicolumn{7}{l}{$\G_e=n,n'\hfill
 (\d n_\zb\d^{-1}=n_\zb, \h n_\zb\h^{-1}=n_\zb^{-1},
  \d^8=(\d^{-1}\h\d\h)^{-4})$}\\
\cline{1-7}
  &\dhdh     & \fenz           & \fee & \fh(\bi)     &\f8(\zv)\\
\cline{1-7}
5 &\eps      & a\frac{dj}{N}   & a'b' &\check{a}     &\tilde{a}\\
6 &n_\zb^{-1}& 0\frac{j}{n}    & ab   &\check{a}     &\tilde{a}\\
  &          & \hf\frac{dj}{n} & ab   &\check{a}+A_0 &\tilde{a}+\frac{dj}{2n}\\
\cline{1-7}
\end{tabular}
\end{center}
\end{table}


\begin{table}
\caption{Spin space-group types on $S$-lattices with $G=8/m$.
  The possible values of $N$ and $j$ are explained in the caption of
  Table \ref{tab:8S}, and the notation $A_0$ is explained in the caption of Table
  \ref{tab:8V}.  Note that $\h^2=\eps$. $\check{a}$ denotes either $0$
  or $\hf$. Spin space-group symbols are $\SL 8^\d/m^\h$ if $\check{a}\equiv
  0$ and $\SL 8_2^\d/a^\h$ if $\check{a}\equiv\hf$.}
\label{tab:8/mS}
\setlength{\extrarowheight}{8pt}
\begin{center}
\begin{tabular}{>{$}l<{$}|>{$}l<{$}>{$}l<{$}>{$}l<{$}>{$}l<{$}>{$}l<{$}}
\cline{1-6}
\multicolumn{6}{l}{$\G_e=1,1',2,2'\hfill (\d^8=\eps)$}\\
\cline{1-6}
  &\dhdh   & \feg(\bi\cv)   & \fh(\bi)   & \f8(\az)\\
\cline{1-6}
1 &\eps    & 0\hf           & \check{a}  &\hf\check{a}\\
2 &2_\zb   & 0\hf           & \check{a}  &\hf{\check{a}}+\qt\\
\cline{1-6}
\multicolumn{6}{l}{$\G_e=222,2'2'2\hfill
 (\h 2_\xb\h^{-1}=2_\xb,\d^{-1}\h\d\h=\d^8=\eps)$}\\
\cline{1-6}
  &\dxd    & \fex & \fey & \fh(\bi)  & \f8(\az)\\
\cline{1-6}
3 &2_\yb   & \hf0 &\hf\hf& \check{a} & \hf\check{a}\\
\cline{1-6}
\multicolumn{6}{l}{$\G_e=n,n'\hfill
 (\h n_\zb\h^{-1}=n_\zb^{-1},\d^8=(\d^{-1}\h\d\h)^{-4})$}\\
\cline{1-6}
  &\dnd      & \d^{-1}\h\d\h  & \fenz      &\fh(\bi)   &\f8(\az)\\
\cline{1-6}
4 &n_\zb     & \eps      & 0\frac{j}{N}    &\check{a}  &\hf\check{a}\\
5 &n_\zb     & n_\zb^{-1}& 0\frac{j}{n}    & \check{a} &\hf\check{a}+\frac{j}{2n}\\
6 &n^{-1}_\zb & \eps     & \hf\frac{j}{n}(n=4)  & \check{a} &\hf\check{a}\\
7 &n^{-1}_\zb  & n_\zb^{-1} & \hf\frac{j}{n}(n=4)  & \check{a}+A_0
   &\hf\check{a}-\hf\frac{j}{n}(n=4)\\
\cline{1-6}
\end{tabular}
\end{center}
\end{table}


\begin{table}
\caption{Spin space-group types on $V$-lattices with $G=8/mmm$.
  The possible values of $a$, $b$, $a'$, $b'$, $N$, $j$, and $d$, as well as the
  notations $A_0$ and $A_1$ are explained in the caption of Table \ref{tab:8V}.
  Lines $7a$ and $7b$ refer to distinct spin space-group types if $\G_e=2'2'2$
  but are scale-equivalent if $\G_e=222$, or $2'2'2'$, for which line $7a$ suffices.
  Note that $\h^2=\hmhm=\eps$. $\hat{a}$,\ $\hat{b}$,\ $\check{a}$, $\ta$
  and $\tb$ denote either 0 or $\hf$.
  Spin space-group symbols are of the form
  $\PL8^\d/m^\h m^\mu m^{\d\mu}$, where $8^\d$ and $m^\h$ are replaced
  as in Table \ref{tab:8/mV} above, and $m^\mu$ and $m^{\d\mu}$ are replaced as in
  Table \ref{tab:8mmV} above.}
\label{tab:8/mmmV}
\setlength{\extrarowheight}{8pt}
\begin{center}
\begin{tabular}{>{$}l<{$}|>{$}l<{$}>{$}l<{$}>{$}l<{$}>{$}l<{$}>{$}l<{$}>{$}l<{$}>{$}l<{$}>{$}l<{$}>{$}l<{$}>{$}l<{$}}
\cline{1-11}
\multicolumn{11}{l}{$\G_e=1,1',2,2',21'\hfill
 (\h^{-1}\mu\h\mu=\h^2=\d^8=\eps)$}\\
\cline{1-11}
  &\mdmd & \dhdh & \mu^2 & \fez & \fee & \fm(\bi) & \fm(\zv)
    &\fh(\bi) & \f8(\zv)\\
\cline{1-11}
1 &\eps  &\eps   & \eps  &ab    &a'b'   &\hat{a}     &\hat{b}
  &\check{a}    &\ta\\
2 &\eps  &\eps   & 2_\zb &0\hf  &\hf b' &\hat{a}     &\hat{b}+\qt
  &\check{a}    &\ta\\
3 &\eps  &2_\zb  &\eps   &\hf0  &a'\hf  &\hat{a}     &\hat{b}
  &\check{a}+A_0&\ta\\
4 &2_\zb &\eps   &\eps   &\hf0  &a'\hf  &\hat{a}+A_0 &\hat{b}
  &\check{a}    &\ta\\
5 &2_\zb &2_\zb  &\eps   &0\hf  &\hf b' &\hat{a}     &\hat{b}
  &\check{a}    &\ta+\qt\\
  &      &       &       &\hf b &a'b'   &\hat{a}+A_0 &\hat{b}
  &\check{a}+A_0&\ta+\frac{b}{2}\\
6 &2_\zb &2_\zb  &2_\zb  &0\hf  &\hf b' &\hat{a}     &\hat{b}+\qt
  &\check{a}    &\ta+\qt\\
\cline{1-11}
\multicolumn{11}{l}{$\G_e=222,2'2'2,2'2'2'\hfill
 (\mu 2_\xb\mu^{-1}=\h 2_\xb\h^{-1}=2_\xb\Rightarrow\mu^2=\eps)$}\\
\cline{1-11}
  &\dxd   &\mdmd   &  \dhdh  &\fex  &\fey &\fezp &\fm(\bi) &\fm(\zv)
    &\fh(\bi) &\f8(\zv)\\
\cline{1-11}
7a&2_\xb  &\eps    & \eps    &0\hf  &\hf0    &--     &\hat{a}
  &\hat{b}        &\check{a}    &\tb\\
7b&       &        &         &ab    &\hf\hf  &--     &\hat{a}
  &\hat{b}        &\check{a}    &\tb\\
8 &2_\yb  &2_\zb   &\eps     &A_0b  &A_1b    &a\hf   &\hat{a}+A_0
  &\hat{b}        &\check{a}    &\tb\\
\cline{1-11}
\multicolumn{11}{l}{$\G_e=n,n',n1'\hfill
 (\dnd=\mu n_\zb\mu^{-1}=n_\zb, \h n_\zb\h^{-1}=n^{-1}_\zb,
  \d^8=(\mdmd)^4, (\dhdh)^{-1}=\mdmd)$}\\
\cline{1-11}
  &\mdmd    & \mu^2  & \fenz & \fee  &\fm(\bi) &\fm(\zv) &\fh(\bi) &\f8(\zv) \\
\cline{1-11}
9 &\eps     &\eps    &a\frac{dj}{N}   & a'b'  &\hat{a}
  &\hat{b}      &\check{a}    &\tb\\
10&\eps     &n_\zb   &0\frac{j}{n}    &a'b'   &\hat{a}
  &\hat{b}+\frac{j}{2n}  &\check{a}   &\tb\\
11&n_\zb    &\eps    &0\frac{j}{n}    &a'b'   &\hat{a}
  &\hat{b}      &\check{a}    &\tb+\frac{j}{2n}\\
  &         &        &\hf\frac{dj}{n} &a'b'   &\hat{a}+A_0
  &\hat{b}      &\check{a}+A_0&\tb+\frac{dj}{2n}\\
12&n_\zb    &n_\zb   &0\frac{j}{n}    &a'b'   &\hat{a}
  &\hat{b}+\frac{j}{2n}   &\check{a}   &\tb+\frac{j}{2n}\\
\cline{1-11}
\end{tabular}
\end{center}
\end{table}


\begin{table}
\caption{Spin space-group types on $S$-lattices with $G=8/mmm$.
  The possible values of $N$ and $j$ are explained in
  the caption of Table \ref{tab:8S}, and the notation $A_0$ is
  explained in the caption of Table \ref{tab:8V}.
  Note that $\h^2=\eps$. $\ta$ and $\tb$ denote either 0 or $\hf$. Spin
  space-group symbols are $\SL 8^\d/m^\h m^\mu m^{\d\mu}$ if
  $\ta=\tb=0$, $\SL 8^\d/m^\h m^\mu c^{\d\mu}$ if $\ta=0$ and
  $\tb=\hf$, $\SL 8^\d/n^\h d^\mu m^{\d\mu}$ if $\ta=\hf$ and $\tb=0$
  and $\SL 8^\d/n^\h d^\mu c^{\d\mu}$ if $\ta=\tb=\hf$. For example, if $\G_e=7$
  , $G_\eps=4mm$ and $\G=(14)2'2'$ the spin-space operations can be chosen to be
  $\d=(14)_\zb$ and $\h=2'_\xb$, $\mu$ is necessarily $\eps$ since $m\in G_\eps$.
  The corresponding line in the Table is 12, if $\ta=0$, $\tb=\hf$ and
  $\Phi_e^{7_\zb}(\az)\equiv\frac{2}{7}$,
  the spin space-group symbol is $S^7_{7_2C}8^{14}m^{2_\xb}mc^{14}$.}
\label{tab:8/mmmS}
\setlength{\extrarowheight}{8pt}
\begin{center}
\begin{tabular}{>{$}l<{$}|>{$}l<{$}>{$}l<{$}>{$}l<{$}>{$}l<{$}>{$}l<{$}>{$}l<{$}>{$}l<{$}>{$}l<{$}>{$}l<{$}>{$}l<{$}}
\cline{1-11}
\multicolumn{11}{l}{$\G_e=1,1',2,2'\hfill (\h^2=\d^8=\eps)$}\\
\cline{1-11}
  &\mdmd & \dhdh & \hmhm & \mu^2 & \feg(\kv) & \fm(\bi) & \fm(\az)
     & \fh(\bi) & \f8(\az)\\
\cline{1-11}
1 &\eps  &\eps   & \eps  & \eps  & 0\hf &\ta &\frac{\ta}{2}+\tb
  &\ta       &\frac{\ta}{2}\\
2 &\eps  &\eps   & \eps  & 2_\zb & 0\hf &\ta &\frac{\ta}{2}+\tb+\qt
  &\ta       &\frac{\ta}{2}\\
3 &\eps  &2_\zb  &2_\zb  &\eps   & 0\hf &\ta &\frac{\ta}{2}+\tb
  &\ta+\hf   &\frac{ta}{2}\\
4 &2_\zb &\eps   & 2_\zb &\eps   & 0\hf &\ta &\frac{\ta}{2}+\tb
  &\ta+\hf   &\frac{\ta}{2}+\qt\\
5 &2_\zb &\eps   &2_\zb  &2_\zb  &0\hf  &\ta &\frac{\ta}{2}+\tb+\qt
  &\ta+\hf   &\frac{\ta}{2}+\qt\\
6 &2_\zb &2_\zb  &\eps   &\eps   &0\hf  &\ta &\frac{\ta}{2}+\tb
  &\ta       &\frac{\ta}{2}+\qt\\
7 &2_\zb &2_\zb  &\eps   &2_\zb  &0\hf  &\ta &\frac{\ta}{2}+\tb+\qt
  &\ta       &\frac{\ta}{2}+\qt\\
\cline{1-11}
\multicolumn{11}{l}{$\G_e=222,2'2'2\hfill
 (\mu2_\xb\mu^{-1}=\d2_\xb\d^{-1}=2_\yb,\h2_\xb\h^{-1}=2_\xb)$}\\
\cline{1-11}
  &\mdmd & \dhdh & \hmhm & \mu^2 & \fex &\fey &\fm(\bi) &\fm(\az)
    &\fh(\bi)  &\f8(\az)\\
\cline{1-11}
8 &2_\zb &\eps   &2_\zb  &2_\zb  &\hf0  &\hf\hf  &\ta
  &\frac{\ta}{2}+\tb+\qt &\ta+\hf   &\frac{\ta}{2}+\qt\\
\cline{1-11}
\multicolumn{11}{l}{$\G_e=n,n'\hfill
 (\h n_\zb\h^{-1}=n^{-1}_\zb,\d n_\zb\d^{-1}=\mu n_\zb\mu^{-1},
  \mu^{-1}\d\mu\d=(\d^{-1}\h\d\h)^{-1}, \h^{-1}\mu\h\mu=\eps,
  \d^8=(\mu^{-1}\d\mu\d)^4)$}\\
\cline{1-11}
  &\dnd &\mdmd & \mu^2 & \fenz &\fm(\bi) &\fm(\az) &\fh(\bi) &\f8(\az)\\
\cline{1-11}
9 &n_\zb       &\eps       &\eps    & 0\frac{j}{N}   &\ta
  &\frac{\ta}{2}+\tb              &\ta      &\frac{\ta}{2}\\
10&n_\zb       &\eps       &n_\zb   & 0\frac{j}{n}   &\ta
  &\frac{\ta}{2}+\frac{j}{2n}+\tb &\ta      &\frac{\ta}{2}\\
11&n_\zb       &n_\zb      &\eps    & 0\frac{j}{n}   &\ta
  &\frac{\ta}{2}+\tb              &\ta      &\frac{\ta}{2}+\frac{j}{2n}\\
12&n_\zb       &n_\zb      &n_\zb   & 0\frac{j}{n}   &\ta
  &\frac{\ta}{2}+\frac{j}{2n}+\tb &\ta      &\frac{\ta}{2}+\frac{j}{2n}\\
13&n_\zb^{-1}  &\eps       &\eps    & \hf\frac{j}{n}(n=4) &\ta
  &\frac{\ta}{2}+\tb              &\ta      &\frac{\ta}{2}\\
14&n_\zb^{-1}  &n_\zb^{-1} &\eps    & \hf\frac{j}{n}(n=4) &\ta+A_0
  &\frac{\ta}{2}+\tb              &\ta+A_0  &\frac{\ta}{2}-\qt\\
15&n_\zb^{-1}  &n_\zb      &\eps    & \hf\frac{j}{n}(n=4) &\ta+A_0
  &\frac{\ta}{2}+\tb              &\ta+A_0  &\frac{\ta}{2}+\qt\\
\cline{1-11}
\end{tabular}
\end{center}
\end{table}


\begin{table}
\caption{Restrictions on the form of $\sv(\kv)$ for any wave vector
  $\kv$ in the magnetic lattice $L$ when $\G_e=2$, $2'$, or $1'$. In each
  case the form of $\sv(\kv)$ depends on the particular values of the
  phases $\feg(\bi\cv)$, where $\g$ is the generator of $\G_e$, and on
  the parities of $\sum n_i$ and $l$, where $\kv=\sum_{i=1}^4 n_i\bi +
  l\cv$. Each entry in the Table contains three values for the form of
  $\sv(\kv)$: the one on the
  left is for $\G_e=2$; the one in the center is for $\G_e=2'$; and
  the one on the right is for $\G_e=1'$.}
\label{tab:sel-2}
\setlength{\extrarowheight}{3pt}
\begin{center}
\begin{tabular}{>{$}c<{$}|>{$}c<{$}|>{$}c<{$}>{$}c<{$}>{$}c<{$}>{$}c<{$}}
\multicolumn{2}{c|}{Lattice spin group}&\multicolumn{2}{c}{$\sum n_i$
  even}&\multicolumn{2}{c}{$\sum n_i$ odd}\\
\multicolumn{1}{c}{Symbols} & \Phi_e^\g(\bi\cv)
  &l \text{ even} &l \text{ odd} &l \text{ even} &l \text{ odd}\\
\cline{1-6}
P^\g_{2c}, S^\g_P & 0\hf& \sz/\sxy/0 &\sxy/\sz/\sv & \sz/\sxy/0 &\sxy/\sz/\sv\\
P^\g_P & \hf 0& \sz/\sxy/0 & \sz/\sxy/0 &\sxy/\sz/\sv &\sxy/\sz/\sv\\
P^\g_S & \hf\hf& \sz/\sxy/0 &\sxy/\sz/\sv &\sxy/\sz/\sv & \sz/\sxy/0 \\
\end{tabular}
\end{center}
\end{table}


\begin{table}
\caption{Restrictions on the form of $\sv(\kv)$ for any wave vector
  $\kv\in L$ when $\G_e=21'$. In each
  case the form of $\sv(\kv)$ depends on the particular values of the
  phase functions for the generators of $\G_e$, and on
  the parities of $\sum n_i$ and $l$, where $\kv=\sum_{i=1}^4 n_i\bi +
  l\cv$. }
\label{tab:sel-21'}
\setlength{\extrarowheight}{3pt}
\begin{center}
\begin{tabular}{>{$}c<{$}|>{$}c<{$}|>{$}c<{$}|>{$}c<{$}>{$}c<{$}>{$}c<{$}>{$}c<{$}}
\multicolumn{3}{c|}{Lattice spin group}&\multicolumn{2}{c}{$\sum n_i$
  even}&\multicolumn{2}{c}{$\sum n_i$ odd}\\
\multicolumn{1}{c}{Symbols} &
  \multicolumn{1}{c}{$\Phi_e^{2_\zb}(\bi\cv)$} &\Phi_e^{\eps'}(\bi\cv)
  &l \text{ even} &l \text{ odd} &l \text{ even} &l \text{ odd}\\
\cline{1-7}
P^{21'}_{2c,P} & 0\hf & \hf 0 & 0 & 0 & \sz & \sxy \\
P^{21'}_{2c,S} & 0\hf & \hf\hf & 0 & \sxy & \sz & 0 \\
P^{21'}_{P,2c} & \hf 0 & 0\hf & 0 & \sz & 0 & \sxy \\
P^{21'}_{P,S} & \hf 0 & \hf\hf & 0 & \sz & \sxy & 0 \\
P^{21'}_{S,2c} & \hf\hf & 0\hf & 0 & \sxy & 0 & \sz \\
P^{21'}_{S,P} & \hf\hf & \hf 0 & 0 & 0 & \sxy & \sz \\
\end{tabular}
\end{center}
\end{table}


\begin{table}
\caption{Restrictions on the form of $\sv(\kv)$ for any wave vector
  $\kv$ in the magnetic lattice $L$ when $\G_e=222$ and $2'2'2$. In each
  case the form of $\sv(\kv)$ depends on the particular values of the
  phase functions for the generators $2^*_\xb$ and $2^*_\yb$ of
  $\G_e$, where the asterisk denotes an optional prime, and on the parities of
  $n_1+n_3$, $n_2+n_4$, and $l$, where $\kv=\sum_{i=1}^4 n_i\bi +
  l\cv$. Each entry in the Table contains two values for the form of
  $\sv(\kv)$: the one on the left is for $\G_e=222$; and the one on
  the right is for $\G_e=2'2'2$.}
\label{tab:sel-222}
\setlength{\extrarowheight}{3pt}
\begin{center}
\begin{tabular}{>{$}c<{$}|>{$}c<{$}|>{$}c<{$}|>{$}c<{$}>{$}c<{$}>{$}c<{$}>{$}c<{$}>{$}c<{$}>{$}c<{$}>{$}c<{$}>{$}c<{$}}
\multicolumn{3}{c|}{}&\multicolumn{4}{c}{$n_1+n_3$
  even}&\multicolumn{4}{c}{$n_1+n_3$ odd}\\
\multicolumn{3}{c|}{Lattice spin group}&\multicolumn{2}{c}{$n_2+n_4$
  even}&\multicolumn{2}{c}{$n_2+n_4$ odd}
&\multicolumn{2}{c}{$n_2+n_4$ even}&\multicolumn{2}{c}{$n_2+n_4$ odd}\\
\multicolumn{1}{c}{Symbols} &
  \multicolumn{1}{c}{$\Phi_e^{2^*_\xb}(\bi\cv)$} &
  \Phi_e^{2^*_\yb}(\bi\cv) &
l\text{ even}& l\text{ odd}& l\text{ even}& l\text{ odd}&
l\text{ even}&l \text{ odd}& l\text{ even}& l\text{ odd}\\
\cline{1-11}
P^{222}_{2c,P,S}, P^{2'2'2}_{2c,P,S}  & 0\hf & \hf 0
& 0/\sz & \sy/\sx & \sx/\sy & \sz/0 & \sx/\sy & \sz/0 & 0/\sz & \sy/\sx\\
P^{2'2'2}_{2c,S,P} & 0\hf & \hf\hf
& 0/\sz & \sz/0 & \sx/\sy & \sy/\sx & \sx/\sy & \sy/\sx & 0/\sz & \sz/0\\
P^{2'2'2}_{P,S,2c}, S^{222}_{\icp,P}, S^{2'2'2}_{\icp,P} & \hf 0& \hf\hf
& 0/\sz & \sx/\sy & \sz/0 & \sy/\sx & \sz/0 & \sy/\sx & 0/\sz & \sx/\sy\\
P^{222}_{\ppp,P},P^{2'2'2}_{\ppp,P} & A_0 0 & A_1 0
& 0/\sz & 0/\sz & \sy/\sx & \sy/\sx & \sx/\sy & \sx/\sy & \sz/0 & \sz/0\\
P^{222}_{\ipp,P}, P^{2'2'2}_{\ipp,P} & A_0\hf & A_1\hf
& 0/\sz & \sz/0 & \sy/\sx & \sx/\sy & \sx/\sy & \sy/\sx & \sz/0 & 0/\sz\\
\end{tabular}
\end{center}
\end{table}


\begin{table}
\caption{Restrictions on the form of $\sv(\kv)$ for any wave vector
  $\kv\in L$ when $\G_e=2'2'2'$. In each case the form of $\sv(\kv)$
  depends on the particular values of the phase functions for the
  generators $2'_\xb$, $2'_\yb$, and $2'_\zb$ of $\G_e$, and on the
  parities of $n_1+n_3$, $n_2+n_4$, and $l$, where $\kv=\sum_{i=1}^4
  n_i\bi + l\cv$.}
\label{tab:sel-2221'}
\setlength{\extrarowheight}{3pt}
\begin{center}
\begin{tabular}{>{$}c<{$}|>{$}c<{$}|>{$}c<{$}|>{$}c<{$}|>{$}c<{$}>{$}c<{$}>{$}c<{$}>{$}c<{$}>{$}c<{$}>{$}c<{$}>{$}c<{$}>{$}c<{$}}
\multicolumn{4}{c|}{}&\multicolumn{4}{c}{$n_1+n_3$
  even}&\multicolumn{4}{c}{$n_1+n_3$ odd}\\
\multicolumn{4}{c|}{Lattice spin group}&\multicolumn{2}{c}{$n_2+n_4$
  even}&\multicolumn{2}{c}{$n_2+n_4$ odd}
&\multicolumn{2}{c}{$n_2+n_4$ even}&\multicolumn{2}{c}{$n_2+n_4$ odd}\\
\multicolumn{1}{c}{Symbols} &
  \multicolumn{1}{c}{$\Phi_e^{2'_\xb}(\bi\cv)$} &
  \multicolumn{1}{c}{$\Phi_e^{2'_\yb}(\bi\cv)$} & \Phi_e^{2'_\zb}(\bi\cv) &
l\text{ even}& l\text{ odd}& l\text{ even}& l\text{ odd}&
l\text{ even}&l \text{ odd}& l\text{ even}& l\text{ odd}\\
\cline{1-12}
P^{2'2'2'}_{\ppp,2c} & A_0 0 & A_1 0
 & 0\hf & 0 & \sz & \sx & 0 & \sy & 0 & 0 & 0 \\
P^{2'2'2'}_{\ppp,S} & A_0 0 & A_1 0
 & \hf\hf & 0 & \sz & 0 & \sx & 0 & \sy & 0 & 0 \\
P^{2'2'2'}_{\ipp,2c} & A_0\hf & A_1\hf
 & 0\hf & 0 & 0 & \sx & 0 & \sy & 0 & 0 & \sz \\
P^{2'2'2'}_{\ipp,S} & A_0\hf & A_1\hf
 & \hf\hf & 0 & 0 & 0 & \sy & 0 & \sx & 0 & \sz \\
\end{tabular}
\end{center}
\end{table}


\begin{table}
\caption{Restrictions on the form of $\sv(\kv)$ for any wave vector
  $\kv\in L$ when $\G_e=n$, $n'$ and $n1'$. In each case the form of $\sv(\kv)$
  depends on the particular value of the phase functions for the
  generators of $\G_e$, on the parity of $\sum n_i$, and on the
  value of $l$, where $\kv=\sum_{i=1}^4 n_i\bi + l\cv$. The conditions
  for obtaining non-extinct $\sv(\kv)$ are listed separately for
  $\G_e=n$, $\G_e=n'$ (even $n$), and $\G_e=n'$ (odd $n$). The
  restrictions on the form of $\sv(\kv)$ for $\G_e=n1'$ are obtained
  from those of $\G_e=n$ with the additional requirement that
  $\sv(\kv)=0$ if: (1) $\Phi_e^{\eps'}(\bi\cv)\equiv0\hf$ and $l$
  is even; (2) $\Phi_e^{\eps'}(\bi\cv)\equiv\hf 0$ and $\sum n_i$
  is even; or (3) $\Phi_e^{\eps'}(\bi\cv)\equiv\hf\hf$ and $\sum n_i+l$
  is even.}
\label{tab:sel-n}
\setlength{\extrarowheight}{3pt}
\begin{center}
\begin{tabular}{>{$}c<{$}|>{$}c<{$}||>{$}c<{$}|>{$}c<{$}||>{$}c<{$}|>{$}c<{$}||>{$}c<{$}>{$}c<{$}}
\multicolumn{2}{c||}{$\G_e=n$} &
\multicolumn{2}{c||}{$\G_e=n'$ (even $n$)} &
\multicolumn{2}{c||}{$\G_e=n'$ (odd $n$)} & & \\
\cline{1-8}
\multicolumn{2}{c||}{$P^n_{n_jc}, S^n_{n_jc}$} &
\multicolumn{2}{c||}{$P^{n'}_{n_jc}, S^{n'}_{n_jc}$} &
\multicolumn{2}{c||}{$P^{n'}_{(2n)_jc}, S^{n'}_{(2n)_jc}$} & & \\
\cline{1-8}
\Phi_e^{n_\zb}(\bi\cv) & lj \text{\ (mod $n$)} &
\Phi_e^{n'_\zb}(\bi\cv) & lj \text{\ (mod $n$)} &
\Phi_e^{n'_\zb}(\bi\cv) & lj \text{\ (mod $2n$)} &
\sum n_i \text{ even} & \sum n_i \text{ odd}\\
\cline{1-8}
0\frac{j}{n} & 0 & 0\frac{j}{n} & \frac{n}2 & 0\frac{j}{2n} & n & \sz & \sz\\
&+1 & & \frac{n}2+1 & & n+2 & \splus & \splus\\
&-1 & & \frac{n}2-1 & & n-2 & \sminus & \sminus\\
& \text{otherwise} & & \text{otherwise} & & \text{otherwise} & 0 & 0\\
\cline{1-8}
\multicolumn{2}{c||}{$P^n_{n_jS}, S^4_{4_jS}$} &
\multicolumn{2}{c||}{$P^{n'}_{n_jS}, S^{4'}_{4_jS}$} &
\multicolumn{2}{c||}{$P^{n'}_{(2n)_jS}$} & & \\
\cline{1-8}
\hf\frac{j}{n} & 0 & \hf\frac{j}{n} & \frac{n}2 &
 \hf\frac{j}{2n} & n & \sz & 0 \\
&+1 & & \frac{n}2+1 & & n+2 & \splus  & 0\\
&-1 & & \frac{n}2-1 & & n-2 & \sminus & 0\\
&\frac{n}{2} &  & 0 &  & 0  & 0 & \sz\\
&\frac{n}{2}+1 & & +1 & & +1 & 0 & \splus\\
&\frac{n}{2}-1 & & -1 & & -1 & 0 & \sminus\\
& \text{otherwise} & & \text{otherwise} & & \text{otherwise} & 0 & 0\\
\cline{1-8}
\multicolumn{2}{c||}{$P^n_{(\frac{n}2)_jP}$} &
\multicolumn{2}{c||}{$P^{n'}_{(\frac{n}2)_jP}$} &
\multicolumn{2}{c||}{$P^{n'}_{n_jP}$} & & \\
\cline{1-8}
\hf\frac{2j}{n}& 0 \text{ or } \frac{n}{2}&
\hf\frac{2j}{n}& 0 \text{ or } \frac{n}{2}&
\hf\frac{j}{n}& 0 \text{ or } n & \sz & 0\\
&\frac{n}{4}+\hf \text{ or } \frac{3n}{4}+\hf&
&\frac{n}{4}+\hf \text{ or } \frac{3n}{4}+\hf&
&\frac{n}{2}+1 \text{ or } \frac{3n}{2}+1& 0 & \splus\\
&\frac{n}{4}-\hf \text{ or } \frac{3n}{4}-\hf&
&\frac{n}{4}-\hf \text{ or } \frac{3n}{4}-\hf&
&\frac{n}2-1 \text{ or } \frac{3n}{2}-1& 0 & \sminus\\
& \text{otherwise} & & \text{otherwise} & & \text{otherwise} & 0 & 0\\
\end{tabular}
\end{center}
\end{table}


\begin{figure}
\caption{The 8-fold star containing the horizontal generating vectors and their
  negatives $\pm\b1\ldots\pm\b4$ is given by solid arrows. The
  horizontal shift $\hv$ [Eq.~\rf{eq:shift}] of the staggered stacking
  vector is denoted by a dashed arrow. The dotted lines indicate the
  orientations of the two types of vertical mirrors and dihedral axes,
  as described in the text.}
\label{fig:star3d}
\begin{picture}(400,400)(-200,-200)
\thicklines
\put(0,0){\vector(1,0){100}}
\put(0,0){\vector(1,1){70.71}}
\put(0,0){\vector(0,1){100}}
\put(0,0){\vector(-1,1){70.71}}
\put(0,0){\vector(-1,0){100}}
\put(0,0){\vector(-1,-1){70.71}}
\put(0,0){\vector(0,-1){100}}
\put(0,0){\vector(1,-1){70.71}}
\dashline[35]{20}(0,0)(120.71,50)
\drawline(110,51)(120.71,50)(115,41)
\thinlines
\dottedline{2}(-190,0)(190,0)
\dottedline{2}(-181.065,-75)(181.065,75)
\path(-200,-200)(-200,200)(200,200)(200,-200)(-200,-200)
\put(100,5){$\b1$}
\put(71,76){$\b2$}
\put(0,105){$\b3$}
\put(-73,76){$\b4$}
\put(-120,5){$\b5=-\b1$}
\put(-94,-82){$\b6=-\b2$}
\put(-23,-112){$\b7=-\b3$}
\put(50,-82){$\b8=-\b4$}
\put(100,47){$\hv$}
\put(150,5){$d$, $m$}
\put(140,70){$d'$, $m'$}
\end{picture}
\end{figure}

\appendix
\newpage

\section{Claculation of the phase functions associated with
  \textnormal{$\G_e$}\\{}
[Not to be included in the printed version of the paper]}
\label{sec:app-a} 

Before starting we note from inspection of Table~\ref{tab:normalsub}
that no quotient group $G/G_\eps$ contains an operation of order 3.
This implies, among other things, that $\G/\G_e$ cannot contain such
an operation and therefore that $\G$ cannot be cubic.  This then
implies that for any possible combination of $\G$ and $\G_e$,
\begin{equation}
  \label{eq:ddgdd}
  \forall \g\in\G_e, \d\in\G:\quad \d^2\g\d^{-2}=\g.
\end{equation}

Relation~\rf{eq:ddgdd} implies that one of three conditions must be
satisfied:
\begin{enumerate}
\item $\d\g\d^{-1}=\g$, or simply $\d$ and $\g$ commute.

\item $\d\g\d^{-1}=\g^{-1}$, where $\g^{-1}\neq\g$. This may happen if
  $\g$ is an $n$-fold rotation ($n>2$), possibly followed by
  time-inversion, and $\d$ is a perpendicular 2-fold rotation.

\item $\g$ is one of a pair of operations in $\G_e$ staisfying
$\d\g\d^{-1}=\g'$ and $\d\g'\d^{-1}=\g$. This may happen
 if the two operations are $2_\xb$ and $2_\yb$, or $2'_\xb$
and $2'_\yb$, and $\d$ is either a 4-fold rotation about the $\zb$
axis or a 2-fold rotation about the diagonal.
\end{enumerate}

We shall use relation \rf{eq:ddgdd}, and the three possibilities for
satisfying it, in order to calculate the constraints imposed on
$\Phi_e^\g(\kv)$ by the spin-space operations, paired with the
generators of the different octagonal point groups. Recall that
we write $\Phi_g^\g(\bi)\=abcd$ instead of fully writing
$\Phi_g^\g(\b1)\=a$, $\Phi_g^\g(\b2)\=b$, $\Phi_g^\g(\b3)\=c$, and
$\Phi_g^\g(\b4)\=d$; we write $\Phi_g^\g(\bi\cv)\=abcde$ to indicate
in addition that $\Phi_g^\g(\cv)\=e$; and occasionally, if the four
phases on the horizontal generating vectors are equal to $a$ we write
$\Phi_g^\g(\bi)\=a$ or $\Phi_g^\g(\bi\cv)\=a\ e$.

\subsection{Constraints imposed by \textnormal{$\d$} (all point groups)}

Let $g_8$ denote the 8-fold generator ($\rr$ or $\rb$), and recall
that $\d\in\G$ denotes the the operation paired with it in the spin
point group. The constraints imposed by the operation $\d$, which must
be satisfied by all the octagonal spin groups, yield the following
results:
\begin{enumerate}
\item[R1.] For any $\g\in\G_e$, the in-plane phases of $\Phi_e^\g$ are
  \begin{equation}
    \label{eq:R1phasesb}
    \Phi_e^\g(\bi)\=
    \begin{cases}
      abab & \text{$V$-lattice},\\
      aaaa & \text{$S$-lattice},
    \end{cases}
  \end{equation}
  where $a$ and $b$ are independently either 0 or $1/2$.
\end{enumerate}

\noindent {\sl\underline{Proof}:\/} Relation \rf{eq:ddgdd} together
with Eq.~\rf{eq:gccGe} yields
\begin{equation}
  \label{eq:R1a}
  \Phi_e^\g(\bi)\=\Phi_e^{\d^2\g\d^{-2}}(g_8^2\bi)\=\Phi_e^\g(g_8^2\bi).
\end{equation}
Thus, for any $\g\in\G_e$
\begin{equation}
  \label{eq:R1b}
  \Phi_e^\g(\b1) \= \Phi_e^\g(\b3) \=a;\quad \Phi_e^\g(\b2) \=
  \Phi_e^\g(\b4) \=b;
\end{equation}
and
\begin{equation}
  \label{eq:R1c}
  \Phi_e^\g(-\bi) \= \Phi_e^\g(\bi) \Longrightarrow
  \Phi_e^\g(\bi) \= 0 {\rm \ or\ } \frac12.
\end{equation}
The last result (due to the linearity of the phase function) implies
that each of the phases $a$ and $b$ in \rf{eq:R1b} can be either $0$
or $1/2$. No further constraints arise from application of
\rf{eq:gccGe} to the vertical stacking vector, on the other hand, for
the staggered stacking vector, using the fact [Table~\ref{tab:symmetry}] that
$g_8^2(\zv+\hv) = (\zv+\hv) + \b4 - \b1$, we obtain
\begin{equation}
  \label{eq:R1d}
  \Phi_e^\g(\zv+\hv)\= \Phi_e^\g(\zv+\hv) + b - a,
\end{equation}
implying that $a\=b$.

Note that as a consequence of R1---along with the fact that for 2-fold
operations all phases are either 0 or $1/2$, and the fact that no two
phase functions $\Phi_e^\g$ can be the same---on $S$-lattices there
can be no more than three operations of order 2 in $\G_e$.

\begin{enumerate}
\item[R2.] For any $\g\in\G_e$, if $\d$ commutes with $\g$
  ($\d\g\d^{-1}=\g$) and the 8-fold generator is $\rr$, then the
  in-plane phases of $\Phi_e^\g$ are
  \begin{equation}
    \label{eq:R2phasesb}
    \Phi_e^\g(\bi)\=
    \begin{cases}
      aaaa & \text{$V$-lattice},\\
      0000 & \text{$S$-lattice},
    \end{cases}
  \end{equation}
  where $a$ is either 0 or $1/2$.

  If the 8-fold generator is $\rb$ the in-plane phases of $\Phi_e^\g$
  are
  \begin{subequations}
  \begin{equation}
    \label{eq:R2barphasesb}
    \Phi_e^\g(\bi)\= aaaa
  \end{equation}
  regardless of the lattice type, and the phase on the stacking vector
  is
  \begin{equation}
    \label{eq:R2barphasesc}
    \Phi_e^\g(\cv)\=
    \begin{cases}
      c & \text{$V$-lattice},\\
      \frac{a}2 +c & \text{$S$-lattice},
    \end{cases}
  \end{equation}
  \end{subequations}
  where $a$ and $c$ are either 0 or $1/2$ but they cannot both be 0.
  As a consequence, on vertical lattices $\g$ is an operation of order
  2, and on staggered lattices $\g$ is of order 2 or 4 depending on
  whether $a\=0$ or $1/2$.
\end{enumerate}

\noindent {\sl\underline{Proof}:\/} From application of
relation~\rf{eq:gccGe} to the horizontal generating vectors we obtain
\begin{equation}
  \label{eq:R2a}
  \Phi_e^\g(\bi)\=\Phi_e^{\d\g\d^{-1}}(g_8\bi)\=\pm\Phi_e^\g(\b{i+1}),
\end{equation}
where the upper (positive) sign is for $g_8=\rr$ and the lower
(negative) sign (due to the linearity of the phase function) is for
$g_8=\rb$. This implies that on $V$-lattices the two phases $a$ and
$b$ in \rf{eq:R1phasesb} are equal.  Application of \rf{eq:gccGe} to
the vertical stacking vector yields (with the same sign convention)
\begin{equation}
  \label{eq:R2b}
  \Phi_e^\g(\zv)\= \pm\Phi_e^\g(\zv).
\end{equation}
This implies for $V$-lattices that if $g_8=\rb$ the phase
$\Phi_e^\g(\zv)\=0$ or $1/2$ and therefore that $\g$ is an operation
of order 2. For the staggered stacking vector, with the aid of
Table~\ref{tab:symmetry}, we obtain
\begin{equation}
  \label{eq:R2c}
  \Phi_e^\g(\zv+\hv)\= \pm\Phi_e^\g(\zv+\hv) \pm a,
\end{equation}
implying for $S$-lattices that if $g_8=\rr$ the phase $a$ in
\rf{eq:R1phasesb} is 0, and if $g_8=\rb$ then $2\Phi_e^\g(\zv+\hv)\=
a$, whose solutions are given by Eq.~\rf{eq:R2barphasesc}, and $\g$ is an
operation of order 2 or 4 depending on whether $a\=0$ or $1/2$.

\begin{enumerate}
\item[R3.] If $\g\in\G_e$, is an operation of order $n>2$,
  $\d\g\d^{-1}=\g^{-1}$, and the 8-fold generator is $\rr$, then the
  lattice must be staggered, $n$ must be 4, and the phases of
  $\Phi_e^\g$ are
  \begin{equation}
    \label{eq:R3phases}
    \Phi_e^\g(\bi)\=\frac12\frac12\frac12\frac12,\quad
    \Phi_e^\g(\zv+\hv)\=\frac14 \text{\ or\ } \frac34.
  \end{equation}

  If the 8-fold generator is $\rb$ the in-plane phases of $\Phi_e^\g$
  are
  \begin{equation}
    \label{eq:R3barphasesb}
    \Phi_e^\g(\bi)\=
    \begin{cases}
      aaaa & \text{$V$-lattice},\\
      0000 & \text{$S$-lattice},
    \end{cases}
  \end{equation}
  where $a$ is either 0 or $1/2$.
\end{enumerate}

\noindent {\sl\underline{Proof}:\/} Application of
\rf{eq:gccGe} to the horizontal generating vectors shows again that on
$V$-lattices the two phases $a$ and $b$ in~\rf{eq:R1phasesb} are equal.
Application of \rf{eq:gccGe} to the vertical stacking vector yields
\begin{equation}
  \label{eq:R3a}
  \Phi_e^\g(\zv)\= \Phi_e^{\g^{-1}}(g_8\zv) \= \mp\Phi_e^\g(\zv),
\end{equation}
where the upper (negative) sign is for $g_8=\rr$ and the lower
(positive) sign is for $g_8=\rb$. This implies that on $V$-lattices if
$g_8=\rr$ the phase $\Phi_e^\g(\zv)$ is 0 or $1/2$. If this is the
case, then because we know that the in-plane phases of $\Phi_e^\g$ are
also either 0 or $1/2$, $\g^2=\eps$ in contradiction to the statement
that the order of $\g$ is greater than 2, and therefore the lattice
cannot be vertical.

For the staggered stacking vector we obtain (with the same sign
convention)
\begin{equation}
  \label{eq:R3b}
  \Phi_e^\g(\zv+\hv)\= \mp\Phi_e^\g(\zv+\hv) \mp a ,
\end{equation}
implying for $S$-lattices that if $g_8=\rb$ the phase $a$ in
\rf{eq:R1phasesb} is 0, and if $g_8=\rr$ then $2\Phi_e^\g(\zv+\hv)\=
a$. If $a\=0$ then all the phases of $\Phi_e^\g$ are either 0 or $1/2$
implying that $\g$ is of order 2, and therefore we must take $a\=1/2$,
in which case $\g$ is of order 4, and $\Phi_e^\g(\zv+\hv)\=1/4$ or
$3/4$.

\begin{enumerate}
\item[R4.] If $2^*_\xb,2^*_\yb\in\G_e$, where the asterisk denotes an
  optional prime, and $\d2^*_\xb\d^{-1}=2^*_\yb$, the directions of
  the $\xb$ and $\yb$ axes in spin-space can be chosen so that the
  in-plane phases of $\Phi_e^{2^*_\xb}$ and $\Phi_e^{2^*_\yb}$ are
  \begin{subequations}
  \begin{equation}
    \label{eq:R4phasesb}
    \begin{cases}
    \Phi_e^{2^*_\xb}(\bi)\=0\frac12 0\frac12,\
    \Phi_e^{2^*_\yb}(\bi)\=\frac12 0\frac12 0 &  \text{$V$-lattice},\\
    \Phi_e^{2^*_\xb}(\bi)\=
    \Phi_e^{2^*_\yb}(\bi)\=\frac12\frac12\frac12\frac12 &  \text{$S$-lattice},
    \end{cases}
  \end{equation}
  and the phases on the stacking vector are
  \begin{equation}
    \label{eq:R4phasec}
    \begin{cases}
    \Phi_e^{2^*_\xb}(\zv)\=
    \Phi_e^{2^*_\yb}(\zv)\=0 \text{\ or\ } \frac12 &  \text{$V$-lattice},\\
    \Phi_e^{2^*_\xb}(\az)\=0,\
    \Phi_e^{2^*_\yb}(\az)\=\frac12 &  \text{$S$-lattice}.
    \end{cases}
  \end{equation}
  \end{subequations}
\end{enumerate}

\noindent {\sl\underline{Proof}:\/} Let $\g_1$ and $\g_2$ denote the
two 2-fold operations, and note that since the phases
$\Phi_e^{\g_1}(\kv)$ and $\Phi_e^{\g_2}(\kv)$ are always either 0 or
$1/2$ the signs of these phases can be ignored. Application of
Eq.~\rf{eq:gccGe} to the horizontal generating vectors yields
\begin{equation}
  \label{eq:R4a}
  \Phi_e^{\g_1}(\bi) \= \Phi_e^{\d\g_1\d^{-1}}(g_8\bi) \=
  \Phi_e^{\g_2}(\pm\b{i+1}),
\end{equation}
This implies through result R1 that on $V$-lattices if
$\Phi_e^{\g_1}(\bi) \=abab$ then $\Phi_e^{\g_2}(\bi) \=baba$, and that
on $S$-lattices $\Phi_e^{\g_1}(\bi) \= \Phi_e^{\g_2}(\bi) \=aaaa$,
where $a$ and $b$ are either 0 or $1/2$.

Application of Eq.~\rf{eq:gccGe} to the vertical stacking vector
yields
\begin{equation}
  \label{eq:R4b}
  \Phi_e^{\g_1}(\zv) \= \Phi_e^{\d\g_1\d^{-1}}(g_8\zv) \=
  \Phi_e^{\g_2}(\zv),
\end{equation}
where we have ignored the sign difference between $\rr\zv$ and
$\rb\zv$. This equality of phases implies that $a$ and $b$ cannot be
equal otherwise $\g_1=\g_2$. We can always choose the orientation of
the $\xb$ and $\yb$ axes in spin space such that the phase function
associated with $2^*_\xb$ is the one whose values on the horizontal
generating vectors are $0\frac120\frac12$.

Finally, application of Eq.~\rf{eq:gccGe} to the staggered stacking
vector, ignoring again the sign difference between $\rr(\zv+\hv)$ and
$\rb(\zv+\hv)$, yields
\begin{equation}
  \label{eq:R4c}
  \Phi_e^{\g_1}(\zv+\hv) \= \Phi_e^{\g_2}(\zv+\hv) + a.
\end{equation}
Since the two phase functions have identical values on the horizontal
sublattice they must differ on the stacking vector. This requires that
$a$ be $1/2$. Here we can always choose the orientation of the $\xb$
and $\yb$ axes in spin space such that the phase function associated
with $2^*_\xb$ is the one whose value on the staggered stacking
vector is $0$.

\subsection{Constraints imposed by \textnormal{$\mu$} (point groups
  $8mm$, $\bar8m2$, and $8/mmm$)}

Using the third line of Table~\ref{tab:symmetry}, summarizing the effect
of the mirror $m$ on the lattice generating vectors, we obtain the
following results:

\begin{enumerate}
\item[M1.] For any $\g\in\G_e$, if $\mu$ commutes with $\g$, the
  in-plane phases of $\Phi_e^\g$ are
  \begin{equation}
    \label{eq:M1phasesb}
    \Phi_e^\g(\bi)\=
    \begin{cases}
      abab & \text{$V$-lattice},\\
      0000 & \text{$S$-lattice},
    \end{cases}
  \end{equation}
  where $a$ and $b$ are independently either 0 or $1/2$.
\end{enumerate}

\noindent {\sl\underline{Proof}:\/} Starting from the general result
R1, we see that no further constraints arise from application of
\rf{eq:gccGe} to the horizontal generating vectors and to the vertical
stacking vector, and so there is no change from the general case for
$V$-lattices. For the staggered stacking vector we obtain
\begin{equation}
  \label{eq:M1a}
  \Phi_e^\g(\zv+\hv)\= \Phi_e^\g(\zv+\hv) - a,
\end{equation}
implying that on $S$-lattices the phase $a$ in \rf{eq:R1phasesb} is 0.

\begin{enumerate}
\item[M2.] If $\g\in\G_e$, is an operation of order $n>2$, and
  $\mu\g\mu^{-1}=\g^{-1}$, then the lattice must be staggered, $n$
  must be 4, and the phases of $\Phi_e^\g$ are
  \begin{equation}
    \label{eq:M2phasesb}
    \Phi_e^\g(\bi)\=\frac12\frac12\frac12\frac12,\quad
    \Phi_e^\g(\zv+\hv)\=\frac14 \text{\ or\ } \frac34.
  \end{equation}
\end{enumerate}

\noindent {\sl\underline{Proof}:\/} From result R1 we know that the
in-plane phases of $\Phi_e^\g$ are either 0 or $1/2$. Application of
\rf{eq:gccGe} to the vertical stacking vector yields an expression
similar to Eq.~\rf{eq:R3a} for $\rr$,
implying that the phase of $\Phi_e^\g(\zv)$ is also 0 or $1/2$. If
this is the case then $\g^2=\eps$ in contradiction to the statement
that the order of $\g$ is greater than 2. The only possibility that is
left is for an $S$-lattice, in which case
application of \rf{eq:gccGe} to the staggered stacking vector yields
an expression similar to Eq.~\rf{eq:R3b} for $\rr$ and therefore to
the same phases obtained in result R3 [Eq.~\rf{eq:R3phases}].

\begin{enumerate}
\item[M3.] If $2^*_\xb,2^*_\yb\in\G_e$, where the asterisk denotes an
  optional prime, $\mu2^*_\xb\mu^{-1}=2^*_\yb$, and the mirror is of
  type $m$, then the lattice must be staggered, the
  in-plane phases of $\Phi_e^{2^*_\xb}$ and $\Phi_e^{2^*_\yb}$ are
  \begin{subequations}
  \begin{equation}
    \label{eq:M3phasesb}
    \Phi_e^{2^*_\xb}(\bi)\=
    \Phi_e^{2^*_\yb}(\bi)\=\frac12\frac12\frac12\frac12,
  \end{equation}
  and the directions of the $\xb$ and $\yb$ axes in spin-space can be
  chosen so that the phases on the staggered stacking vector are
  \begin{equation}
    \label{eq:M3phasec}
    \Phi_e^{2^*_\xb}(\zv+\hv)\=0,\
    \Phi_e^{2^*_\yb}(\zv+\hv)\=\frac12.
  \end{equation}
  \end{subequations}
\end{enumerate}

\noindent {\sl\underline{Proof}:\/} Let $\g_1$ and $\g_2$ denote the
two 2-fold operations, and note that since the phases
$\Phi_e^{\g_1}(\kv)$ and $\Phi_e^{\g_2}(\kv)$ are always either 0 or
$1/2$ the signs of these phases can be ignored. Application of
Eq.~\rf{eq:gccGe} to the horizontal generating vectors yields
\begin{equation}
  \label{eq:M3a}
  \Phi_e^{\g_1}(\bi) \=  \Phi_e^{\g_2}(\b{i+2k}),
\end{equation}
for some integer $k$ that depends on $i$. Together with result R1,
Eq.~\rf{eq:M3a} implies that the two phase functions are identical on
the horizontal sublattice. Application of Eq.~\rf{eq:gccGe} to the
vertical stacking vector establishes that the two phase functions are
identical everywhere in contradiction with the fact that
$\g_1\neq\g_2$, and therefore that the lattice cannot be vertical.
Application of Eq.~\rf{eq:gccGe} to the staggered stacking vector
yields
\begin{equation}
  \label{eq:M3b}
  \Phi_e^{\g_1}(\zv+\hv) \= \Phi_e^{\g_2}(\zv+\hv) + a.
\end{equation}
Since the two phase functions have identical values on the horizontal
sublattice they must differ on the stacking vector, requiring that
$a$ be $1/2$. We then choose the orientation of the $\xb$
and $\yb$ axes in spin space such that the phase function associated
with $2^*_\xb$ is the one whose value on the staggered stacking
vector is $0$.

\subsection{Constraints imposed by \textnormal{$\alpha$} (point groups
  $822$ and $\bar82m$)}

Using the fourth line of Table~\ref{tab:symmetry}, summarizing the
effect of the dihedral rotation $d$ on the lattice generating vectors,
we obtain the following results:

\begin{enumerate}
\item[D1.] For any $\g\in\G_e$, if $\alpha$ commutes with $\g$ the
  in-plane phases are the general ones given in result R1
  [Eq.~\rf{eq:R1phasesb}]. If the lattice is vertical, the phase of
  $\Phi_e^\g(\zv)$ is independently 0 or $1/2$, implying that $\g$ is
  an operation of order 2. If the lattice is staggered the phase on
  the stacking vector is
  \begin{equation}
    \label{eq:D1phasesc}
    \Phi_e^\g(\zv+\hv)\=\frac{a}2 + c,
  \end{equation}
  where $a$ is the in-plane phase in~\rf{eq:R1phasesb}, and $c\=0$ or
  $1/2$ but $a$ and $c$ cannot both be 0. Consequently, $\g$ is an
  operation of order 2 or 4, depending on whether $a\=0$ or $1/2$.
\end{enumerate}

\noindent {\sl\underline{Proof}:\/} Application of Eq.~\rf{eq:gccGe}
to the horizontal generating vectors yields no further constraints
beyond the general result R1. Application of Eq.~\rf{eq:gccGe} to the
vertical stacking vector yields
\begin{equation}
  \label{eq:D1pre-a}
  \Phi_e^{\g}(\zv) \= -\Phi_e^{\g}(\zv),
\end{equation}
implying that the phase of $\Phi_e^\g(\zv)$ is 0 or $1/2$. If this is
the case then $\g$ is necessarily an operation of order 2. Application
of Eq.~\rf{eq:gccGe} to the vertical stacking vector yields
\begin{equation}
  \label{eq:D1a}
  \Phi_e^{\g}(\zv+\hv) \= -\Phi_e^{\g}(\zv+\hv) + \Phi_e^{\g}(2\hv)
  -a.
\end{equation}
It follows from the general result R1 that
$\Phi_e^{\g}(2\hv)\=Phi_e^{\g}(\b1+\b2+\b3-\b4)\=0$, and therefore
that $2\Phi_e^{\g}(\zv+\hv) \=a$, whose solutions are given by
Eq.~\rf{eq:D1phasesc}. If $a\=0$ then $\g$ is again of order 2; If
$a\=1/2$ then $\g$ is of order 4.

\begin{enumerate}
\item[D2.] If $\g\in\G_e$, is an operation of order $n>2$,
  $\alpha\g\alpha^{-1}=\g^{-1}$, and the lattice is vertical there are
  no additional constraints on the phase function $\Phi_e^\g$. If the
  lattice is staggered then the in-plane phases of $\Phi_e^\g$ are all
  0.
\end{enumerate}

\noindent {\sl\underline{Proof}:\/} This can easily be seen by
applying Eq.~\rf{eq:gccGe} to the generating vectors while noting that
$\Phi_e^{\g}(2\hv)\=0$.

\begin{enumerate}
\item[D3.] If $2^*_\xb,2^*_\yb\in\G_e$, where the asterisk denotes an
  optional prime, and $\alpha2^*_\xb\alpha^{-1}=2^*_\yb$, then the
  lattice must be staggered, the in-plane phases of $\Phi_e^{2^*_\xb}$
  and $\Phi_e^{2^*_\yb}$ are
  \begin{equation}
    \label{eq:D3phasesb}
    \Phi_e^{2^*_\xb}(\bi)\=
    \Phi_e^{2^*_\yb}(\bi)\=\frac12\frac12\frac12\frac12,
  \end{equation}
  and the directions of the $\xb$ and $\yb$ axes in spin-space can be
  chosen so that the phases on the staggered stacking vector are
  \begin{equation}
    \label{eq:D3phasec}
    \Phi_e^{2^*_\xb}(\cv)\=0,\
    \Phi_e^{2^*_\yb}(\cv)\=\frac12.
  \end{equation}
\end{enumerate}

\noindent {\sl\underline{Proof}:\/} This result is established in the
same way as result M3 for the mirror $m$ due to the fact that the sign
of the phases can be ignored and the fact that $\Phi_e^{\g}(2\hv)\=0$.

\subsection{Constraints imposed by \textnormal{$\eta$} (point groups
  $8/m$ and $8/mmm$)}

Finally, using the fifth line of Table~\ref{tab:symmetry}, summarizing the
effect of the horizontal mirror $h$ on the lattice generating vectors,
we obtain the following results:

\begin{enumerate}
\item[H1.] For any $\g\in\G_e$, if $\eta$ commutes with $\g$, then the
  phase of $\Phi_e^\g$ on the stacking vector is
  \begin{equation}
    \label{eq:H1phasesc}
    \Phi_e^\g(\cv)\=0 \text{ or } \frac12,
  \end{equation}
  which implies that $\g$ is an operation of order 2.
\end{enumerate}

\noindent {\sl\underline{Proof}:\/} Application of Eq.~\rf{eq:gccGe}
to the horizontal generating vectors yields no further constraints
beyond the general result R1. Application of Eq.~\rf{eq:gccGe} to the
vertical stacking vector yields Eq.~\rf{eq:D1pre-a}, and application
of Eq.~\rf{eq:gccGe} to the staggered stacking vector yields
\begin{equation}
  \label{eq:H1b}
  \Phi_e^{\g}(\zv+\hv) \= -\Phi_e^{\g}(\zv+\hv) + \Phi_e^{\g}(2\hv).
\end{equation}
Because $\Phi_e^{\g}(2\hv)\=0$, Eqs.~\rf{eq:D1pre-a} and \rf{eq:H1b}
imply that for both lattice types the phase on the stacking vector is
either 0 or $1/2$.

\begin{enumerate}
\item[H2.] If $\g\in\G_e$, is an operation of order $n>2$, and
  $\eta\g\eta^{-1}=\g^{-1}$ there are no additional constraints on the
  phase function $\Phi_e^\g$.
\end{enumerate}

\noindent {\sl\underline{Proof}:\/} This can easily be seen by
applying Eq.~\rf{eq:gccGe} to the generating vectors while noting that
$\Phi_e^{\g}(2\hv)\=0$.

\begin{enumerate}
\item[H3.] If $2^*_\xb,2^*_\yb\in\G_e$, where the asterisk denotes an
  optional prime, then they must both commute with $\eta$ and their
  phase functions are only constrained by results R1 and H1.
\end{enumerate}

\noindent {\sl\underline{Proof}:\/} Let us denote the two operations
by $\g_1$ and $\g_2$ and assume that rather than commuting with $\eta$
they satisfy $\eta\g_1\eta^{-1}=\g_2$, then application of Eq.~\rf{eq:gccGe}
to the generating vectors of both lattice types, while noting that
$\Phi_e^{\g}(2\hv)\=0$, establishes that both phase functions are
identical everywhere in contradiction with the fact that
$\g_1\neq\g_2$.

\newpage
\section{Octagonal spin space-group types\\{}
[Not to be included in the printed version of the paper]}
\label{sec:app-b}

The following Tables list all the 3-dimensional octagonal spin
space-group types, explicitly identifying the spin-space operations
$\d$, $\mu$, $\eta$, and $\alpha$ appearing in the spin point-group
generators. There are a total of 16 tables, one for each combination
of point group $G$ and lattice type. The first few columns of each
table give the structure of the spin point group $\GS$ by listing
$\G_e$, $G_\eps$, $\G$, and the quotient group $G/G_\eps$. Following
these is a column that explicitly lists the generators $(\rr,\d)$,
$(\rb,\d)$, $(m,\mu)$, $(d,\alpha)$, and $(h,\eta)$. The last column
in each table refers to a line in the corresponding spin space-group
table in the main part of the paper
(Tables~\ref{tab:8V}--\ref{tab:8/mmmS}), which gives the possible
values of all the phase functions for the generators of the spin point
group, as well as a rule for generating the spin space-group symbol.

\begin{longtable}{|>{$}c<{$}|>{$}c<{$}|>{$}c<{$}|>{$}c<{$}|>{$}c<{$}|>{$}c<{$}|}
\caption[]{Explicit list of octagonal spin space-group types with $G=8$
  on $V$-lattices. The last column refers to line numbers in Table
  \ref{tab:8V}, where the possible phase functions are listed, and
  rules are given to generate the spin space-group symbol.}\\
\hline \G_e&G_\eps&G/G_\eps&\G&\textnormal{generator}&line\\
\hline \endfirsthead\\
\caption[]{\textnormal{continued}}\\
\hline \G_e&G_\eps&G/G_\eps&\G&\textnormal{generator}&line\\
\hline \endhead \hline \multicolumn{6}{r}{\textnormal{continued on next page}} \endfoot \hline \endlastfoot
1&8&1&1&(\rr,\eps)&1\\ 
\cline{2-6} &4&2&2^*&(\rr,\ssr2)&1\\
\cline{4-6} &&&1'&(\rr,\eps')&1\\
\cline{2-6} &2&4&4^*&(\rr,\ssr4)&1\\
\cline{2-6} &1&8&8^*&(\rr,\ssr8)&1\\
\hline 2&8&1&2&(\rr,\eps)&1\\ 
\cline{2-6} &4&2&4^*&(\rr,\ssr4)&1\\
\cline{4-6} &&&2^*2^*2&(\rr,2^*_\xb)&1\\
\end{longtable}

\begin{longtable}{|>{$}c<{$}|>{$}c<{$}|>{$}c<{$}|>{$}c<{$}|>{$}c<{$}|>{$}c<{$}|}

\caption[]{Explicit list of octagonal spin space-group types with $G=8$
  on $S$-lattices. The last column refers to line numbers in Table
  \ref{tab:8S}, where the possible phase functions are listed, and
  rules are given to generate the spin space-group symbol.}\\
\hline \G_e&G_\eps&G/G_\eps&\G&\textnormal{generator}&line\\
\hline \endfirsthead\\
\caption[]{\textnormal{continued}}\\
\hline \G_e&G_\eps&G/G_\eps&\G&\textnormal{generator}&\Phi_e^{n^*_\zb}(\kv)\\
\hline \endhead \hline \multicolumn{6}{r}{\textnormal{continued on next page}} \endfoot \hline \endlastfoot
1&8&1&1&(\rr,\eps)&1\\
\cline{2-6} &4&2&2^*&(\rr,\ssr2)&1\\
\cline{4-6} &&&1'&(\rr,\eps')&1\\
\cline{2-6} &2&4&4^*&(\rr,\ssr4)&1\\
\cline{2-6} &1&8&8^*&(\rr,\ssr8)&1\\
\hline 2&8&1&2&(\rr,\eps)&1\\
\cline{2-6} &4&2&4^*&(\rr,\ssr4)&1\\
\cline{4-6} &&&2^*2^*2&(\rr,2^*_\xb)&1\\
\cline{4-6} &&&21'&(\rr,\eps')&1\\
\cline{2-6} &2&4&8^*&(\rr,\ssr8)&1\\
\cline{2-6} &1&8&16^*&(\rr,\ssr{16})&2\\
\hline 1'&8&1&1'&(\rr,\eps)&1\\
\cline{2-6} &4&2&21'&(\rr,\sr2)&1\\
\cline{2-6} &2&4&41'&(\rr,\sr4)&1\\
\cline{2-6} &1&8&81'&(\rr,\sr8)&1\\
\hline 2'&8&1&2'&(\rr,\eps)&1\\
\cline{2-6} &4&2&2'2^*2'^*&(\rr,2^*_\xb)&1\\
\cline{4-6} &&&21'&(\rr,\eps')&1\\
\hline n&8&1&n&(\rr,\eps)&4\\
\cline{2-6}
&4&2&(2n)^*&(\rr,\ssr{2n})&8\\
\cline{4-6} &&&n1'&(\rr,\eps')&4\\
\cline{4-6} &&&n2^*2^*&(\rr,2^*_\xb)&4\\
\cline{2-6}
&2&4&(4n)^*&(\rr,\ssr{4n})&6\\
\cline{2-6}
&1&8&(8n)^*&(\rr,\ssr{8n})&5\\
\hline n'&8&1&n'&(\rr,\eps)&4\\
\cline{2-6} &4&2&n1'&(\rr,\eps')&4\\
\hline n'&8&1&n1'&(\rr,\eps)&4\\
\cline{2-6} &4&2&(2n)1'&(\rr,\sr{n})&8\\
\cline{2-6} &2&4&(4n)1'&(\rr,\sr{4n})&6\\
\cline{2-6} &1&8&(8n)1'&(\rr,\sr{8n})&5\\
\hline 2^*2^*2&4&2&4^*2^\dagger2^{*\dagger}&(\rr,\ssr4)&3\\
\end{longtable}

\begin{longtable}{|>{$}c<{$}|>{$}c<{$}|>{$}c<{$}|>{$}c<{$}|>{$}c<{$}|>{$}c<{$}|}

\caption[]{Explicit list of octagonal spin space-group types with
  $G=\bar 8$
  on $V$-lattices. The last column refers to line numbers in Table
  \ref{tab:bar8V}, where the possible phase functions are listed, and
  rules are given to generate the spin space-group symbol.}\\
\hline \G_e&G_\eps&G/G_\eps&\G&\textnormal{generator}&line\\
\hline \endfirsthead\\
\caption[]{\textnormal{continued}}\\
\hline \G_e&G_\eps&G/G_\eps&\G&\textnormal{generator}&line\\
\hline \endhead \hline \multicolumn{5}{r}{\textnormal{continued on next
page}} \endfoot \hline \endlastfoot 1&\bar 8&1&1&(\rb,\eps)&1\\
\cline{2-6} &4&2&2^*&(\rb,\ssr2)&1\\
\cline{4-6} &&&1'&(\rb,\eps')&1\\
\cline{2-6} &2&4&4^*&(\rb,\ssr4)&1\\
\cline{2-6} &1&8&8^*&(\rb,\ssr8)&1\\
\hline 2&\bar 8&1&2&(\rb,\eps)&1\\
\cline{2-6} &4&2&4^*&(\rb,\ssr4)&1\\
\cline{4-6} &&&2^*2^*2&(\rb,2^*_\xb)&1\\
\cline{4-6} &&&21'&(\rb,\eps')&1\\
\cline{2-6} &2&4&8^*&(\rb,\ssr8)&1\\
\hline 2'&\bar 8&1&2'&(\rb,\eps)&1\\
\cline{2-6} &4&2&2'2^*2'^*&(\rb,2^*_\xb)&1\\
\cline{4-6} &&&21'&(\rb,\eps')&1\\
\cline{2-6}& 2&4&41'&(\rb,4^*_\zb)&1\\
\cline{2-6}& 1&8&81'&(\rb,8^{*}_\zb)&1\\
\hline 1'&\bar 8&1&1'&(\rb,\eps)&1\\
\cline{2-6} &4&2&21'&(\rb,\sr2)&1\\
\cline{2-6} &2&4&41'&(\rb,\sr4)&1\\
\cline{2-6} &1&8&81'&(\rb,\sr8)&1\\
\hline 21'&\bar 8&1&21'&(\rb,\eps)&1\\
\cline{2-6} &4&2&41'&(\rb,\sr4)&1\\
\cline{4-6} &&&2'2'2'&(\rb,2_\xb)&1\\
\cline{2-6} &2&4&81'&(\rb,\sr8)&1\\

\hline n&4&2&n2^*2^*&(\rb,2^*_\xb)&4\\
\hline n'&4&2&n'2^*n'^*&(\rb,2^*_\xb)&4\\
\hline n1'&4&2&n221'&(\rb,2_\xb)&4\\

\hline 222&\bar 8&1&222&(\rb,\eps)&2a\\
\cline{2-6} &4&2&2'2'2'&(\rb,\eps')&2a\\
\cline{4-6} &&&4^*2^\dagger2^{*\dagger}&(\rb,\ssr4)&3\\

\hline 2'2'2&\bar 8&1&2'2'2&(\rb,\eps)&2\\
\cline{2-6} &4&2&2'2'2'&(\rb,\eps')&2\\
\cline{4-6} &&&4^*2^\dagger2^{*\dagger}&(\rb,\ssr4)&3\\

\hline 2'2'2'&4&2&4221'&(\rb,\sr4)&3\\
\end{longtable}

\begin{longtable}{|>{$}c<{$}|>{$}c<{$}|>{$}c<{$}|>{$}c<{$}|>{$}c<{$}|>{$}c<{$}|}

\caption[]{Explicit list of octagonal spin space-group types with
  $G=\bar 8$
  on $S$-lattices. The last column refers to line numbers in Table
  \ref{tab:bar8S}, where the possible phase functions are listed, and
  rules are given to generate the spin space-group symbol.}\\
\hline \G_e&G_\eps&G/G_\eps&\G&\textnormal{generator}&line\\
\hline \endfirsthead\\
\caption[]{\textnormal{continued}}\\
\hline \G_e&G_\eps&G/G_\eps&\G&\textnormal{generator}&line\\
\hline \endhead \hline \multicolumn{6}{r}{\textnormal{continued on next
page}} \endfoot \hline \endlastfoot 1&\bar 8&1&1&(\rb,\eps)&1\\
\cline{2-6} &4&2&2^*&(\rb,\ssr2)&1\\
\cline{4-6} &&&1'&(\rb,\eps')&1\\
\cline{2-6} &2&4&4^*&(\rb,\ssr4)&1\\
\cline{2-6} &1&8&8^*&(\rb,\ssr8)&1\\
\hline 2&\bar 8&1&2&(\rb,\eps)&1\\
\cline{2-6} &4&2&4^*&(\rb,\ssr4)&1\\
\cline{4-6} &&&2^*2^*2&(\rb,2^*_\xb)&1\\
\cline{4-6} &&&21'&(\rb,\eps')&1\\
\cline{2-6} &2&4&8^*&(\rb,\ssr8)&1\\
\hline 2'&\bar 8&1&2'&(\rb,\eps)&1\\
\cline{2-6} &4&2&2'2^*2'^*&(\rb,2^*_\xb)&1\\
\cline{4-6} &&&21'&(\rb,\eps')&1\\
\cline{2-6} &2&4&41'&(\rr,4^*_\zb)&1\\
\cline{2-6} &2&4&41'&(\rr,8^{*}_\zb)&1\\
\hline 1'&\bar 8&1&1'&(\rb,\eps)&1\\
\cline{2-6} &4&2&21'&(\rb,\sr2)&1\\
\cline{2-6} &2&4&41'&(\rb,\sr4)&1\\
\cline{2-6} &1&8&81'&(\rb,\sr8)&1\\

\hline n&\bar 8&1&n&(\rb,\eps)&4\\
\cline{2-6} &4&2&(2n)^*&(\rb,\ssr{2n})&4\\
\cline{4-6} &&&n2^*2^*&(\rb,2^*_\xb)&3\\
\cline{4-6} &&&n1'&(\rb,\eps')&4\\
\hline n'&\bar 8&1&n'&(\rb,\eps)&4\\
\cline{2-6} &4&2&n'2^*n'^*&(\rb,2^*_\xb)&3\\

\hline 2^*2^*2&4&2&2'2'2'&(\rb,\eps')&2\\
\end{longtable}

\begin{longtable}{|>{$}c<{$}|>{$}c<{$}|>{$}c<{$}|>{$}l<{$}|>{$}c<{$}|}

\caption[]{Explicit list of octagonal spin space-group types with $G=8mm$
  on $V$-lattices. The last column refers to line numbers in Table
  \ref{tab:8mmV}, where the possible phase functions are listed, and
  rules are given to generate the spin space-group symbol.}\\
\hline
G_\eps&G/G_\eps&\G&\textnormal{generators}&line\\
\hline \endfirsthead\\
\caption[]{\textnormal{continued}}\\
\hline G_\eps&G/G_\eps&\G&\textnormal{generators}&line\\
\hline \endhead \hline \multicolumn{5}{r}{\textnormal{continued on next page}} \endfoot \hline \endlastfoot
\multicolumn{5}{|l|}{$\G_e=1$}\\
\hline 8mm&1&1&(\rr,\eps)(m,\eps)&1\\
\cline{1-5} 8&m&2&(\rr,\eps)(m,2_\zb)&1\\
\cline{3-5} &&2'&(\rr,\eps)(m,2'_\zb)&1\\
\cline{3-5} &&1'&(\rr,\eps)(m,\eps')&1\\
\cline{1-5} 4mm&2&2^*&(\rr,2^*_\zb)(m,\eps)&1\\
\cline{3-5} &&1'&(\rr,\eps')(m,\eps)&1\\
\cline{1-5} 4m'm'&2&2^*&(\rr,2^*_\zb)(m,2^*_\zb)&1\\
\end{longtable}\clearpage
\begin{longtable}{|>{$}c<{$}|>{$}c<{$}|>{$}c<{$}|>{$}l<{$}|>{$}c<{$}|}

\caption[]{Explicit list of octagonal spin space-group types with $G=8mm$
  on $S$-lattices. The last column refers to line numbers in Table
  \ref{tab:8mmS}, where the possible phase functions are listed, and
  rules are given to generate the spin space-group symbol.}\\
\hline G_\eps&G/G_\eps&\G&\textnormal{generators}&line\\
\hline \endfirsthead\\
\caption[]{\textnormal{continued}}\\
\hline G_\eps&G/G_\eps&\G&\textnormal{generators}&line\\
\hline \endhead \hline \multicolumn{5}{r}{\textnormal{continued on next page}} \endfoot \hline \endlastfoot
\multicolumn{5}{|l|}{$\G_e=1$}\\
\hline 8mm&1&1&(\rr,\eps)(m,\eps)&1\\
\cline{1-5} 8&m&2&(\rr,\eps)(m,2_\zb)&1\\
\cline{3-5} &&2'&(\rr,\eps)(m,2'_\zb)&1\\
\cline{3-5} &&1'&(\rr,\eps)(m,\eps')&1\\
\cline{1-5} 4mm&2&2^*&(\rr,2^*_\zb)(m,\eps)&1\\
\cline{3-5} &&1'&(\rr,\eps')(m,\eps)&1\\
\cline{1-5} 4m'm'&2&2^*&(\rr,2^*_\zb)(m,2^*_\zb)&1\\
\end{longtable}
\clearpage


\begin{longtable}{|>{$}c<{$}|>{$}c<{$}|>{$}c<{$}|>{$}l<{$}|>{$}c<{$}|>{$}c<{$}|>{$}c<{$}|>{$}c<{$}|>{$}c<{$}|}

\caption[]{Explicit list of octagonal spin space-group types with
  $G=\bar 8m2$
  on $V$-lattices. The last column refers to line numbers in Table
  \ref{tab:bar8m2V}, where the possible phase functions are listed, and
  rules are given to generate the spin space-group symbol.}\\
\hline G_\eps&G/G_\eps&\G&\textnormal{generators}&line\\
\hline \endfirsthead\\
\caption[]{\textnormal{continued}}\\
\hline G_\eps&G/G_\eps&\G&\textnormal{generators}&line\\
\hline \endhead \hline \multicolumn{5}{r}{\textnormal{continued on next page}} \endfoot \hline \endlastfoot
\multicolumn{5}{|l|}{$\G_e=1$}\\
\hline \bar 8m2&1&1&(\rb,\eps)(m,\eps)&1\\
\cline{1-5} \bar 8&2&2^*&(\rb,\eps)(m,2^*_\zb)&1\\
\cline{3-5} &&1'&(\rb,\eps)(m,\eps')&1\\
\cline{1-5} 4mm&2&2^*&(\rb,2^*_\zb)(m,\eps)&1\\
\cline{3-5} &&1'&(\rb,\eps')(m,\eps)&1\\
\cline{1-5} 42'2'&2&2^*&(\rb,2^*_\zb)(m,2^*_\zb)&1\\
\cline{3-5} &&1'&(\rb,\eps')(m,\eps')&1\\
\end{longtable}
\clearpage

\begin{longtable}{|>{$}c<{$}|>{$}c<{$}|>{$}c<{$}|>{$}l<{$}|>{$}c<{$}|}

\caption[]{Explicit list of octagonal spin space-group types with
  $G=\bar 8m2$
  on $S$-lattices. The last column refers to line numbers in Table
  \ref{tab:bar8m2S}, where the possible phase functions are listed, and
  rules are given to generate the spin space-group symbol.}\\
\hline G_\eps&G/G_\eps&\G&\textnormal{generators}&line\\
\hline \endfirsthead\\
\caption[]{\textnormal{continued}}\\
\hline G_\eps&G/G_\eps&\G&\textnormal{generators}&line\\
\hline \endhead \hline \multicolumn{5}{r}{\textnormal{continued on next page}} \endfoot \hline \endlastfoot
\multicolumn{5}{|l|}{$\G_e=1$}\\
\hline \bar 8m2&1&1&(\rb,\eps)(m,\eps)&1\\
\cline{1-5} \bar \bar 8&2&2^*&(\rb,\eps)(m,2^*_\zb)&1\\
\cline{3-5} &&1'&(\rb,\eps)(m,\eps')&1\\
\cline{1-5} 4mm&2&2^*&(\rb,2^*_\zb)(m,\eps)&1\\
\cline{3-5} &&1'&(\rb,\eps')(m,\eps)&1\\
\cline{1-5} 42'2'&2&2^*&(\rb,2^*_\zb)(m,2^*_\zb)&1\\
\cline{3-5} &&1'&(\rb,\eps')(m,\eps')&1\\
\end{longtable}

\begin{longtable}{|>{$}c<{$}|>{$}c<{$}|>{$}c<{$}|>{$}l<{$}|>{$}c<{$}|}

\caption[]{Explicit list of octagonal spin space-group types with $G=822$
  on $V$-lattices. The last column refers to line numbers in Table
  \ref{tab:822V}, where the possible phase functions are listed, and
  rules are given to generate the spin space-group symbol.}\\
\hline G_\eps&G/G_\eps&\G&\textnormal{generators}&line\\
\hline \endfirsthead\\
\caption[]{\textnormal{continued}}\\
\hline G_\eps&G/G_\eps&\G&\textnormal{generators}&line\\
\hline \endhead \hline \multicolumn{5}{r}{\textnormal{continued on next page}} \endfoot \hline \endlastfoot
\multicolumn{5}{|l|}{$\G_e=1$}\\
\hline 822&1&1&(\rr,\eps)(d,\eps)&1\\
\cline{1-5} 8&m&2^*&(\rr,\eps)(d,2^*_\zb)&1\\
\cline{3-5} &&1'&(\rr,\eps)(d,\eps')&1\\
\cline{1-5} 422&2&2^*&(\rr,2^*_\zb)(d,\eps)&1\\
\cline{3-5} &&1'&(\rr,\eps')(d,\eps)&1\\
\cline{1-5} 42'2'&2&2^*&(\rr,2^*_\zb)(d,2^*_\zb)&1\\
\cline{3-5} &&1'&(\rr,\eps')(d,\eps')&1\\
\end{longtable}
\clearpage


\begin{longtable}{|>{$}c<{$}|>{$}c<{$}|>{$}c<{$}|>{$}l<{$}|>{$}c<{$}|}

\caption[]{Explicit list of octagonal spin space-group types with $G=822$
  on $S$-lattices. The last column refers to line numbers in Table
  \ref{tab:822S}, where the possible phase functions are listed, and
  rules are given to generate the spin space-group symbol.}\\
\hline G_\eps&G/G_\eps&\G&\textnormal{generators}&line\\
\hline \endfirsthead\\
\caption[]{\textnormal{continued}}\\
\hline G_\eps&G/G_\eps&\G&\textnormal{generators}&line\\
\hline \endhead \hline \multicolumn{5}{r}{\textnormal{continued on next page}} \endfoot \hline \endlastfoot
\multicolumn{5}{|l|}{$\G_e=1$}\\
\hline 822&1&1&(\rr,\eps)(d,\eps)&1\\
\cline{1-5} 8&2&2^*&(\rr,\eps)(d,2^*_\zb)&1\\
\cline{3-5} &&1'&(\rr,\eps)(d,\eps')&1\\
\cline{1-5} 422&2&2^*&(\rr,2^*_\zb)(d,\eps)&1\\
\cline{3-5} &&1'&(\rr,\eps')(d,\eps)&1\\
\cline{1-5} 42'2'&2&2^*&(\rr,2^*_\zb)(d,2^*_\zb)&1\\
\cline{3-5} &&1'&(\rr,\eps')(d,\eps')&1\\
\end{longtable}
\clearpage

\begin{longtable}{|>{$}c<{$}|>{$}c<{$}|>{$}c<{$}|>{$}l<{$}|>{$}c<{$}|}

\caption[]{Explicit list of octagonal spin space-group types with
  $G=\bar 82m$
  on $V$-lattices. The last column refers to line numbers in Table
  \ref{tab:bar82mV}, where the possible phase functions are listed, and
  rules are given to generate the spin space-group symbol.}\\
\hline G_\eps&G/G_\eps&\G&\textnormal{generators}&line\\
\hline \endfirsthead\\
\caption[]{\textnormal{continued}}\\
\hline G_\eps&G/G_\eps&\G&\textnormal{generators}&line\\
\hline \endhead \hline \multicolumn{5}{r}{\textnormal{continued on next page}} \endfoot \hline \endlastfoot
\multicolumn{5}{|l|}{$\G_e=1$}\\
\hline \bar 8m2&1&1&(\rb,\eps)(d,\eps)&1\\
\bar 8&2&2^*&(\rb,\eps)(d,2^*_\zb)&1\\
&&1'&(\rb,\eps)(d,\eps')&1\\
4mm&2&2^*&(\rb,2^*_\zb)(d,\eps)&1\\
&&1'&(\rb,\eps')(d,\eps)&1\\
42'2'&2&2^*&(\rb,2^*_\zb)(d,2^*_\zb)&1\\
&&1'&(\rb,\eps')(d,\eps')&1\\
\end{longtable}

\begin{longtable}{|>{$}c<{$}|>{$}c<{$}|>{$}c<{$}|>{$}l<{$}|>{$}c<{$}|}

\caption[]{Explicit list of octagonal spin space-group types with
  $G=\bar 82m$
  on $S$-lattices. The last column refers to line numbers in Table
  \ref{tab:bar82mS}, where the possible phase functions are listed, and
  rules are given to generate the spin space-group symbol.}\\
\hline G_\eps&G/G_\eps&\G&\textnormal{generators}&line\\
\hline \endfirsthead\\
\caption[]{\textnormal{continued}}\\
\hline G_\eps&G/G_\eps&\G&\textnormal{generators}&line\\
\hline \endhead \hline \multicolumn{5}{r}{\textnormal{continued on next page}} \endfoot \hline \endlastfoot
\multicolumn{5}{|l|}{$\G_e=1$}\\
\hline \bar 82m&1&1&(\rb,\eps)(d,\eps)&1\\
\cline{1-5} \bar \bar 8&2&2^*&(\rb,\eps)(d,2^*_\zb)&1\\
\cline{3-5} &&1'&(\rb,\eps)(d,\eps')&1\\
\cline{1-5} 422&2&2^*&(\rb,2^*_\zb)(d,\eps)&1\\
\cline{3-5} &&1'&(\rb,\eps')(d,\eps)&1\\
\cline{1-5} 4m'm'&2&2^*&(\rb,2^*_\zb)(d,2^*_\zb)&1\\
\cline{3-5} &&1'&(\rb,\eps')(d,\eps')&1\\
\end{longtable}


\begin{longtable}{|>{$}c<{$}|>{$}c<{$}|>{$}c<{$}|>{$}l<{$}|>{$}c<{$}|}
\caption[]{Explicit list of octagonal spin space-group types with
  $G=8/m$ on $V$-lattices. The last column refers to line numbers in
  Table~\ref{tab:8/mV}, where the possible phase functions are listed,
  and rules are given to generate the spin space-group symbol.}\\ 
\hline G_\eps&G/G_\eps&\G&\textnormal{generators}& line\\\hline
\endfirsthead\\
\caption[]{\textnormal{continued}}\\
\hline G_\eps&G/G_\eps&\G&\textnormal{generators}& line\\
\hline \endhead \hline \multicolumn{5}{r}{\textnormal{continued on next page}} \endfoot \hline \endlastfoot
\multicolumn{5}{|l|}{$\G_e=1$}\\ 
\hline 8/m&1&1&(\rr,\eps)(h,\eps)&1\\
\hline 8&2&2^*&(\rr,\eps)(h,\ssr2)&1\\
\cline{3-5}&&1'&(\rr,\eps)(h,\eps')&1\\
\hline\bar 8&2&2^*&(\rr,\ssr2)(h,\ssr2)&1\\
\cline{3-5}&&1'&(\rr,\eps')(h,\eps')&1\\
\hline 4/m&2&2^*&(\rr,\ssr2)(h,\eps)&1\\
\cline{3-5}&&1'&(\rr,\eps')(h,\eps)&1\\
\hline
\end{longtable}

\begin{longtable}{|>{$}c<{$}|>{$}c<{$}|>{$}c<{$}|>{$}l<{$}|>{$}c<{$}|}
\caption[]{Explicit list of octagonal spin space-group types with
  $G=8/m$ on $S$-lattices. The last column refers to line numbers in
  Table~\ref{tab:8/mS}, where the possible phase functions are listed,
  and rules are given to generate the spin space-group symbol.}\\ 
\hline G_\eps&G/G_\eps&\G&\textnormal{generators}& line\\\hline
\endfirsthead\\
\caption[]{\textnormal{continued}}\\
\hline G_\eps&G/G_\eps&\G&\textnormal{generators}& line\\
\hline \endhead \hline \multicolumn{5}{r}{\textnormal{continued on next page}} \endfoot \hline \endlastfoot
\multicolumn{5}{|l|}{$\G_e=1$}\\ 
\hline 8/m&1&1&(\rr,\eps)(h,\eps)&1\\
\hline 8&2&2^*&(\rr,\eps)(h,\ssr2)&1\\
\cline{3-5}&&1'&(\rr,\eps)(h,\eps')&1\\
\hline\bar 8&2&2^*&(\rr,\ssr2)(h,\ssr2)&1\\
\cline{3-5}&&1'&(\rr,\eps')(h,\eps')&1\\
\hline 4/m&2&2^*&(\rr,\ssr2)(h,\eps)&1\\
\cline{3-5}&&1'&(\rr,\eps')(h,\eps)&1\\
\hline

\end{longtable}

\begin{longtable}{|>{$}c<{$}|>{$}c<{$}|>{$}c<{$}|>{$}l<{$}|>{$}c<{$}|}

\caption[]{Explicit list of octagonal spin space-group types with
  $G=8/mmm$ on $V$-lattices. The last column refers to line numbers in
  Table~\ref{tab:8/mmmV}, where the possible phase functions are listed,
  and rules are given to generate the spin space-group symbol.}\\ 
\hline G_\eps&G/G_\eps&\G&\textnormal{generators}& line\\\hline
\endfirsthead\\
\caption[]{\textnormal{continued}}\\
\hline G_\eps&G/G_\eps&\G&\textnormal{generators}& line\\
\hline \endhead \hline \multicolumn{5}{r}{\textnormal{continued on next page}} \endfoot \hline \endlastfoot
\multicolumn{5}{|l|}{$\G_e=1$}\\ \hline

8/mmm&1&1&(\rr,\eps)(m,\eps)(h,\eps)&1\\
\hline 8mm&2&2^*&(\rr,\eps)(m,\eps)(h,\ssr2)&1\\
\cline{3-5} &&1'&(\rr,\eps)(m,\eps)(h,\eps')&1\\
\hline 822&2&2^*&(\rr,\eps)(m,\ssr2)(h,\ssr2)&1\\
\cline{3-5} &&1'&(\rr,\eps)(m,\eps')(h,\eps')&1\\
\hline 8/8&2&2^*&(\rr,\eps)(m,\ssr2)(h,\eps)&1\\
\cline{3-5} &&1'&(\rr,\eps)(m,\eps')(h,\eps)&1\\
\hline
\end{longtable}


\begin{longtable}{|>{$}c<{$}|>{$}c<{$}|>{$}l<{$}|>{$}c<{$}|}
\caption[]{Explicit list of octagonal spin space-group types with
  $G=8/mmm$ on $S$-lattices. The last column refers to line numbers in
  Table~\ref{tab:8/mmmS}, where the possible phase functions are listed,
  and rules are given to generate the spin space-group symbol.}\\ 
\hline G_\eps&\G&\textnormal{generators}& line\\\hline
\endfirsthead\\
\caption[]{\textnormal{continued}}\\
\hline G_\eps&\G&\textnormal{generators}& line\\
\hline \endhead \hline \multicolumn{4}{r}{\textnormal{continued on next page}} \endfoot \hline \endlastfoot
\multicolumn{4}{|l|}{$\G_e=1$}\\ \hline
8/mmm&1&(\rr,\eps)(m,\eps)(h,\eps)&1\\
\hline 8mm&2^*&(\rr,\eps)(m,\eps)(h,\ssr2)&1\\
\cline{2-4} &1'&(\rr,\eps)(m,\eps)(h,\eps')&1\\
\hline 822&2^*&(\rr,\eps)(m,\ssr2)(h,\ssr2)&1\\
\cline{2-4} &1'&(\rr,\eps)(m,\eps')(h,\eps')&1\\
\hline 8/m&2^*&(\rr,\eps)(m,\ssr2)(h,\eps)&1\\
\cline{2-4} &1'&(\rr,\eps)(m,\eps')(h,\eps)&1\\
\end{longtable}

\end{document}